\documentclass[12pt,a4paper]{article}

\usepackage[T1]{fontenc}
\usepackage[latin1]{inputenc}
\usepackage[nosort]{cite}
\usepackage{color}
\usepackage[bulletsep]{collref}
\usepackage{graphicx}
\usepackage{bbm}
\usepackage{amsmath}
\usepackage{amssymb}
\usepackage{physics}
\usepackage{mathtools}
\usepackage{subfig}
\usepackage{verbatim}
\usepackage{multirow}
\usepackage{tikz}
\usepackage{amsthm}
\usepackage{fancyhdr}
\usepackage{wrapfig}
\usepackage{hyperref}
\usepackage{dsfont}
\usepackage{delimset} 
\usepackage{shuffle} 
\usepackage{mathtools}

\makeatletter \@addtoreset{equation}{section} \makeatother

\makeatletter
\let\old@startsection=\@startsection
\let\oldl@section=\l@section
\renewcommand{\@startsection}[6]{\old@startsection{#1}{#2}{#3}{#4}{#5}{#6\mathversion{bold}}}
\renewcommand{\l@section}[2]{\oldl@section{\mathversion{bold}#1}{#2}}
\makeatother

\makeatletter
\let\old@makecaption=\@makecaption
\def\@makecaption{\small\old@makecaption}
\makeatother

\let\oldPhi=\Phi
\let\oldPsi=\Psi
\let\oldGamma=\Gamma
\let\oldDelta=\Delta
\let\oldSigma=\Sigma
\let\oldTheta=\Theta
\let\oldPi=\Pi
\let\oldUpsilon=\Upsilon
\renewcommand{\Phi}{\mathnormal{\oldPhi}}
\renewcommand{\Psi}{\mathnormal{\oldPsi}}
\renewcommand{\Gamma}{\mathnormal{\oldGamma}}
\renewcommand{\Sigma}{\mathnormal{\oldSigma}}
\renewcommand{\Delta}{\mathnormal{\oldDelta}}
\renewcommand{\Theta}{\mathnormal{\oldTheta}}
\renewcommand{\Pi}{\mathnormal{\oldPi}}
\renewcommand{\Upsilon}{\mathnormal{\oldUpsilon}}

\newcommand{\superN}{\mathcal{N}}

\renewcommand{\Im}{\mathop{\mathrm{Im}}}

\newcommand{\gen}[1]{\mathrm{#1}}
\newcommand{\levo}[1]{ \gen{\widehat #1}}
\newcommand{\genfield}[1]{\mathbb{#1}}
\newcommand{\levofield}[1]{ \genfield{\widehat #1}}

\newcommand{\Li}{\mathrm{Li}}

\newcommand{\Duv}{\mathcal{D}_{uv}}
\newcommand{\Dvu}{\mathcal{D}_{vu}}

\newcommand{\Eval}{s} 
\makeatletter
\newlength{\apb@width}
\newcommand{\autoparbox}[2][c]{\settowidth{\apb@width}{#2}\parbox[#1]{\apb@width}{#2}}
\newcommand{\includegraphicsbox}[2][]{\autoparbox{\includegraphics[#1]{#2}}}
\makeatother

\ifx\genfrac\sdflkaj

\else

\fi
\newcommand{\sfrac}[2]{{\textstyle\frac{#1}{#2}}}
\newcommand{\half}{\sfrac{1}{2}}
\newcommand{\ihalf}{\sfrac{i}{2}}

\newcommand{\grp}[1]{\mathrm{#1}}

\makeatletter
\def\mr@ignsp#1 {\ifx\:#1\@empty\else #1\expandafter\mr@ignsp\fi}%
\newcommand{\multiref}[1]{\begingroup
\xdef\mr@no@sparg{\expandafter\mr@ignsp#1 \: }%
\def\mr@comma{}%
\@for\mr@refs:=\mr@no@sparg\do{\mr@comma\def\mr@comma{,}\ref{\mr@refs}}%
\endgroup}
\makeatother

\newcommand{\hypref}[2]{\ifx\href\asklfhas #2\else\href{#1}{#2}\fi}
\newcommand{\Secref}[1]{Section~\multiref{#1}}
\newcommand{\secref}[1]{section~\multiref{#1}}
\newcommand{\Appref}[1]{Appendix~\multiref{#1}}
\newcommand{\appref}[1]{appendix~\multiref{#1}}

\newcommand{\tabref}[1]{table~\multiref{#1}}

\renewcommand{\eqref}[1]{(\multiref{#1})}

\ifx\href\asklfhas\newcommand{\href}[2]{#2}\fi

\newcommand{\be}{\begin{eqnarray}}
\newcommand{\ee}{\end{eqnarray}}

\newcommand{\PDE}{\mathrm{PDE}}

\begin{document}

\thispagestyle{empty}

\begin{flushright}\footnotesize
\texttt{HU-EP-21/54}\\
\texttt{SAGEX-21-38-E}
\end{flushright}
\vspace{.2cm}

\begin{center}%
{\LARGE\textbf{\mathversion{bold}%
Yangian Ward Identities for \\
Fishnet Four-Point Integrals}\par}

\vspace{1cm}
{\textsc{Luke Corcoran${}^a$, Florian Loebbert${}^a$, Julian Miczajka${}^b$ }}
\vspace{8mm} \\
\textit{%
${}^a$Institut f\"{u}r Physik, Humboldt-Universit\"{a}t zu Berlin, \\
Zum Gro{\ss}en Windkanal 6, 12489 Berlin, Germany\\[5pt]
${}^b$Max-Planck-Institut f\"{u}r Physik, \\
F\"{o}hringer Ring 6, 80805 M\"{u}nchen, Germany
}
\vspace{.5cm}

\texttt{ \{corcoran,loebbert\}@physik.hu-berlin.de\\ miczajka@mpp.mpg.de}
%

\par\vspace{15mm}

\textbf{Abstract} \vspace{5mm}

\begin{minipage}{13cm}
We derive and study Yangian Ward identities for the infinite class of four-point ladder integrals and their Basso--Dixon generalisations. 
These symmetry equations follow from interpreting the respective Feynman integrals as correlation functions in the bi-scalar fishnet theory. 
Alternatively, the presented identities can be understood as anomaly equations for a momentum space conformal symmetry. 
The Ward identities take the form of inhomogeneous extensions of the partial differential equations defining the Appell hypergeometric functions. We employ a manifestly conformal tensor reduction in order to express these inhomogeneities in compact form, which are given by linear combinations of Basso--Dixon integrals with shifted dimensions and propagator powers. The Ward identities naturally generalise to a one-parameter family of $D$-dimensional integrals representing correlators in the generalised fishnet theory of Kazakov and Olivucci. When specified to two spacetime dimensions, the Yangian Ward identities decouple. Using separation of variables, we explicitly  bootstrap the solution for the conformal 2D box integral. The result is a linear combination of Yangian invariant products of Legendre functions, which reduce to elliptic $K$ integrals for an isotropic choice of propagator powers. We comment on differences in the transcendentality patterns in two and four dimensions and their relations to discontinuities. 
\end{minipage}
\end{center}

\newpage 

\tableofcontents
\bigskip
\hrule


\section{Introduction and Overview}

While quantum field theory represents one of the most advanced frameworks for our current description of nature, precise computational predictions still challenge whole communities of researchers. In particular, calculations at higher loop orders require more and more sophisticated technology and mathematical insight.
A rare example of quantum field theory structures in four dimensions, which are known analytically to all loop orders, are the scalar four-point fishnet integrals. Their analytic expressions were conjectured first by Basso and Dixon in \cite{Basso:2017jwq}, and as such the respective Feynman diagrams are also referred to as \textit{Basso--Dixon graphs}:
\begin{equation}
\includegraphicsbox{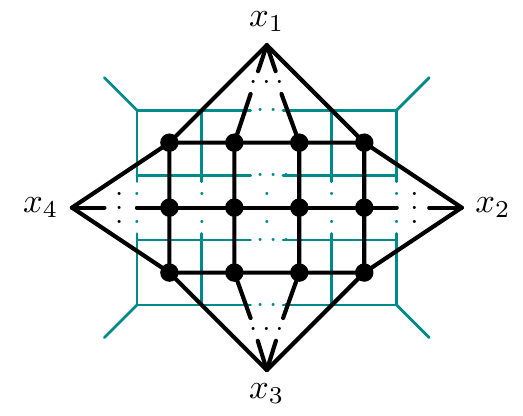}.
\end{equation}
Here the black diagram can be interpreted as a position space $(x_j)$ correlation function while the green diagram represents its dual momentum space $(p_j)$ graph with momenta defined via  $x_j-x_{j+1}=p_j$. 

When these diagrams are shrunk to a one-dimensional lattice, the corresponding \emph{ladder integrals} simplify substantially. Results for the conformal ladder graphs have been known since the works of \cite{Usyukina:1993ch} and represent a paradigm of Feynman integration and its connections to mathematics.
Up to a conformal weight, the ladder integrals are given by single-valued polylogarithms of the conformal variables $z$ and $\bar{z}$. The full Basso--Dixon graphs on the other hand take the form of particular determinants of precisely these ladder functions.

Remarkably, the fishnet integrals are in one-to-one correspondence with certain observables%
\footnote{They can alternatively be interpreted as correlation functions or as off-shell scattering amplitudes.}
 in a very special quantum field theory, whose name is inherited from the structure of its Feynman graphs. The so-called \emph{fishnet theory} was first proposed in \cite{Gurdogan:2015csr} and is defined by the bi-scalar Lagrangian
\begin{equation}\label{eq:LFN}
\mathcal{L}_{\text{FN}}=N_c\tr \left(-\partial_\mu X \partial^{\mu} \bar{X}-\partial_\mu Z \partial^{\mu} \bar{Z}+\xi^2 XZ\bar{X}\bar{Z}\right).
\end{equation}
Here $X$ and $Z$ denote scalar $\grp{SU}(N_c)$ matrix fields, and $\bar{X}$ and $\bar{Z}$ their Hermitian conjugates. Notably, the quartic interaction term $\text{tr}( XZ\bar{X}\bar{Z})$ lacks a Hermitian conjugate partner, rendering the theory defined by \eqref{eq:LFN} non-unitary. The upshot is a vast simplification of the Feynman diagrams in the planar limit. Omitting  matrix structures and factors of $N_\text{c}$, the position space Feynman rules are simply given by 
\begin{equation}\label{eq:rules}
\includegraphicsbox{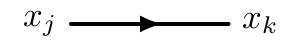} =\frac{1
}{x_{jk}^2},
\qquad\qquad
\includegraphicsbox{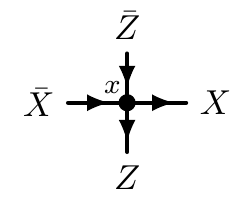} 
= \xi^2 \int \dd^4x,
\end{equation}
where $x_{jk}^\mu=x_j^\mu-x_k^\mu$.
Because of the chiral nature of the interaction vertex (we drop arrows in all other graphs), the possible planar Feynman diagrams one can draw for a given physical quantity are severely limited. For example, diagrams renormalising the mass of the scalars $X, Z$ and the coupling $\xi^2$ are simply absent. Although double-trace couplings are generated under the RG flow, there are fixed points where the theory becomes a true (logarithmic) conformal field theory~\cite{Fokken:2013aea,Grabner:2017pgm}. 

This conformality appears to be a major agent in the integrability of the theory. For example, diagrams contributing to the anomalous dimension of the BMN vacuum operator $\text{tr} (X^L)$ take an iterative wheel-like structure, and the associated graph-building operator has been shown to be intimately related to a non-compact conformal spin chain \cite{Gromov:2017cja}. 

Another example is given by the correlation functions of \cite{Chicherin:2017cns,Chicherin:2017frs}, which in the planar limit are represented by single conformal Feynman integrals in perturbation theory. This afforded the unique opportunity to study the integrability of a class of correlation functions directly at the level of Feynman integrals. In particular, the following fishnet Feynman graphs were shown to be \textit{Yangian invariant}, that is annihilated by a particular representation of the conformal Yangian algebra $Y(\mathfrak{so}(1,5))$:
\begin{equation}
\includegraphicsbox{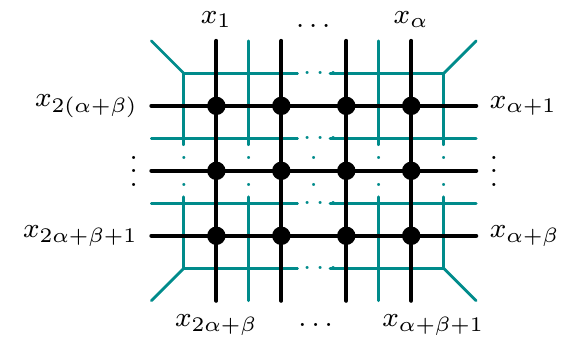}.
\end{equation}
Yangian invariance lies at the heart of quantum integrability, and in this case reflects the invariance of the Feynman integrals under \textit{two} independent copys of the conformal algebra. These are given by the standard representations of the conformal generators in coordinate and dual momentum space, whose closure gives rise to the Yangian \cite{Drummond:2009fd,Loebbert:2020glj}. The simplest correlator of fishnet type is represented by the well-known conformal four-point box integral
 \begin{equation}\label{eq:4ptcorrelator}
 \langle\text{tr}(Z(x_1)\bar{X}(x_2)\bar{Z}(x_3)X(x_4))\rangle=\includegraphicsbox[scale=1]{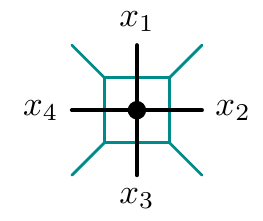}=\frac{1}{x_{13}^2x_{24}^2}\frac{\text{BW}(z,\bar{z})}{z-\bar{z}},
\end{equation}
which evaluates essentially to the famous Bloch--Wigner function \cite{Usyukina:1993ch,Zagier:2007knq} of the conformal variables $z$ and $\bar{z}$ defined by
$z\bar{z}={x_{12}^2x_{34}^2}/{x_{13}^2x_{24}^2}$,
and
$(1-z)(1-\bar{z})={x_{14}^2x_{23}^2}/{x_{13}^2x_{24}^2}$.
%
The next simplest correlator is represented by the six-point double box integral with generic external kinematics:
\begin{equation}
\left\langle\text{tr}(Z(x_1)Z(x_2)\bar{X}(x_3) \bar{Z}(x_4) \bar{Z}(x_5)X(x_6))\right\rangle
=\includegraphicsbox{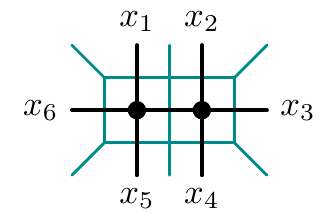}.
\label{eq:6ptcorrelator}
\end{equation}
 Up to a conformal weight, this integral is a function of 9 cross ratios $\psi_{3,3}(u_1,\dots,u_9)$
for which in some region of kinematic space a formal series representation was found~\cite{Ananthanarayan:2020ncn}. In a 7-cross ratio limit this integral has been expressed in terms of elliptic polylogarithms~\cite{Bourjaily:2017bsb,Kristensson:2021ani}. The fully off-shell case has not yet been given as such, but should in principle be expressible in terms of the same function class~\cite{Bloch:2021hzs}. 
The sub-family of fishnet Feynman integrals with one-dimensional lattice structure, which generalises the above box and double box integral to $2(n+1)$ external points, has been dubbed \emph{train tracks} due to the shape of their graphs in momentum space (green graph):
\begin{equation}
\includegraphicsbox{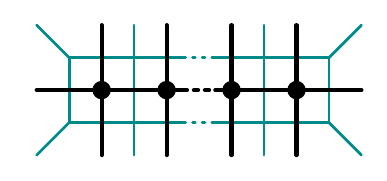}.
\label{eq:TrainTracks}
\end{equation}
For on-shell kinematics, these integrals were conjectured to be associated with Calabi--Yau geometries \cite{Bourjaily:2018ycu}.

The Yangian invariance of the above fishnet integrals implies differential equations in the cross ratios for the respective conformal functions. Since integrability is typically highly constraining, this fact initiated the idea to bootstrap Feynman integrals using only these differential equations, and a few extra discrete symmetries \cite{Loebbert:2019vcj}. This was achieved with great success for the $D$-dimensional conformal box integral as well as in various other one-loop setups \cite{Loebbert:2020glj,Loebbert:2020hxk,Corcoran:2020epz}. Moreover, this approach triggered interesting developments in the theory of Mellin--Barnes integrals \cite{Ananthanarayan:2020ncn,Ananthanarayan:2020fhl,Ananthanarayan:2020xpd}. 
At two loops, however, it is yet unclear how to employ the Yangian bootstrap to determine the double box \eqref{eq:6ptcorrelator} in terms of appropriate function classes. 

In general, the higher-point fishnet correlators depend on more and more external points and thus on more and more conformal cross ratios. Despite their Yangian invariance it remains unclear how to express them in terms of known functions. In fact, exploring the set of functions relevant to higher loop integrals is currently an active area of research.
The above Basso--Dixon correlators represent a very interesting kinematical simplification of generic square fishnets to four-point correlators:
\begin{equation}\label{eq:4ptlimit}
\includegraphicsbox{FigBDGraphSomeLabels.pdf} \longrightarrow \includegraphicsbox{FigBDGraphLimitsSomeLabelsIntro.pdf}.
\end{equation}
In particular, this coincidence limit implies that the resulting conformal four-point integral merely depends on two conformal variables, while the number of loop integrations may be arbitrarily high. Under this limit the above train tracks \eqref{eq:TrainTracks} are turned into the ladder integrals, which furnish the building blocks for the elegant Basso--Dixon determinant formula.

It is a natural question to ask for the reason behind the simplicity of these four-point integrals and indeed integrability appears to be responsible. Basso--Dixon integrals represent fishnet correlators with specific open boundary conditions and a fruitful example for the application of integrability techniques to correlation functions in conformal field theory. These include the BMN/flux tube pictures of \cite{Basso:2017jwq}, and the generalisation of Sklyanin's separation of variables \cite{Sklyanin:1995bm} to non-compact, higher-rank spin chains \cite{Derkachov:2019tzo,Derkachov:2021ufp}. Despite this, the eventual analytic proof for the Basso--Dixon formula \cite{Basso:2021omx} did not directly appeal to techniques of integrability, but rather involved exact methods of summation/integration.

Although after the coincidence limit the Basso--Dixon integrals (beyond the box) are not Yangian \emph{invariant}, taking the four-point limit \eqref{eq:4ptlimit} of the invariance equation for the left hand side implies that the Yangian contains some information about the integrals on the right hand side. It is the purpose of this paper to investigate what this information is. 
By performing this somewhat technical limit, we find that the Yangian level-one momentum generator $\levo{P}^\mu$ does not annihilate the Basso--Dixon correlators, but rather returns a specific linear combination of analogous correlators, where one of the fields is replaced by a descendant $\Phi \rightarrow \partial^{\mu}_{x_i}\Phi$. At the level of Feynman integrals, this leads to vector integrals contributing on the right hand side of the Yangian Ward identity. 
We devised a conformal tensor reduction to express the Yangian equations as formal identities for the conformal Basso--Dixon functions $\phi_{\alpha\beta}(u,v)$, where $\alpha$ and $\beta$ is the width and height of the Basso--Dixon diagram respectively. Schematically these take the form
\begin{equation}
\Duv \phi_{\alpha\beta}=\mathrm{d}^+ \mathcal{A}\phi_{\alpha\beta}.
\label{eq:FormalDuvOnphi}
\end{equation}
Here $\Duv$ is a differential operator in the conformal cross ratios $u$ and $v$ (or alternatively $z,\bar z$), $\mathrm{d}^+$ is a dimension-raising operator which shifts the dimension $D\rightarrow D+2$, and $\mathcal{A}$ is a linear combination of operators which raise specific propagator powers of the integral. For the ladders we provide an algorithmic way to describe the operators $\mathcal{A}$. 
For the case of the simple box integral the inhomogeneity on the right hand side of the above Ward identity vanishes.

We emphasise that the identity \eqref{eq:FormalDuvOnphi} is formal: $\mathrm{d}^+ \mathcal{A}$ is not defined on the conformal function $\phi_{\alpha\beta}$, but rather on the overall Feynman integral. However, both $\phi_{\alpha\beta}$ and $\mathrm{d}^+ \mathcal{A}\phi_{\alpha\beta}$ belong to the same family of integrals with Basso--Dixon topology and generic propagator powers. We can thus understand the above Yangian equation in analogy to contiguous relations for hypergeometric functions. A simple example is the following identity for the Gau{\ss} hypergeometric series ${}_2F_1$:
\begin{equation}
\partial_z \, {}_2F_1(a,b,c|z)=\frac{a b}{c}\,  {}_2F_1(a+1,b+1,c+1|z),
\label{eq:contiguous}
\end{equation}
which relates the action of a differential operator to a parameter shift in analogy to \eqref{eq:FormalDuvOnphi}. In fact, hypergeometric functions like the above are well known to describe Feynman integrals with spacetime dimensions and propagator powers entering into the parametric arguments of these functions.

The identities \eqref{eq:FormalDuvOnphi} generalise naturally to analogous correlators in the $D$-dimensional fishnet theory of \cite{Kazakov:2018qbr}. This theory is both non-unitary and non-local, but appears to retain integrability in the planar limit. Moreover, the corresponding Feynman graphs also enjoy a Yangian symmetry. For $D=2$ we show that the Ward identities split up. The corresponding homogeneous Ward identity for the box becomes a pair of \textit{ordinary} differential equations in $z$ and $\bar{z}$, which are easily solved by separation of variables. The result takes the form of a single-valued combination of Legendre $P$ and $Q$ functions, which reduce to elliptic $K$ integrals for an isotropic choice of propagator powers. Comparing the interplay between discontinuities and transcendentality properties of the respective solutions reveals interesting differences to the box integral in four dimensions.

The paper is organised as follows. In \secref{sec:FishnetToBasso} we recall the Yangian Ward identities for fishnet Feynman graphs, and from these we derive the Ward identities for the Basso--Dixon integrals. In \secref{correlatorfeynman} we describe how to express these identities in terms of Feynman integrals. In \secref{sec:tensordecomp} we describe a method to rewrite the appearing vector integral coefficients in terms of higher dimensional scalar integrals, and conformalise the resulting expressions. Using this, in \secref{sec:YangWard} we express the Yangian Ward identities as single operatorial equations for the Basso--Dixon functions. In \secref{sec:generalD} we explain the generalisation to the $D$-dimensional fishnet theory \cite{Kazakov:2018qbr}. Finally, in \secref{sec:separation} we describe the separability of the equations in $D=2$.

We have verified all of the identities shown in this paper by an explicit numerical integration of the appearing conformal integrals, when analytic expressions were not available. For the numerical integration of the conformal Feynman parameter integrals we employed the Mathematica function NIntegrate. The precision of the agreement depends on the loop order of the appearing integrals; the checks at one, two, three, and four-loop orders have typical relative precisions of $10^{-9}, 10^{-6}, 10^{-3}$, and $10^{-1}$, respectively. We have checked all of the relations for at least 10 different numerical configurations in Euclidean kinematics, for which the integrals vary over multiple magnitudes.

\section{From Fishnets to Basso--Dixon Correlators}\label{sec:FishnetToBasso}

In this section we demonstrate how the previously observed Yangian symmetry of $n$-point fishnet integrals can be turned into a symmetry statement for four-point Basso--Dixon integrals.

\subsection{Yangian Invariant Fishnets}
\label{sec:YangInvFish}

Our starting point is the Yangian symmetry of fishnet Feynman integrals \cite{Chicherin:2017cns,Chicherin:2017frs}. In particular, we are interested in square fishnets given by a lattice of propagators with $\alpha\times \beta$ integration vertices:
\begin{equation}
\tilde{I}_{\alpha\beta}=\includegraphicsbox{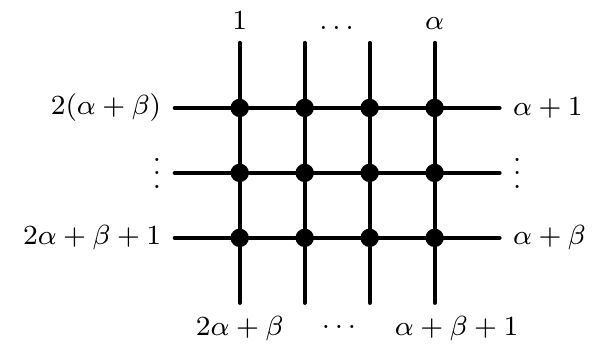}
\end{equation}
These are annihilated by the conformal Yangian level-zero and level-one generators  $\gen{J}^A$ and $\levo{J}^A$, which act as differential operators on the coordinates $x^\mu_j\in\mathbb{R}^4$ for $ j=1,2,\dots, 2(\alpha+\beta)$:
\begin{equation}
\gen{J}^A \tilde{I}_{\alpha\beta}=0,
\qquad\qquad
\levo{J}^A  \tilde{I}_{\alpha\beta}=0.
\label{eq:YangInvEqs}
\end{equation}
 The level-zero generators $\gen{J}^A$ take the form
 \begin{equation}
 \gen{J}^A=\sum_{j=1}^n\gen{J}^A_j,
 \end{equation}
where $n=2(\alpha+\beta)$ denotes the total number of external points. The densities $\gen{J}^A_j$ read
\begin{align}
\gen{P}^{\mu}_j &= -i \partial_j^{\mu}, &
\gen{L}_j^{\mu \nu} &= i x_j^{ \mu} \partial_j^{ \nu} 
	- ix^{ \nu}_j \partial_j^{\mu}, \nonumber \\
\gen{D}_j &= -i x_{j  \mu} \partial_j^{ \mu} - i \Delta_j , &
\gen{K}^{\mu}_j &= -i \brk*{ 2 x_j^{\mu} x_j^{ \nu} 
		- \eta ^{\mu  \nu} x_j ^2 }\partial_{j, \nu}
	- 2i \Delta_j x_j^{\mu} ,
\label{eq:genrep}
\end{align}
where $\partial_j^{\mu}\coloneqq \partial_{x_j}^\mu$ These densities also furnish the building blocks for the corresponding level-one generators, defined as
\begin{equation}
\levo{J}^A
=\half f^A{}_{BC} \sum_{k=1}^n  \sum_{j=1}^{k-1} \gen{J}^C_j \gen{J}^B_k+\sum_{j=1}^n \Eval_j \gen{J}^A_j.
\label{eq:LevoDef}
\end{equation}
Here the $f^A{}_{BC}$ denote the structure constants of the conformal algebra $\mathfrak{so}(1,5)$. The summation indices run over the external fields of the Feynman integral. The symbols $s_j$ represent the so-called evaluation parameters which we choose to guarantee Yangian invariance according to the rules of \cite{Chicherin:2017cns}:
\begin{equation}
s_j=(\underbracket{0,\dots,0}_{\alpha},\underbracket{-1,\dots,-1}_{\beta},\underbracket{-2,\dots,-2}_\alpha,\underbracket{-3,\dots,-3}_{\beta})_j,
\qquad j=1,\dots,n.
\end{equation}
Since we consider external legs corresponding to complex scalar matrix fields $\Phi_a\in \{ X, Z\}$, we will typically be interested in scaling dimensions entering the representation of the conformal generators being $\Delta_j=1$ for $j=1,\dots,n$. 

Conformal symmetry, i.e.\ invariance under the level-zero generators $\gen{J}^A$, implies that the above integrals can be written in the form
\begin{equation}
\tilde{I}_{\alpha\beta}=V_{\alpha\beta}\, \tilde{\phi}_{\alpha\beta}(u_j),
\end{equation}
with $V_{\alpha\beta}$ being a prefactor that carries the conformal weight of the integral and $\tilde{\phi}_{\alpha\beta}$ denoting a conformal function which only depends on a number of conformal cross ratios $u_j$.

To get a better idea of what the level-one invariance equation in \eqref{eq:YangInvEqs} means, let us specify the level-one generator $\levo{J}^A$ to the level-one momentum generator:
\begin{equation}
\levo{P}^\mu = \sfrac{i}{2} \sum_{j<k=1}^n
	\brk*{\gen{P}_j^\mu \gen{D}_k + \gen{P}_{j\nu} \gen{L}_k^{\mu\nu} 
		- (j\leftrightarrow k) }  
	+ \sum_{j=1}^n s_j \gen{P}_j^\mu 
	=
		 \sum_{j<k=1}^n\levo{P}^{\mu}_{jk} 
		 	+ \sum_{j=1}^n s_j \gen{P}_j^\mu .
\label{eq:Phatexpl}
\end{equation}
Here, using the densities for the conformal generators given in \eqref{eq:genrep}, one finds
\begin{align}
\levo{P}^{\mu}_{jk} = \sfrac{i}{2}\brk*{ T^{\nu \mu \rho} \partial_{j, \rho} \partial_{k, \nu}
+ \Delta_j \partial_{k}^{\mu} - \Delta_k  \partial_{j}^{\mu} } ,
\qquad
T^{\nu \mu \rho}=x_{jk}^\nu \eta^{\mu \rho}+x_{jk}^\rho \eta^{\mu \nu}-x_{jk}^\mu \eta^{\nu \rho} .
\label{eq:Phatjk}
\end{align}
Now the invariance of the above diagrams under the level-one generators can be written as
\begin{equation}
0=\levo{P}^\mu \tilde{I}_{\alpha\beta}=V_{\alpha\beta} \sum_{j<k=1}^n \frac{x_{jk}^\mu}{x_{jk}^2}\PDE_{jk}\,\tilde{\phi}_{\alpha\beta}(u_j),
\label{eq:YangianInvarianceEquation}
\end{equation}
where $\PDE_{jk}$ denotes some differential operators in the conformal cross ratios $u_j$. At least for lower numbers of points it can be argued that the vectors $x_{jk}^\mu/{x_{jk}^2}$ are independent such that Yangian invariance implies a system of differential equations for the conformal function \cite{Loebbert:2019vcj}: 
\begin{equation}
\PDE_{jk} \,\tilde{\phi}_{\alpha\beta}=0, \qquad 1\leq j<k\leq n. 
\end{equation}
We stress that these are \emph{homogeneous} partial differential equations, a fact that will change when considering coincidence limits of the external points.
Note also that in the homogeneous case the Yangian equations for the remaining level-one generators are easily obtained by commuting the level-one momentum generator with the level-zero generators of the conformal algebra and using that $\tilde I_{\alpha\beta}$ is a level-zero invariant.

\subsection{Four-Point Coincidence Limit}
\label{sec:CoincLimit}
The purpose of this paper is to determine the imprint that the Yangian invariance equation \eqref{eq:YangianInvarianceEquation} for the fishnet correlators leaves on Basso--Dixon correlators, which emerge from the former in a coincident point limit. While the limit can be carried out straightforwardly for the correlators, there is a subtlety in the limit of the invariance equation, which is due to the fact that the Yangian generators in their differential form do not commute with the coincidence limit. 
 
In order to tackle these issues with the coincidence limit, it is useful to interpret the above Yangian symmetry of Feynman integrals as Ward identities for correlation functions in the bi-scalar fishnet theory with complex scalars $X$ and $Z$. Then we can distinguish the representation of the conformal algebra on coordinates $\gen{J}^A$ and the representation on the fields $\genfield{J}^A$.\footnote{We are very grateful to Dennis M\"uller for important insights on these points.}
On a single leg, the field and coordinate representation are essentially related by a minus sign,%
\footnote{This ensures consistent commutation relations due to $\genfield{J}^A \genfield{J}^B \Phi = -\genfield{J}^A\gen{J}^B \Phi = \gen{J}^B \gen{J}^A \Phi$ (see \cite{Muller:2018okc} for more details).}
\begin{equation}
\genfield{J}^A \Phi(x) = -\gen{J}^A(\Delta) \Phi(x),
\label{eq:genfieldtogen}
\end{equation}
where the scaling dimension $\Delta$ in the coordinate representation generator is dictated by the scaling dimension of the field $\Phi$. 

However, when acting on multiple fields, the field representation carries an additional label that encodes on which of the \emph{fields} the operator acts (in contrast to the coordinate representation which carries a label that encodes which \emph{coordinate} the operator acts on), i.e.~
\begin{equation}
\genfield{J}^A_k (\Phi_1(x_1) \dots \Phi_n(x_n)) = -\Phi_1(x_1)\dots (\gen{J}^A_k \Phi_k(x_k)) \dots \Phi_n(x_n).
\end{equation}
The distinction between these two representations becomes non-trivial as soon as we consider products of fields that are evaluated at the same coordinate. As an example, consider the action of an operator in the field representation on the product of two fields that depend on the same coordinate:
\begin{equation}
\genfield{J}_1^A (\Phi_1(x_1) \Phi_2(x_1)) = -(\gen{J}_1^A \Phi_1(x_1)) \Phi_2(x_1).
\end{equation}
In contrast, an operator in the coordinate representation can not distinguish between the two fields in this product and by the Leibniz rule naturally acts on both fields%
\footnote{Note that the scaling dimensions only ever appear multiplicatively in the conformal generators \eqref{eq:genrep}. Hence, while they could be shuffled around on the right hand side of this equation, we have picked a physically sensible representation for this identity.}
\begin{equation}
\gen{J}_1^A(\Delta_1+\Delta_2)(\Phi_1(x_1)\Phi_2(x_1)) = (\gen{J}_1^A(\Delta_1) \Phi_1(x_1)) \Phi_2(x_1) + \Phi_1(x_1) (\gen{J}_1^A(\Delta_2) \Phi_2(x_1)).
\end{equation}
Therefore, the field representation allows us to act separately on different fields that depend on the same coordinate. As we will discuss next, this provides us with a natural extension of the above Yangian symmetry of square fishnet diagrams to diagrams with coincident external legs.
 
 Before the coincidence limit, the Yangian level-one symmetry may be written in two equivalent forms of the following level-one Ward identity for the fishnet correlator:
\begin{equation}
\levofield{J}^A \includegraphicsbox{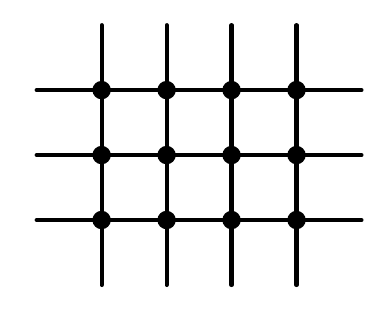} =\levo{J}^A \includegraphicsbox{FigBDGraphNoLabels.pdf}=0.
\label{eq:YangianWard}
\end{equation}
Here the level-one generator in the field representation takes the form 
\begin{equation}
\levofield{J}^A=\levofield{J}^A_\text{bi}-\sum_{j=1}^n \Eval_j \genfield{J}^A_j
=\half f^A{}_{BC} \sum_{k=1}^n  \sum_{j=1}^{k-1} \genfield{J}^C_j \genfield{J}^B_k-\sum_{j=1}^n \Eval_j \genfield{J}^A_j.
\label{eq:defllevofield}
\end{equation}
Naturally, since all fields in the fishnet correlators depend on separate coordinates, the field and coordinate representation are virtually indistinguishable.

Having formulated the Yangian Ward identities for generic square fishnet correlators, we would like to understand what they imply for the four-point Basso--Dixon (BD) limit of these correlators. This limit consists of taking the external points $x_j$ on each of the four sides of the square fishnet to be coincident according to
\begin{align}
&x_j\to x_1, \qquad \text {for} \quad j=1,\dots, \alpha,
\\ \notag
&x_j\to x_2, \qquad \text {for} \quad j=\alpha+1,\dots, \alpha+\beta,
\\ \notag
&x_j\to x_3, \qquad \text {for} \quad j=\alpha+\beta+1,\dots, 2\alpha+\beta,
\\ \notag
&x_j\to x_4, \qquad \text {for} \quad j=2\alpha+\beta+1,\dots, 2(\alpha+\beta).
\end{align}
Graphically the limit is illustrated by the following figure:
\begin{equation}
\includegraphicsbox{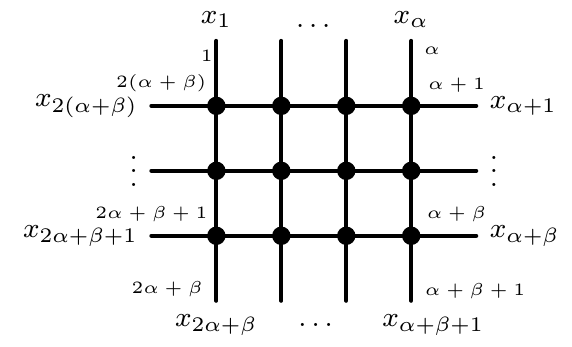}
\quad\to\quad
\includegraphicsbox{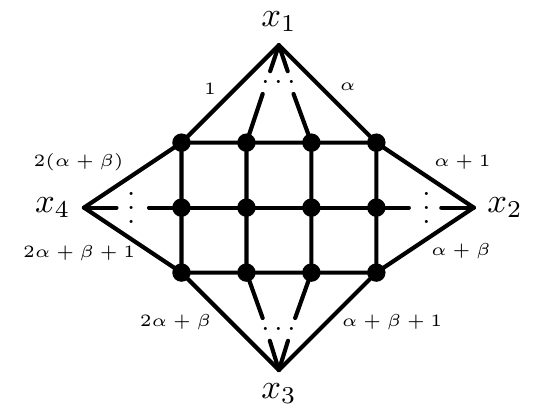}.
\end{equation}

Importantly, while the limit of the Yangian Ward identity in the coordinate representation suffers from the subtleties described at the beginning of this section, it trivially commutes with the level-one generators in the field representation:
\begin{equation}
\levofield{J}^A  \includegraphicsbox[scale=1]{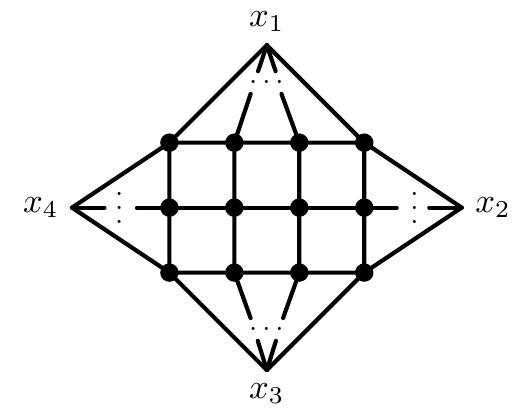} = 0.
\end{equation} 
However, translating the Ward identity into a differential equation for Feynman integrals becomes more subtle. 
To illustrate this point, consider the level-one generator acting on the product of two fields whose coordinate arguments are taken to be coincident:
\begin{equation}
\levofield{J}^A \Phi_1(x_1)\Phi_2(x_1)=\half f^A{}_{BC}[\genfield{J}^C_1\Phi_1(x_1)][\genfield{J}^B_2\Phi_2(x_1)]\neq\levo{J}^A \Phi_1(x_1)\Phi_2(x_1).
\end{equation}
The last inequality implies that we cannot simply replace the field representation $\genfield{J}$ of the conformal (Yangian) generators by the coordinate representation $\gen{J}$. However, we can replace the generators in the middle term of the above equation, where each conformal generator acts on a single field: 
\begin{equation}
\half f^A{}_{BC}[\genfield{J}^C_1\Phi_1(x_1)][\genfield{J}^B_2\Phi_2(x_1)]
=
\half f^A{}_{BC}[\gen{J}^C_1\Phi_1(x_1)][\gen{J}^B_1\Phi_2(x_1)].
\end{equation}
When inserting the above identities into a correlator, this will lead to correlators including descendant fields. Since the conformal generators are represented by first-order differential operators with vector indices, we will thus find true vector integrals contributing to the Ward identity, where these derivatives act on single propagators, but cannot be pulled in front of the whole integral. 

Notably, in other contributions to the Yangian Ward identity, the field representation of the level-one generators can be replaced by the coordinate representation, for instance 
\begin{align}
&\half f^A{}_{BC}[\genfield{J}^C_1\Phi_1(x_1)]\Phi_2(x_1)[\genfield{J}^B_3\Phi_3(x_2)]
+
\half f^A{}_{BC}\Phi_1(x_1)[\genfield{J}^C_2\Phi_2(x_1)][\genfield{J}^B_3\Phi_3(x_2)]
\nonumber\\
&\qquad=
\half f^A{}_{BC}[\gen{J}^C_1\Phi_1(x_1)\Phi_2(x_1)][\gen{J}^B_2\Phi_3(x_2)],
\end{align}
which is due to the fact that the generators $\gen{J}^A_j$ are first order differential operators and where we assume that the $\Delta_j$ are equal for fields evaluated at equal points.

Similarly, local contributions to the Yangian level-one generator, like the terms multiplying the evaluation parameters $\Eval_a$, can be replaced by the coordinate space generators as e.g.\ 
\begin{equation}
\sum_{a=1}^{\alpha+1} \Eval_a \genfield{J}^A _a \Phi_1(x_1)\dots \Phi_\alpha(x_1)\Phi_{\alpha+1}(x_2)
=
\brk*{\Eval_1 \gen{J}^A _1+\Eval_2 \gen{J}^A _2} \Phi_1(x_1)\dots \Phi_\alpha(x_1)\Phi_{\alpha+1}(x_2),
\end{equation}
as long as the evaluation parameters associated with fields situated at the same point $x_1$ are equal, i.e.\ in the above example $s_1=\dots=s_\alpha$.

With these points in mind, we can now write the above Ward identity for an $\alpha\times \beta$ Basso--Dixon graph in the form\footnote{We always consider single-trace correlation functions. Here we omit the trace for brevity.}
\begin{align}
0&=\levofield{J}^A |_{1n}\langle\Phi_1(x_1)\Phi_2(x_1)\dots \Phi_{n-1}(x_4)\Phi_n(x_4)\rangle
=
\levo{J}^A |_{14}\langle\Phi_1(x_1)\Phi_2(x_1)\dots \Phi_{n-1}(x_4)\Phi_n(x_4)\rangle
\nonumber\\
+&
\half \sum_{a<b=1}^\alpha f^A{}_{BC} \langle\Phi_1(x_1)\dots [\gen{J}_1^C\Phi_a(x_1)]\dots [\gen{J}_1^B\Phi_b(x_1)]\dots\Phi_\alpha(x_1)\Phi_{\alpha+1}(x_2)\dots \Phi_n(x_4)\rangle
\nonumber\\
+&
\half \sum_{a<b=\alpha+1}^{\alpha+\beta} f^A{}_{BC} \langle\Phi_1(x_1)\dots\Phi_\alpha(x_1)\Phi_{\alpha+1}(x_2)\dots  [\gen{J}_2^C\Phi_a(x_2)]\dots [\gen{J}_2^B\Phi_b(x_2)] \dots
 \Phi_n(x_4)\rangle
\nonumber\\
+& (\text{two similar}).
\end{align}
Here the notation $|_{1n}$ indicates the range of the summations in the definition of the level-one generator \eqref{eq:defllevofield}.
This is the consequence of Yangian symmetry for the infinite class of Basso--Dixon integrals.

In order to understand what this identity means explicitly, let us again specify the generators $\levofield{J}^A$ and $\levo{J}^A$ to the level-one momentum operator $\levofield{P}^\mu$ and $\levo{P}^\mu$, respectively, see \eqref{eq:Phatexpl}.
With these expressions the above level-one Ward identity becomes
\begin{align}\label{eq:Phatgen}
0&=\levo{P}^\mu |_{14}\langle\Phi_1(x_1)\Phi_2(x_1)\dots \Phi_{n-1}(x_4)\Phi_n(x_4)\rangle
\\
+&
\ihalf \sum_{a<b=1}^\alpha 
\big\{
\langle\Phi_1(x_1)\dots [\Delta_a\Phi_a(x_1)]\dots [\partial_1^\mu \Phi_b(x_1)]\dots\Phi_\alpha(x_1)\Phi_{\alpha+1}(x_2)\dots \Phi_n(x_4)\rangle
\nonumber\\
&\qquad\quad
- \langle\Phi_1(x_1)\dots [\partial_1^\mu\Phi_a(x_1)]\dots [\Delta_b \Phi_b(x_1)]\dots\Phi_\alpha(x_1)\Phi_{\alpha+1}(x_2)\dots \Phi_n(x_4)\rangle
\big\}
\nonumber\\
+&
\ihalf \sum_{a<b=\alpha+1}^{\alpha+\beta} 
\big\{
\langle\Phi_1(x_1)\dots\Phi_\alpha(x_1)\Phi_{\alpha+1}(x_2)\dots  [\Delta_a\Phi_a(x_2)]\dots [\partial_2^\mu \Phi_b(x_2)] \dots \Phi_{\alpha+\beta}(x_2)\dots \Phi_n(x_4)\rangle
\nonumber\\
&\qquad\quad
-\langle\Phi_1(x_1)\dots\Phi_\alpha(x_1)\Phi_{\alpha+1}(x_2)\dots  [\partial_2^\mu \Phi_a(x_2)]\dots [\Delta_b \Phi_b(x_2)] \dots \Phi_{\alpha+\beta}(x_2)\dots \Phi_n(x_4)\rangle
\big\}
\nonumber\\
+& (\text{two similar}).
\nonumber
\end{align}

Notably, due to the anti-symmetrisation $j\leftrightarrow k$ in \eqref{eq:Phatexpl} and the symmetry of $T^{\nu\mu\rho}$ in $\nu$ and $\rho$, the $T^{\nu\mu\rho}$-contribution to $\levo{P}^\mu$ drops out of the last four lines of the above equations.

 If we assume that the above fields have all distinct scaling dimensions $\Delta_a$, the above is the final form of our Ward identity. Note, however, that this does not correspond to the above bi-scalar fishnet theory! However, this finds application in the case of the $D$-dimensional generalisation of the fishnet theory of \cite{Kazakov:2018qbr}, see the discussion in \secref{sec:generalD}.
 
 In the four-dimensional fishnet theory we consider ordinary scalar fields with $\Delta_a=1$. Then the coincidence limit implies that the scaling dimensions $\Delta_j$ entering the coordinate representation $\levo{P}^\mu |_{14}$ in the above expression take the values
\begin{align}
\Delta_1&= \Delta_3 = \sum_{a=1}^\alpha \Delta_a=\alpha, & \Delta_2&=\Delta_4=\sum_{a=1}^\beta \Delta_a=\beta.
\end{align}
This is consistent with the choice of scaling dimensions that implies level-zero invariance of the Basso--Dixon integrals:
\begin{equation}
\gen{J}^A(\Delta_j) I_{\alpha \beta}=0.
\end{equation}
$I_{\alpha\beta}$ are defined explicitly in \eqref{eq:ScalarIntegrals}. Using $\Delta_a=1$ (and multiplying by an overall factor $2i$), we evaluate one sum of each double sum to find
\begin{align}
&2i\levo{P}^\mu |_{14}\langle\Phi_1(x_1)\Phi_2(x_1)\dots \Phi_{n-1}(x_4)\Phi_n(x_4)\rangle=
\label{eq:PhatIdentityCorr}
\\
&
 \sum_{a=1}^\alpha (2a-\alpha-1)\big\{
 \langle\Phi_1(x_1)\dots [\partial_1^\mu \Phi_a(x_1)]\dots\Phi_\alpha(x_1)\Phi_{\alpha+1}(x_2)\dots \Phi_n(x_4)\rangle
\nonumber\\
&\qquad +
\langle\Phi_1(x_1)\dots 
 \Phi_{\alpha+\beta+1}(x_3)\dots  [\partial_3^\mu \Phi_{a+\alpha+\beta}(x_3)] \dots \Phi_{2\alpha+\beta}(x_3)
 \dots \Phi_n(x_4)\rangle
 \big\}
 \nonumber\\
+&
 \sum_{a=1}^{\beta} (2a-\beta-1)
 \big\{
 \langle\Phi_1(x_1)\dots\Phi_\alpha(x_1)\Phi_{\alpha+1}(x_2)\dots  [\partial_2^\mu \Phi_{a+\alpha}(x_2)] \dots \Phi_{\alpha+\beta}(x_2)\dots \Phi_n(x_4)\rangle
 \nonumber\\
 &\qquad
 +\langle\Phi_1(x_1)\dots \Phi_{2\alpha+\beta}(x_3)
 \Phi_{2\alpha+\beta+1}(x_4)\dots  [\partial_4^\mu \Phi_{a+2\alpha+\beta}(x_4)] \dots \Phi_{2(\alpha+\beta)}(x_4)
\big\}.
\nonumber
\end{align}
Hence, the Yangian Ward identity implies that acting with the four-point coordinate space level-one momentum generator on a Basso--Dixon graph yields a combination of correlators of $n-1$ scalar fields $\Phi_a$ and a single descendent field $\partial^\mu \Phi_b(x)$. 
We will refer to the left hand side of \eqref{eq:PhatIdentityCorr} as the \textit{differential} contribution to the Ward identity, and the right hand side as the \textit{vector} contribution. In the next section we explicitly evaluate both sides of the equation as a constraint on conformal functions.

Note that one can obtain the analogue of the above vector contribution to the Yangian relations for the other level-one generators from the conformal algebra. As opposed to the Yangian equation for the invariants discussed in \secref{sec:YangInvFish}, see \eqref{eq:YangianInvarianceEquation}, we now have a non-trivial right hand side to deal with. For instance we could extract the action of $\levo{D}$ and $\levo{L}^{\mu\nu}$ from contractions of the following algebra relation evaluated on $I_{\alpha\beta}$, using that $\levo{P}^\nu I_{\alpha\beta}$ is given by \eqref{eq:PhatIdentityCorr}:
\begin{align}
2i \brk*{\eta^{\mu\nu} \levo{D}-\levo{L}^{\mu\nu}}I_{\alpha\beta}
=\comm{\gen{K}^\mu}{\levo{P}^\nu}I_{\alpha\beta}
=\gen{K}^\mu \levo{P}^\nu I_{\alpha\beta}
\end{align}
Here we have used $\gen{K}^\mu I_{\alpha\beta}=0$. Hence, for example the anomalous Ward identity for the level-one dilatation operator $\levo{D}$ can be obtained from the $\levo{P}^\nu$ identity by action with $\gen{K}^\mu$ and contraction with $\eta^{\mu\nu}$. From the explicit form of the inhomogeneity for the integrals $I_{\alpha\beta}$ (see  \eqref{eq:alphabetavectordecomp} in the below \secref{sec:WI}), which is split into conformally invariant functions multiplied by certain weight vectors, it is clear that the special conformal generator acts non-trivially only on the weight vectors. Therefore, all of the level-one Ward identities contain the same information about the conformal functions.

The Yangian generators at level two and higher can be obtained by iterative commutation of the lower level generators. Since the $I_{\alpha\beta}$ are not annihilated by e.g.\ the level-one generators, the analysis of the respective Ward identities will be more involved. However, they are automatically satisfied once level-one symmetry is assumed and hence provide no independent information.

\subsection{Examples}
To illustrate \eqref{eq:PhatIdentityCorr} let us display a few simple cases graphically. 

\paragraph{Box.}
In the simplest case of the box integral with $\alpha=\beta=1$ the vector contribution on the right hand side of \eqref{eq:PhatIdentityCorr} vanishes identically and we recover the statement \cite{Loebbert:2019vcj}
\begin{align}
\levo{P}^\mu |_{14}\includegraphicsbox{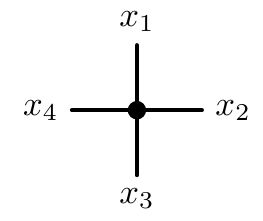} = 0.
\end{align}
\paragraph{Double Ladder.}
The next simplest example is the double ladder%
\footnote{A more correct but less compact terminology would be \emph{double-rung ladder}. We avoid to use the also common name \emph{double box} for this diagram in order not to confuse it with the six-point double box integral of \eqref{eq:6ptcorrelator}.}
 integral with $\alpha=2, \beta=1$, for which the Yangian Ward identity reads
\begin{align}
2i \levo{P}^\mu|_{14} \includegraphicsbox{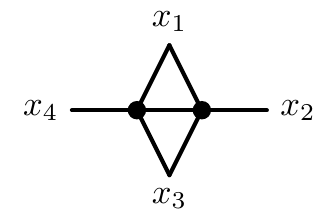} =  -&\includegraphicsbox{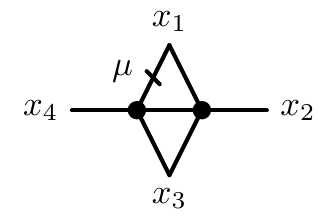}+\includegraphicsbox{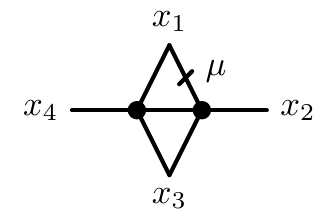}\notag\\
-& \includegraphicsbox{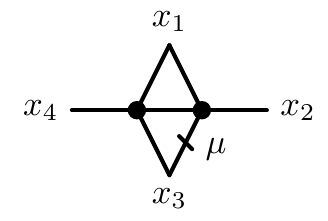}+\includegraphicsbox{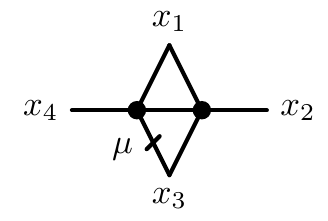}.
\label{eq:DoubleLadderExampleFigures}
\end{align}
Here we use slashed lines to denote propagators that carry an additional derivative. 

\paragraph{Window.} As a final example, the Yangian Ward identity for the simplest integral with a two-dimensional lattice of integration points, the window integral,\footnote{This diagram is referred to as the window because of its form in momentum space. 
In dual momentum/position space it resembles a shuriken.} reads
\begin{align}
2i \levo{P}^\mu|_{14}\includegraphicsbox{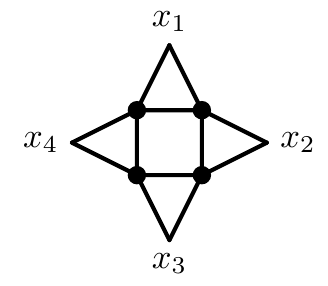} = -& \includegraphicsbox{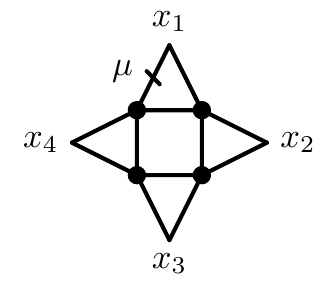}+\includegraphicsbox{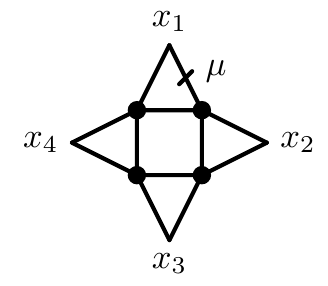}\notag\\
-&\includegraphicsbox{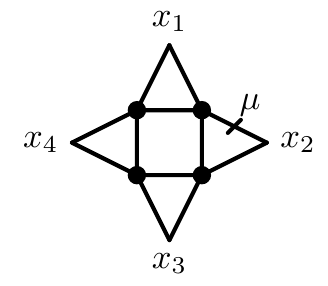}+\includegraphicsbox{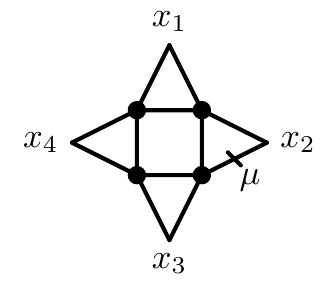}\mp& (\text{4 more}).
\end{align}
\section{From Correlators to Feynman Integrals}\label{correlatorfeynman}

In order to understand the above relations in terms of Feynman integrals, let us explicitly evaluate the different contributions. Due to dual conformal level-zero symmetry, we can write the Basso--Dixon correlator in the form

\begin{equation}
I_{\alpha\beta}
=\left\langle\text{tr}(Z^\alpha(x_1)\bar{X}^\beta(x_2)\bar{Z}^\alpha(x_3)X^\beta(x_4))\right\rangle
=
 \includegraphicsbox[scale=.7]{FigBDGraphLimitsSomeLabels.pdf}
=\frac{\xi^{2\alpha\beta}}{x_{13}^{2\alpha}x_{24}^{2\beta}}
\phi_{\alpha\beta}(u,v),
\label{eq:I4albe}
\end{equation}
where $u$ and $v$ are the four-point conformal cross ratios 
\begin{equation}
u=\frac{x_{12}^2x_{34}^2}{x_{13}^2x_{24}^2}, 
\qquad\qquad
 v=\frac{x_{14}^2x_{23}^2}{x_{13}^2x_{24}^2}.
\label{eq:crossratios4}
\end{equation}
$I_{\alpha\beta}$ can be expressed as a single $\alpha\beta$-loop Feynman integral
\begin{equation}
\label{eq:ScalarIntegrals}
I_{\alpha\beta}=\frac{\xi^{2\alpha\beta}}{\pi^{2\alpha\beta}}\int \prod_{l,m=1}^{\alpha,\beta}\dd^4 x_{l,m}\left(\prod_{j=1}^{\alpha}\prod_{k=0}^\beta \frac{1}{(x_{j,k}-x_{j,k+1})^2}\prod_{j=0}^\alpha\prod_{k=1}^{\beta}\frac{1}{(x_{j,k}-x_{j+1,k})^2}\right),
\end{equation}
where $x_{i,0}=x_1, x_{\alpha+1,j}=x_2, x_{i,\beta+1}=x_3, x_{0,j}=x_4$, for $i=1,2,\dots,\alpha$ and $j=1,2,\dots,\beta$. 

In the Yangian Ward identity we also encounter a version of this integral where one external leg carries an additional derivative, i.e. 
\begin{equation}
\label{eq:VectorIntegrals}
I_{\alpha\beta}^{\mu,n}=\frac{\xi^{2\alpha\beta}}{\pi^{2\alpha\beta}}\int \prod_{l,m=1}^{\alpha,\beta}\dd^4 x_{l,m}\left( \frac{(x_{n,1}-x_1)^\mu}{(x_{n,1}-x_1)^2}\prod_{j=1}^{\alpha}\prod_{k=0}^\beta \frac{1}{(x_{j,k}-x_{j,k+1})^2}\prod_{j=0}^\alpha\prod_{k=1}^{\beta}\frac{1}{(x_{j,k}-x_{j+1,k})^2}\right),
\end{equation}
where $n=1,2,\dots,\alpha$. Integrals containing derivatives acting on points different from $x_1$ are related to
$I_{\alpha\beta}^{\mu,n}$ via permutations of the external points. For convenience we henceforth omit the dependence on the fishnet theory coupling constant $\xi^2$  in expressions for $I_{\alpha\beta}$ and $I_{\alpha\beta}^{\mu,n}$.

\subsection{Basso--Dixon Formula}\label{sec:BassoDixon}
Notably, the functions $\phi_{\alpha\beta}(u,v)$ are all known in closed form. They are best expressed in terms of the auxiliary variables $z$ and $\bar{z}$, defined by
\begin{equation}
z\bar{z}=u,\hspace{1cm} (1-z)(1-\bar{z})=v.
\end{equation}
For $\beta=1$ we have
\begin{equation}\label{eq:ladder}
\phi_{\alpha1}(z,\bar{z})=-\frac{1}{z-\bar{z}}L_\alpha\brk*{\sfrac{z}{z-1},\sfrac{\bar{z}}{\bar{z}-1}},
\end{equation}
where $L_\alpha$ denotes the ladder functions, which are single-valued polylogarithms of $z$ and $\bar{z}$, see  \cite{Usyukina:1993ch,Dixon:2012yy}:
\begin{equation}
L_{\alpha}(z,\bar{z})=\sum_{r=0}^\alpha \frac{(-1)^r (2\alpha-r)!}{r!(\alpha-r)!\alpha!} \log(z\bar z)^r
\brk*{\Li_{2\alpha-r}\left(z\right)-\Li_{2\alpha-r}\left(\bar{z}\right)}.
\end{equation}
For general $\alpha,\beta$, the functions $\phi_{\alpha\beta}$ can be expressed as a determinant of these ladder integrals, which was conjectured first in \cite{Basso:2017jwq} and proven recently in \cite{Basso:2021omx}. The result is
\begin{equation}
\phi_{\alpha\beta}=\text{det}M^{\alpha\beta}, \hspace{1cm} M^{\alpha\beta}_{ij}=c^{\alpha\beta}_{ij}\phi_{\beta-\alpha-1+i+j,1},
\end{equation}
where the coefficients $c_{ij}^{\alpha\beta}$ are defined according to
\begin{equation}
c^{\alpha\beta}_{ij}=\Bigg\{\begin{array}{lr}
        \prod_{k=j+1}^i p_{jk}(p_{jk}-1), & \text{for } i>j,\\
        1, & \text{for } i=j,\\
        \prod_{k=i+1}^j [p_{jk}(p_{jk}-1)]^{-1}, & \text{for } i<j,    \end{array}
\end{equation}
with
\begin{equation}
p_{jk}=\beta-\alpha-1+j+k.
\end{equation}
For example, for $\alpha=\beta=2$ we have
\begin{equation}
\phi_{22}(z,\bar{z})=\begin{vmatrix}
\phi_{11} & 2\phi_{21} \\ 
\frac{1}{6}\phi_{21} & \phi_{31}
\end{vmatrix}=\phi_{11}\phi_{31}-\frac{1}{3}(\phi_{21})^2.
\end{equation}
\subsection{Ward Identity}
\label{sec:WI}

Here we analyse in detail the left and right hand side of the Yangian Ward identity for generic Basso--Dixon integrals 
\eqref{eq:PhatIdentityCorr}.

\paragraph{Differential Part of Ward Identity (Left Hand Side).}
Using the following values for the scaling dimensions and evaluation parameters entering the level-one momentum generator
\begin{equation}
\Delta_j=(\alpha,\beta,\alpha,\beta)_j,
\qquad
\Eval_j=-(0,1,2,3)_j,
\qquad
j=1,\dots,4,
\label{eq:DeltasAndss}
\end{equation}
we act with $\levo{P}^\mu$ on \eqref{eq:I4albe} and find the general expression
\begin{align}
2i\levo{P}^\mu I_{\alpha\beta}=\frac{-4}{x_{13}^{2\alpha}x_{24}^{2\beta}}
\brk[s]*{
\brk*{\frac{x_{12}^\mu}{x_{12}^2}+\frac{x_{34}^\mu}{x_{34}^2}}u \Duv^{\alpha\beta}\phi_{\alpha\beta}(u,v)
+\brk*{\frac{x_{23}^\mu}{x_{23}^2}+\frac{x_{41}^\mu}{x_{41}^2}}v \Dvu^{\alpha\beta}\phi_{\alpha\beta}(u,v)
},
\label{eq:pmuhatBD}
\end{align}
with the differential operator
\begin{equation}
\Duv^{\alpha\beta}=
\alpha \beta+(\alpha+\beta+1)v\partial_v+\brk*{(\alpha+\beta+1)u-\sfrac{\alpha+\beta}{2}}\partial_u+v^2\partial_v^2+(u-1)u\partial_u^2+2uv\partial_u\partial_v.
\end{equation}
In the following we will use the shorthand 
\begin{equation}
H^\mu_{\alpha\beta}\coloneqq 2i\levo{P}^\mu I_{\alpha\beta}.
\end{equation}
Using conformal symmetry one can argue that the coefficients of the vectors $x_{jk}^\mu/x_{jk}^2$ are in fact independent in the Yangian Ward identity \cite{Loebbert:2019vcj}. Let us thus investigate these coefficients on the right hand side of the equation \eqref{eq:PhatIdentityCorr} in the following.
For completeness we note that $\Duv^{\alpha\beta}$ and $\Dvu^{\alpha\beta}$ are the special cases $\gamma=\gamma'=(\alpha+\beta)/2$ of the differential operators that are known to annihilate the Appell hypergeometric function $F_4$:
\begin{align}
\Dvu^{\alpha\beta\gamma\gamma'}=&\brk*{\alpha \beta+(\alpha+\beta+1)u \partial_u  +\brk*{(\alpha+\beta+1)v-\gamma'} \partial_v +  u^2 \partial^2_u  +(v-1)v \partial^2_v +2 v u \partial_u \partial_v  } , \nonumber \\
\Duv^{\alpha\beta\gamma\gamma'}=&\brk*{\alpha \beta+(\alpha+\beta+1)v \partial_v  +\brk*{(\alpha+\beta+1)u-\gamma} \partial_u +  v^2 \partial^2_v  +(u-1)u \partial^2_u +2 v u \partial_v \partial_u  }  .
\label{eq:AppellDiffOps}
\end{align}
Notably, the conformal box integral for generic propagator powers satisfies homogenous Ward identities, and so is annihilated by $\Dvu^{\alpha\beta\gamma\gamma'}$ and $\Duv^{\alpha\beta\gamma\gamma'}$.  Therefore it can be expressed in terms of $F_4$, with the parameters $\alpha, \beta, \gamma, \gamma'$ relating to the four propagator powers \cite{Loebbert:2019vcj}.

\paragraph{Vector Part of Ward Identity (Right Hand Side).}
Just like the scalar integrals \eqref{eq:ScalarIntegrals}, all vector integrals of the form \eqref{eq:VectorIntegrals} satisfy (level-zero) conformal Ward identities:
\begin{align}
\gen{D}I_{\alpha\beta}^{\mu,n} (x_1,\dots,x_4) &= 0,\label{eq:InhomWard1}\\
\left(\gen{K}^\mu \eta_{\nu\rho} + 2 i (\delta^{\mu}_\nu x_{1,\rho} - \delta^{\mu}_\rho x_{1,\nu})\right)I_{\alpha\beta}^{\rho,n}(x_1,\dots,x_4) &= -2 i \delta^{\mu}_\nu I_{\alpha\beta}(x_1,\dots,x_4),
\label{eq:InhomWard2}
\end{align}
with scaling dimensions $\Delta_i = (\alpha+1, \beta, \alpha,\beta)$. These identities can be derived straightforwardly by commuting the homogeneous Ward identities for the scalar integrals with the extra derivatives contained in the vector integrals. 
As we explain in more detail in \Appref{app:vectordecomp}, the general solution to \eqref{eq:InhomWard1} and \eqref{eq:InhomWard2} takes the form 
\begin{equation}
x_{13}^{2\alpha}x_{24}^{2\beta}I_{\alpha\beta}^{\mu,n}=-\frac{x_{12}^\mu}{x_{12}^2}F_{2,n}^{\alpha\beta}(u,v)-\frac{x_{13}^\mu}{x_{13}^2}F_{3,n}^{\alpha\beta}(u,v)-\frac{x_{14}^\mu}{x_{14}^2}F_{4,n}^{\alpha\beta}(u,v),
\label{eq:alphabetavectordecomp}
\end{equation}
where the coefficient functions further need to satisfy
\begin{align}\label{eq:Fsum}
F_{2,n}^{\alpha\beta}(u,v) + F_{3,n}^{\alpha\beta}(u,v) + F_{4,n}^{\alpha\beta}(u,v) = I_{\alpha\beta}(u,v),
\end{align}
for each $n=1,2,\dots,\alpha$.

\subsection{Examples: Double Ladder and Window}
We present here explicitly the form of the Yangian Ward identity for the double ladder and window integrals. In appendix \ref{app:ExtraWard} we present analogous results for triple and quadruple ladder.
\paragraph{Double Ladder.} We consider the correlator  \eqref{eq:I4albe} for $\alpha=2, \beta=1$
\begin{equation}
\left\langle\text{tr}(Z(x_1)Z(x_1)\bar{X}(x_2) \bar{Z} (x_3) \bar{Z}(x_3)X(x_4))\right\rangle
=
\includegraphicsbox{FigDoubleLadder.pdf},
\label{eq:4ptcorrelatordoubleladder}
\end{equation}
which is represented by the well-known double ladder integral
\begin{equation}
I_{21}=\int\frac{\dd^4 x_a}{\pi^2}\frac{\dd^4 x_{b}}{\pi^2}\frac{1}{(x_{a1}^2x_{a3}^2x_{a4}^2)x_{ab}^2(x_{b1}^2x_{b2}^2x_{b3}^2)}=\frac{1}{x_{13}^4x_{24}^2}\phi_{21}(u,v).
\label{eq:doubleladder}
\end{equation}
We would like to understand what the Yangian Ward identities \eqref{eq:YangianWard} imply for the double ladder integral. Therefore we specialise \eqref{eq:PhatIdentityCorr} to compute the action of the level-one generator $\levo{P}^\mu$ on the four-point correlator \eqref{eq:4ptcorrelatordoubleladder}: 
\begin{equation}
H^\mu_{21}
\coloneqq 2i\levo{P}^\mu |_{14}\left\langle\text{tr}(Z^2(x_1)\bar{X}(x_2) \bar{Z}^2(x_3)X(x_4))\right\rangle.
\end{equation}
The Yangian Ward identity implies that
\begin{align}
H^\mu_{21}=
& +\left\langle\text{tr}(Z(x_1)[\partial_{1}^\mu Z(x_1)]\bar{X}(x_2) \bar{Z}^2(x_3)X(x_4))\right\rangle
\nonumber\\
&- \left\langle\text{tr}([\partial_{1}^\mu Z(x_1)]Z(x_1)\bar{X}(x_2) \bar{Z}^2(x_3)X(x_4))\right\rangle
\nonumber\\
&+\left\langle\text{tr}(Z^2(x_1)\bar{X}(x_2) \bar{Z} (x_3) [\partial_{3}^\mu \bar{Z}(x_3)]X(x_4))\right\rangle
\nonumber \\
&- \left\langle\text{tr}( Z^2(x_1)\bar{X}(x_2) [\partial_{3}^\mu \bar{Z} (x_3)] \bar{Z}(x_3)X(x_4))\right\rangle,
\label{eq:RHSdoubleladder}
\end{align}
which is another representation of \eqref{eq:DoubleLadderExampleFigures}. Explicitly, we see that evaluating the four-point coordinate space level-one momentum generator on the four-point correlator \eqref{eq:4ptcorrelatordoubleladder} yields a linear combination of correlation functions involving a single descendant field. Let us focus on the  single contribution
\begin{equation}
2I_{21}^{\mu,2}= \left\langle\text{tr}(Z(x_1)[\partial_{1}^\mu Z(x_1)]\bar{X}(x_2) \bar{Z}^2(x_3) X(x_4))\right\rangle,
\end{equation}
where the vector integral $I_{21}^{\mu,2}$ is a specialisation of \eqref{eq:VectorIntegrals}:
\begin{align}
2I_{21}^{\mu,2}&=\int\frac{\dd^4 x_a}{\pi^2}\frac{\dd^4 x_{b}}{\pi^2}\frac{1}{x_{a1}^2x_{a3}^2x_{a4}^2x_{ab}^2}\left(\partial_{1}^\mu \frac{1}{x_{b1}^2}\right)\frac{1}{x_{b2}^2x_{b3}^2}\notag\\&=2\int \frac{\dd^4 x_a}{\pi^2} \frac{\dd^4 x_{b}}{\pi^2}\frac{x_{ b1}^\mu}{ x_{a1}^2 x_{a3}^2 x_{a4}^2x_{ab}^2 x_{b1}^4x_{b2}^2 x_{b3}^2 }.
\label{eq:vecdoubleladderdef}
\end{align}
 We can then represent $H^\mu_{21}$ in terms of antisymmetrisations of $I_{21}^{\mu,2}$:
\begin{equation}
H^\mu_{21}=(2I^{\mu,2}_{21}-x_2\leftrightarrow x_4)-x_1\leftrightarrow x_3.
\end{equation}
Moreover, using \eqref{eq:alphabetavectordecomp} 
 we find the following vector decomposition for $I_{21}^{\mu,2}$:
\begin{equation}
x_{13}^4x_{24}^2I_{21}^{\mu,2}=-\frac{x_{12}^\mu}{x_{12}^2}F_2(u,v)-\frac{x_{13}^\mu}{x_{13}^2}F_3(u,v)-\frac{x_{14}^\mu}{x_{14}^2}F_4(u,v).
\label{eq:21vectordecomp}
\end{equation}
By an explicit Feynman parametrisation we obtained integral expressions for the conformal functions $F_i(u,v)$ which were useful for numerical checks, see \appref{app:feynmanparam}. From \eqref{eq:Fsum} the vector coefficients $F_i(u,v)$ further satisfy
\begin{equation}
\phi_{21}(u,v)=F_2(u,v)+F_3(u,v)+F_4(u,v).
\label{eq:Fsum2}
\end{equation}
This can also be seen by contracting both sides of \eqref{eq:21vectordecomp} with $-x_1^\mu$ and sending $x_1\rightarrow\infty$ with a conformal transformation. Under the transpositions of points $x_1\leftrightarrow x_3$ and $x_2\leftrightarrow x_4$ the cross ratios  are exchanged $u\leftrightarrow v$. Using this fact and \eqref{eq:21vectordecomp} we calculate $H^\mu_{21}$ in terms of the $F_i$ as
\begin{align}
x_{13}^4x_{24}^2H^\mu_{21}=
&-2\left(\frac{x_{12}^\mu}{x_{12}^2}+\frac{x_{34}^\mu}{x_{34}^2}\right)\brk[s]*{F_2(u,v)-F_4(v,u)}
\nonumber\\
&-2\left(\frac{x_{23}^\mu}{x_{23}^2}+\frac{x_{41}^\mu}{x_{41}^2}\right)\brk[s]*{F_2(v,u)-F_4(u,v)}.
\label{eq:H21mu}
\end{align}
Comparing \eqref{eq:H21mu} and \eqref{eq:pmuhatBD} we recover the following constraint on the components of the vector decomposition $F_i$:
\begin{equation}
2u\Duv^{21} \phi_{21}=F_2(u,v)-F_4(v,u),
\label{eq:Yangianconstraint}
\end{equation}
and the same equation with $u$ and $v$ swapped. So the Yangian differential operator acting on the conformal double ladder is a combination of the coefficient functions in the vector decomposition \eqref{eq:21vectordecomp}. The left hand side of \eqref{eq:Yangianconstraint} can be calculated exactly by acting with $2u\Duv^{21}$ on the ladder function \eqref{eq:ladder}. We obtained explicit Feynman parametrisations of $F_i$ in order to check agreement with the right hand side numerically, see \eqref{eq:F2} and \eqref{eq:F4}. 
\paragraph{Window.}
For $\alpha=\beta=2$, we consider the correlator \eqref{eq:I4albe} which reads
\begin{equation}
\left\langle\text{tr}(Z^2(x_1)\bar{X}^2(x_2)\bar{Z}^2(x_3)X^2(x_4))\right\rangle =
\includegraphicsbox{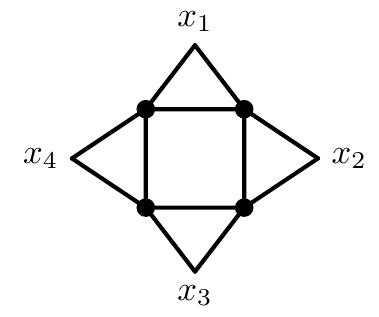}.
\end{equation}
The corresponding window integral takes the form
\begin{equation}\label{eq:window}
I_{22}=\int \frac{\dd^4 x_a}{\pi^2}\frac{\dd^4 x_{b}}{\pi^2}\frac{\dd^4 x_{c}}{\pi^2}\frac{\dd^4 x_{d}}{\pi^2}\frac{1}{(x_{a1}^2x_{a4}^2)x_{ab}^2(x_{b1}^2x_{b2}^2)x_{bc}^2(x_{c2}^2x_{c3}^2)x_{cd}^2(x_{d3}^2x_{d4}^2)x_{da}^2}=\frac{1}{x_{13}^4x_{24}^4}\phi_{22}(u,v).
\end{equation}
Let us write \eqref{eq:PhatIdentityCorr} specialised to this case in full:
\begin{align}
&H^\mu_{22}
=2i\levo{P}^\mu |_{14}\left\langle\text{tr}(Z^2(x_1)\bar{X}^2(x_2)\bar{Z}^2(x_3)X^2(x_4))\right\rangle
\label{eq:RHSwindow}
\\
&= \left\langle\text{tr}(Z(x_1)[\partial_{1}^\mu Z(x_1)]\bar{X}^2(x_2) \bar{Z}^2(x_3)X^2(x_4))\right\rangle- \left\langle\text{tr}([\partial_{1}^\mu Z(x_1)]Z(x_1)\bar{X}^2(x_2) \bar{Z}^2(x_3)X^2(x_4))\right\rangle
\nonumber\\
&+\left\langle\text{tr}(Z^2(x_1)\bar{X}(x_2)[\partial^\mu_{2}\bar{X}(x_2)] \bar{Z}^2 (x_3)X^2(x_4))\right\rangle-\left\langle\text{tr}(Z^2(x_1)[\partial^\mu_{2}\bar{X}(x_2)] \bar{X}(x_2)\bar{Z}^2 (x_3)X^2(x_4))\right\rangle\nonumber\\
&+\left\langle\text{tr}(Z^2(x_1)\bar{X}^2(x_2) \bar{Z} (x_3) [\partial_{3}^\mu \bar{Z}(x_3)]X^2(x_4))\right\rangle- \left\langle\text{tr}( Z^2(x_1)\bar{X}^2(x_2) [\partial_{3}^\mu \bar{Z} (x_3)] \bar{Z}(x_3)X^2(x_4))\right\rangle
\nonumber\\
&+\left\langle\text{tr}(Z^2(x_1)\bar{X}^2(x_2) \bar{Z}^2 (x_3)X(x_4) [\partial_{4}^\mu X(x_4)])\right\rangle-\left\langle\text{tr}(Z^2(x_1)\bar{X}^2(x_2) \bar{Z}^2 (x_3) [\partial_{4}^\mu X(x_4)]X(x_4))\right\rangle.
\nonumber
\end{align}
 We focus on the first correlators in the second and third lines of \eqref{eq:RHSwindow} which read
\begin{align}
2I_{22}^{\mu,2}=& \left\langle\text{tr}(Z(x_1)[\partial_{1}^\mu Z(x_1)]\bar{X}^2(x_2)\bar{Z}^2(x_3)X^2(x_4))\right\rangle,\label{eq:windowfirstcor}\\
2\tau I_{22}^{\mu,2}=& \left\langle\text{tr}(Z^2(x_1)\bar{X}(x_2)[\partial^\mu_{2}\bar{X}(x_2)] \bar{Z}^2 (x_3)X^2(x_4))\right\rangle \label{eq:windowsecondcor}. \end{align}
Here $I_{22}^{\mu,2}$ is the vector integral
\begin{align}
&I_{22}^{\mu,2}=\int \frac{\dd^4 x_a}{\pi^2}\frac{\dd^4 x_{b}}{\pi^2}\frac{\dd^4 x_{c}}{\pi^2}\frac{\dd^4 x_{d}}{\pi^2}\frac{x_{b1}^\mu}{(x_{a1}^2x_{a4}^2)x_{ab}^2(x_{b1}^4x_{b2}^2)x_{bc}^2(x_{c2}^2x_{c3}^2)x_{cd}^2(x_{d3}^2x_{d4}^2)x_{da}^2},
\end{align}
and $\tau$ is the cycle which maps $x_i\rightarrow x_{i+1}$, where we identify $x_5\equiv x_1$. In fact all integrals appearing in \eqref{eq:RHSwindow} can be recovered from $I_{22}^{\mu,2}$ by a permutation of points, which at most exchanges $u$ and $v$. Computing $H_{22}^\mu$ in full we find
\begin{equation}
H_{22}^\mu=2(I_{22}^{\mu,2}+\tau I_{22}^{\mu,2}-x_2\leftrightarrow x_4)-x_{1}\leftrightarrow x_3.
\label{eq:H22mu}
\end{equation}
We expand $I_{22}^{\mu,2}$ in terms of its vector components 
\begin{equation}
x_{13}^4x_{24}^4I_{22}^\mu=-\frac{x_{12}^\mu}{x_{12}^2}W_2(u,v)-\frac{x_{13}^\mu}{x_{13}^2}W_3(u,v)-\frac{x_{14}^\mu}{x_{14}^2}W_4(u,v).
\label{eq:windowvectordecomp}
\end{equation}
Under the cycle $\tau$ and transpositions $x_{1}\leftrightarrow x_3, x_2\leftrightarrow x_4$ we have $u\leftrightarrow v$. Using this and \eqref{eq:windowvectordecomp} we compute \eqref{eq:H22mu} in terms of the $W_i(u,v)$ to be
\begin{equation}
x_{13}^4x_{24}^4H^\mu_{22}=-4\left(\frac{x_{12}^\mu}{x_{12}^2}+\frac{x_{34}^\mu}{x_{34}^2}\right)(W_2(u,v)-W_4(v,u))-4\left(\frac{x_{23}^\mu}{x_{23}^2}+\frac{x_{41}^\mu}{x_{41}^2}\right)(W_2(v,u)-W_4(u,v)).
\label{eq:H22mu2}
\end{equation}
 Comparing \eqref{eq:H22mu2} and \eqref{eq:pmuhatBD} we have the constraint
 \begin{equation}
 u\Duv^{22}\phi_{22}=W_2(u,v)-W_4(v,u),
 \label{eq:Duvphiwindow}
 \end{equation}
 and the same equation with $u\leftrightarrow v$. The structure of \eqref{eq:Yangianconstraint} and \eqref{eq:Duvphiwindow} persists for general Basso--Dixon correlators: $u\Duv^{\alpha\beta}\phi_{\alpha\beta}$ is equal to a linear combination of coefficients in the vector decomposition of vector integrals related to $\phi_{\alpha\beta}$. For higher $\alpha, \beta$ more vector integrals $I^{\mu, n}_{\alpha\beta}$, and hence more independent vector coefficients, contribute to the Ward Identity. We present the Ward identity for the triple and quadruple ladder in \appref{app:ExtraWard}. In the next section we show how one can rewrite these identities formal equations for the functions $\phi_{\alpha\beta}$.


\section{Conformal Decomposition of Vector Integrals}
\label{sec:tensordecomp}
In the previous section we saw that the Yangian differential operator $u\Duv^{\alpha\beta}$ acting on the conformal function $\phi_{\alpha\beta}$ is equal to a combination of coefficients of the conformal expansion \eqref{eq:alphabetavectordecomp} of certain vector integrals. Such coefficients can be written as linear combinations of higher dimensional scalar integrals with modified propagator powers 
\cite{Tarasov:1996br,ReFiorentin:2015kri,Kalin:2019qup}. We exploit this fact to rewrite the vector coefficients as such, and reformulate the Yangian equations in the form of \eqref{eq:Yangianconstraint} as a formal identity%
\footnote{See the comments around \eqref{eq:contiguous} for an interpretation.} for $\phi_{\alpha\beta}$. We first review the basic ingredients of this formalism necessary for our purposes. 

\subsection{Vector Decomposition in General}
We consider an $\ell$-loop Feynman graph $G$ in
position space\footnote{Alternatively, we can understand this as dual momentum space. Internal points are integrated over, and we still refer to these as loop momenta.}  with $N$ massless lines and arbitrary propagator powers $\nu_1,\dots,\nu_N$ in dimension $D$ 
\begin{equation}
I^{\ell,G}_{\nu,D}=\int \prod_{i=1}^\ell \frac{\dd^D x_{a_i}}{\pi^{D/2}}\prod_{\langle jk\rangle\in G}\frac{1}{x_{jk}^{2\nu_{jk}}},
\end{equation}
where $\langle jk\rangle$ is the line joining points $x_j$ and $x_k$ in $G$. We associate to each line $\langle jk\rangle$ a Feynman parameter $\alpha_l$ for $l=1,\dots,N$ and set $\nu_{jk}\rightarrow \nu_l$. Such an integral has a \textit{Schwinger parametrisation} 
\begin{equation}
I^{\ell,G}_{\nu,D}=\left(\prod_{j=1}^N\int_{0}^\infty \frac{\dd \alpha_j \alpha_j^{\nu_j-1}}{\Gamma_{\nu_j}}\right)\frac{\exp(-\mathcal{F}(x_{ij}^2)/\mathcal{U})}{\mathcal{U}^{D/2}},
\label{eq:schwinger}
\end{equation}
where  $\mathcal{U}$ and $\mathcal{F}(x_{ij}^2)$ are polynomials in the Feynman parameters $\alpha_1,\dots,\alpha_N$, which depend on the toplogy of the Feynman graph $G$. There is a standard procedure to decompose integrals with tensor insertions of the loop momenta in the numerator into a linear combination of higher dimensional scalar integrals with modified propagator powers. Here we are interested in the case with only a single vector insertion
\begin{equation}
I^{\ell,G}_{\nu,D}(x_{a_mn}^\mu)=\int \prod_{i=1}^\ell \frac{\dd^D x_{a_i}}{\pi^{D/2}}x_{a_mn}^\mu\prod_{\langle jk\rangle\in G}\frac{1}{x_{jk}^{2\nu_{jk}}},
\label{eq:vectorintgeneral}
\end{equation}
where $x_{a_mn}^\mu=x_{a_m}^\mu-x_{n}^\mu$ is the difference between an integration point  $x_{a_m}^\mu$, $m=1,\dots,\ell$, and an external point $x_n^\mu$, $n=1,2,3,4$.  An algorithm to compute this decomposition explicitly is presented in \cite{Kalin:2019qup}, and here we give the main ingredients needed for a single vector insertion. 

We first form an $\ell\times \ell$ symmetric matrix $B$ and an $\ell$-vector $c$. $B_{ij}$ and $c_i$ are indexed by the integration points $a_1,\dots, a_\ell$. $B_{ij}$ is minus the sum of the Feynman parameters corresponding to the legs connecting internal point $i$ and internal point $j$ for $i\neq j$, and it is the sum of the Feynman parameters corresponding to all legs connected to internal point $i$ for $i=j$. $c_i$ is the sum of Feynman parameters connecting internal point $i$ to some external point $x_k^\mu$, multiplied by that external point. In terms of $B$ and $c$ we have
\begin{equation}
\mathcal{U}=\det B, \qquad \frac{\mathcal{F}}{\mathcal{U}}=d\hspace{0.07cm}-c^TB^{-1}\hspace{0.05cm}c,
\end{equation}
where $d$ is the sum over squares of external points $x_i^2$, each multiplied by the sum of the Feynman parameters corresponding to the legs connected to that external point. An example for the double ladder will be given in the next section.

In order to compute the vector decomposition of \eqref{eq:vectorintgeneral} one should compute the quantity
\begin{equation}\label{eq:adj}
(B^{-1}\cdot c)_{a_m}-x_n^\mu=-\frac{1}{\mathcal{U}}\sum_{l\neq n}x_{nl}^\mu  \mathcal{P}_{l,m,n}(\alpha),
\end{equation}
where $ \mathcal{P}_{l,m,n}(\alpha)$ are polynomials in the Feynman parameters. Then at the level of the Schwinger parametrisation we have
\begin{equation}\label{eq:Pnum}
I^{\ell,G}_{\nu,D}(x_{a_mn}^\mu)=-\sum_{l\neq n}x_{nl}^\mu \left(\prod_{j=1}^N\int_{0}^\infty \frac{\dd \alpha_j \alpha_j^{\nu_j-1}}{\Gamma_{\nu_j}}\right)\frac{ \mathcal{P}_{l,m,n}(\alpha)\exp(-\mathcal{F}(x_{ij}^2)/\mathcal{U})}{\mathcal{U}^{(D+2)/2}},
\end{equation}
from which we see the vector coefficients 
of $I^{\ell,G}_{\nu,D}(x_{a_mn}^\mu)$ are linear combinations of scalar integrals of dimension $D+2$, with propagator powers modified appropriately by the polynomials $ \mathcal{P}_{l,m,n}$.
\paragraph{Example: Double Ladder Vector Decomposition.}
As a simple example we consider the double ladder in $D$ dimensions with generic propagator powers.
This is given by the formula 
\begin{equation}\label{eq:DoubleLadderGenFormula}
I^{\nu,D}_{21}=\int\frac{\dd^D x_a}{\pi^{D/2}}\frac{\dd^D x_{b}}{\pi^{D/2}}\frac{1}{(x_{a1}^{2\nu_1}x_{a3}^{2\nu_5}x_{a4}^{2
\nu_6})x_{ab}^{2\nu_7}(x_{b1}^{2\nu_2}x_{b2}^{2\nu_3}x_{b3}^{2\nu_4})}=
\includegraphicsbox{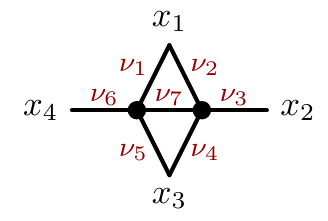}.
\end{equation}
For this integral we have
 \begin{equation}
B_{ij}=\begin{pmatrix}
 \alpha _1+\alpha _5+\alpha _6+\alpha _7 & -\alpha _7 \\
 -\alpha _7 & \alpha _2+\alpha _3+\alpha _4+\alpha _7 \\
\end{pmatrix}, \qquad
c_i=\begin{pmatrix}
 \alpha _1 x_1^\mu+\alpha _5 x_3^\mu+\alpha _6 x_4^\mu \\
 \alpha _2 x_1^\mu+\alpha _3 x_2^\mu+\alpha _4 x_3^\mu \\
\end{pmatrix},
\end{equation}
\begin{equation}
d=x_1^2(\alpha_1+\alpha_2)+x_2^2\alpha_3+x_3^2(\alpha_4+\alpha_5)+x_4^2\alpha_6.
\end{equation}
This integral can be expressed in terms of $\mathcal{U}$ and $\mathcal{F}$ polynomials as
\begin{equation}
I^{\nu,D}_{21}=\left(\prod_{j=1}^7\int_{0}^\infty \frac{\dd \alpha_j \alpha_j^{\nu_j-1}}{\Gamma_{\nu_j}}\right)\frac{\exp(-\mathcal{F}/\mathcal{U})}{\mathcal{U}^{D/2}}.
\label{eq:doubleladderschwinger}
\end{equation}
$\mathcal{U}$ and $\mathcal{F}$ are determined only by the topology of the graph in momentum space, and are given by
\begin{align}\label{eq:DoubleLadderU}
\mathcal{U}=&\alpha _1 \alpha _2+\alpha _1 \alpha _3+\alpha _1 \alpha _4+\alpha _2 \alpha _5+\alpha _3 \alpha _5+\alpha _4 \alpha _5+\alpha _2 \alpha _6+\alpha
   _3 \alpha _6\notag\\
   &+\alpha _4 \alpha _6+\alpha _1 \alpha _7+\alpha _2 \alpha _7+\alpha _3 \alpha _7+\alpha _4 \alpha _7+\alpha _5 \alpha _7+\alpha _6
   \alpha _7,
\end{align}
\begin{align}
\mathcal{F}=&x_{12}^2(\alpha _1 \alpha _2 \alpha _3 +\alpha _2 \alpha _3 \alpha _5 +\alpha _2 \alpha _3 \alpha _6 +\alpha _1 \alpha _3 \alpha
   _7 +\alpha _2 \alpha _3 \alpha _7)+x_{13}^2(\alpha _1 \alpha _2 \alpha _4 \notag\\
  &+\alpha _1 \alpha _2 \alpha _5 +\alpha _1
   \alpha _3 \alpha _5 +\alpha _1 \alpha _4 \alpha _5 +\alpha _2 \alpha _4 \alpha _5 +\alpha _2 \alpha _4 \alpha _6
 +\alpha _1 \alpha _4 \alpha _7\notag\\
   &+\alpha _2 \alpha _4 \alpha _7 +\alpha _1 \alpha _5 \alpha _7 +\alpha _2
   \alpha _5 \alpha _7)+x_{14}^2(\alpha _1 \alpha _2 \alpha _6 +\alpha _1 \alpha _3 \alpha _6 +\alpha _1 \alpha _4 \alpha _6
   \notag\\
   &+\alpha _1 \alpha _6 \alpha _7 +\alpha _2 \alpha _6 \alpha _7 )+x_{23}^2(\alpha _1 \alpha _3 \alpha _4 +\alpha _3
   \alpha _4 \alpha _5 +\alpha _3 \alpha _4 \alpha _6 +\alpha _3 \alpha _4 \alpha _7 \notag\\
   &+\alpha _3 \alpha _5 \alpha _7)+x_{34}^2(\alpha _5 \alpha _6 \alpha _7 +\alpha _2 \alpha _5 \alpha _6+\alpha _3 \alpha _5 \alpha _6 +\alpha _4
   \alpha _5 \alpha _6+\alpha _4 \alpha _6 \alpha _7)\notag\\
   &+\alpha _3 \alpha _6 \alpha _7 x_{24}^2.
\end{align}
$\mathcal{U}$ corresponds to the 15 spanning trees of the double ladder graph in momentum space. $\mathcal{F}$ corresponds to the 31 spanning 2--forests, each of which is assigned a momentum invariant corresponding to the momentum flowing through each of the trees in the forest, cf.\ e.g.\ \cite{Panzer:2015ida}. We are interested in a scalar decomposition of \eqref{eq:vecdoubleladderdef}
\begin{align}
I_{21}^{\mu,2}&=\int\frac{\dd^4 x_a}{\pi^{2}}\frac{\dd^4 x_{b}}{\pi^{2}}\frac{x_{b1}^\mu}{(x_{a1}^{2}x_{a3}^{2}x_{a4}^{2})x_{ab}^{2}(x_{b1}^{4}x_{b2}^{2}x_{b3}^{2})}
\label{eq:I21dec}
\\ \notag
&=\frac{1}{x_{13}^4x_{24}^2}\left(-\frac{x_{12}^\mu}{x_{12}^2}F_2(u,v)-\frac{x_{13}^\mu}{x_{13}^2}F_3(u,v)-\frac{x_{14}^\mu}{x_{14}^2}F_4(u,v)\right),
\end{align}
i.e.\ we want to determine the functions $F_i(u,v)$ as linear combinations of higher dimensional double ladder integrals with modified propagator powers. We compute \eqref{eq:adj} explicitly in this case
\begin{align}
\mathcal{U}[(B^{-1}\cdot c)_b-x_1^\mu]&= -x_{12}^\mu (\alpha_1\alpha_3+\alpha_3\alpha_5+\alpha_3\alpha_6+\alpha_3\alpha_7)\notag\\&-x_{13}^{\mu}(\alpha_1\alpha_4+\alpha_4\alpha_5+\alpha_4\alpha_6+\alpha_4\alpha_7+\alpha_5\alpha_7)-x_{14}^\mu \alpha_6\alpha_7.
\end{align}
Combining \eqref{eq:adj}, \eqref{eq:Pnum}, and \eqref{eq:doubleladderschwinger} leads to formulas for the vector double ladder integral coefficients in \eqref{eq:I21dec}:
\begin{align}\label{eq:F2F3F4}
&F_{2}(u,v)= x_{13}^4x_{24}^2x_{12}^2\brk*{I^6_{2,2,2,1,1,1,1}+I^6_{1,2,2,1,2,1,1}+I^6_{1,2,2,1,1,2,1}+I^6_{1,2,2,1,1,1,2}},\\
&F_3(u,v)=x_{13}^6x_{24}^2\brk*{I^6_{1,2,1,1,2,1,2}+I^6_{2,2,1,2,1,1,1}+I^6_{1,2,1,2,2,1,1}+I^6_{1,2,1,2,1,2,1}+I^6_{1,2,1,2,1,1,2}}\notag,\\
&F_4(u,v)=x_{13}^4x_{24}^2x_{14}^2I^{6}_{1,2,1,1,1,2,2}.\notag
\end{align}
Interestingly $F_4(u,v)$ is given by a single manifestly conformal 6-dimensional scalar integral. Note that $F_{2}(u,v)$ and $F_3(u,v)$ are not manifestly conformal when expressed as \eqref{eq:F2F3F4}.

\subsection{Conformalisation}\label{sec:conformalise}
In general, performing a vector decomposition on the integrals appearing on the right hand side of the Yangian Ward identities will lead to expressions for the vector coefficients which are not manifestly conformal, such as $F_2$ and $F_3$ in $\eqref{eq:F2F3F4}$. However, they can always be rewritten using manifestly conformal 6-dimensional integrals, as will be described in the following. To explain the method, let us consider some conformal function of four external points
\begin{align}
f(x_1,x_2,x_3,x_4) = f(u,v),
\end{align}
with the usual conformal four-point cross ratios  \eqref{eq:crossratios4}. If we consider the limit of one point going to infinity, e.g.~$x_1$, we see that the cross ratios  remain finite and the function is given by
\begin{align}
\lim_{x_1\rightarrow \infty} f(x_1,x_2,x_3,x_4) = f(\tilde{u},\tilde{v})
\end{align}
with the reduced cross ratios  
\begin{align}
\tilde{u} = \frac{x_{34}^2}{x_{24}^2}, \qquad \tilde{v} = \frac{x_{23}^2}{x_{24}^2}.
\end{align}
The same function of reduced cross ratios can be reached by performing a translation 
\begin{align}
f(x_1,x_2,x_3,x_4) &= f(0,x_2-x_1,x_3-x_1,x_4-x_1)\coloneqq f(0,y_2,y_3,y_4), 
\label{eq:InversionTrickI}
\end{align}
and then an inversion on $y_2, y_3, y_4$
\begin{equation}
 f(0,y_2,y_3,y_4)\rightarrow  f\left(0,\frac{y_2}{y_2^2},\frac{y_3}{y_3^2},\frac{y_4}{y_4^2}\right)=f(\tilde{u},\tilde{v}),
 \label{eq:InversionTrickII}
\end{equation}
where the last equivalence uses the identity 
\begin{align}\label{eq:InversionSquaredDiff}
\left(\frac{y_i^\mu}{y_{i}^2} - \frac{y_j^\mu}{y_{j}^2}\right)^2 = \frac{y_{ij}^2}{y_{i}^2 y_{j}^2}.
\end{align} 
Since all of these operations are linear, we can perform them for all integrals in \eqref{eq:F2F3F4} separately, keeping in mind that they will lead to an identity between integrals only when summed together. To turn this observation into a recipe for conformalisation, note that the steps in \eqref{eq:InversionTrickI} and \eqref{eq:InversionTrickII} are invertible, whereas sending one point to infinity is not. Hence, starting from a non-manifestly conformal sum of integrals, we can reach a manifestly conformal expression for the same quantity via the following steps:
\begin{itemize}
\item Send one of the points $x_j$ to infinity. 
\item Perform an inversion on the remaining points and the integration variables.
\item Restore the eliminated point by translating the remaining external points by $x_j$.
\end{itemize}
\paragraph{Example: Double Ladder Conformalisation.} As an example, consider the integral $x_{13}^4 x_{24}^2 x_{12}^2 I^6_{1,2,2,1,2,1,1}$, which appears in the sum \eqref{eq:F2F3F4}. Taking the limit $x_1\rightarrow \infty$, we are left with
\begin{align}
\lim_{x_1\rightarrow \infty} x_{13}^4 x_{24}^2 x_{12}^2 I^6_{1,2,2,1,2,1,1} = \int \dd^6 x_a \dd^6 x_b \frac{x_{24}^2}{x_{2b}^4 x_{3b}^2 x_{3a}^4 x_{4a}^2 x_{ab}^2}.
\end{align}
Following the above recipe, we now perform an inversion 
\begin{align}
x_j^\mu \rightarrow \frac{x_j^\mu}{x_j^2}
\end{align}
on this integral. To rewrite it in terms of squared differences of $x$-variables, we also perform the substitution $x_{a,b}^\mu \rightarrow \frac{x_{a,b}^\mu}{x_{a,b}^2}$ for which 
\begin{align}
\text{d}^6 x_{a,b} \rightarrow \frac{\dd^6 x_{a,b}}{x_{a,b}^{12}}.
\end{align}
Then using \eqref{eq:InversionSquaredDiff} the integral becomes
\begin{align}
\int \dd^6 x_a \dd^6 x_b \frac{x_{24}^2}{x_{2b}^4 x_{3b}^2 x_{3a}^4 x_{4a}^2 x_{ab}^2} \rightarrow \int \dd^6 x_a \dd^6 x_b \frac{x_2^2 x_3^6 x_{24}^2}{x_a^4 x_b^4 x_{2b}^4 x_{3b}^2 x_{3a}^4 x_{4a}^2 x_{ab}^2}.
\end{align}
Now we restore the point $x_1$ by substituting $x_j\rightarrow x_j - x_1$ and arrive at the final replacement
\begin{align}
x_{13}^4 x_{24}^2 x_{12}^2 I^6_{1,2,2,1,2,1,1} \rightarrow x_{13}^6 x_{24}^2 x_{12}^2 I^6_{2,2,2,1,2,1,1}.
\end{align}
Graphically, the steps we have performed can be represented by 
\begin{align}
x_{13}^4 x_{24}^2 x_{12}^2\includegraphicsbox{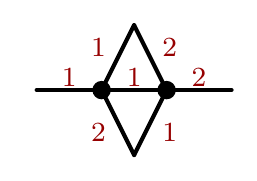}\longrightarrow x_{24}^2 \includegraphicsbox{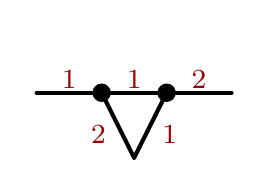} \longrightarrow  x_{13}^6 x_{24}^2 x_{12}^2 \includegraphicsbox{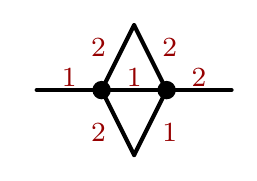}.
\end{align}
Let us stress again that this is not an identity between integrals but only valid if the integral on the left-hand side appears as a summand in some conformal expression. The full conformalised result for $F_2$ and $F_3$ in \eqref{eq:F2F3F4} reads
\begin{align}\label{eq:conformalF2F3}
&F_2(u,v)=x_{12}^2 x_{24}^2 x_{13}^4(I^6_{2,1,2,1,1,1,2}+x_{13}^2I^6_{2,2,2,1,2,1,1}+x_{14}^2I^6_{2,2,2,1,1,2,1}), \\ \notag
&F_3(u,v)= x_{24}^2 x_{13}^6(I^6_{1,2,1,1,2,1,2}+I^6_{2,1,1,2,1,1,2}+x_{13}^2I^6_{2,2,1,2,2,1,1}+x_{14}^2I^6_{2,2,1,2,1,2,1}).
\end{align}
\section{Yangian Ward Identities}\label{sec:YangWard}

In the previous section we discussed how to make a tensor decomposition of the vector integrals appearing in the right hand side of the Yangian Ward identities, and conformalise the resulting expressions. Here we present the refined Yangian Ward identities for the ladders and the window integral.

\subsection{Box}
For completeness we write down the homogeneous Yangian Ward indentities for the box integral 
 \begin{equation}\label{eq:BoxFig}
I_{11}=\int \frac{\dd^4 x_a}{\pi^2}\frac{1}{x_{a1}^2x_{a2}^2x_{a3}^2x_{a4}^2}=\frac{1}{x_{13}^2x_{24}^2}\phi_{11}(u,v) = \includegraphicsbox{FigBox.pdf},
\end{equation}
which take the form
\begin{equation}
u\Duv^{11}\phi_{11}=0, \qquad v\Dvu^{11}\phi_{11}=0.
\end{equation}
These identities were used in \cite{Loebbert:2019vcj} to bootstrap the Bloch--Wigner function (divided by $z-\bar{z}$)
\begin{equation}\label{eq:BlochWigner}
\phi_{11}(z,\bar{z})=\frac{2 \text{Li}_2(z)-2 \text{Li}_2(\bar{z})+(\log z+\log\bar{z})(\log(1-z)-\log(1-\bar{z}))}{z-\bar{z}}.
\end{equation}
This was extended to Minkowskian kinematics in \cite{Corcoran:2020epz}.
\subsection{Double Ladder}
Let us introduce operators $A_j$ which raise the propagator power $\nu_j$ by 1, and also a dimension-raising operator $\mathrm{d}^+$ which increases the dimension of an integral $D\rightarrow D+2$ and adds a factor of $1/\pi$ per loop.
Then using \eqref{eq:F2F3F4} the Yangian constraint \eqref{eq:Yangianconstraint} for the double ladder can be written as
 \begin{align}\label{eq:YangPDEnoshift}
 \left[2u\Duv^{21}-x_{12}^2\gen{d}^+ A_3(A_2(A_1+A_5+A_6+A_7)-A_1A_7)\right]\phi_{21}=0,\\ \notag
\left[2v\Dvu^{21}-x_{14}^2\gen{d}^+ A_6(A_1(A_2+A_3+A_4+A_7)-A_2A_7)\right]\phi_{21}=0,
 \end{align}
where the propagators are labelled as their powers in \eqref{eq:DoubleLadderGenFormula}:
 \begin{equation}\label{eq:DoubleLadderFig}
 \includegraphicsbox{FigDoubleLadderFeynmanParameters.pdf}.
 \end{equation}
 The above equations \eqref{eq:YangPDEnoshift} are mapped into each other under $x_2\leftrightarrow x_4$, which corresponds to $u\leftrightarrow v$. One can obtain two more equations of this type by swapping $x_1\leftrightarrow x_3$. There are two ways to proceed to simplify \eqref{eq:YangPDEnoshift}. Firstly, one can use the manifestly conformal version \eqref{eq:conformalF2F3} of $F_2(u,v)$. This leads directly to
\begin{align}\label{eq:WardDoubleLadder1}
&[2u\Duv^{21} - x_{12}^2 \mathrm{d}^+ A_{1,2,3}(x_{13}^2A_5+x_{14}^2A_6)]\phi_{21}=0, \\ \notag
&[2v\Dvu^{21} - x_{14}^2 \mathrm{d}^+ A_{1,2,6}(x_{13}^2A_4+x_{12}^2A_3)]\phi_{21}=0.
\end{align}
 Secondly, \eqref{eq:YangPDEnoshift} can be simplified by acting with $\partial_{x_{12}^2}$ on the Schwinger parametrisation
 \begin{align}\label{eq:dx12schwinger}
&\partial_{x_{12}^2}\left(\prod_{j=1}^7\int_{0}^\infty \frac{\dd \alpha_j \alpha_j^{\nu_j-1}}{\Gamma_{\nu_j}}\right)\frac{\exp(-\mathcal{F}/\mathcal{U})}{\mathcal{U}^{2}}=-\left(\prod_{j=1}^7\int_{0}^\infty \frac{\dd \alpha_j \alpha_j^{\nu_j-1}}{\Gamma_{\nu_j}}\right)\frac{\partial_{x_{12}^2}\mathcal{F}\exp(-\mathcal{F}/\mathcal{U})}{\mathcal{U}^{3}}\\
&=-\left(\prod_{j=1}^7\int_{0}^\infty \frac{\dd \alpha_j \alpha_j^{\nu_j-1}}{\Gamma_{\nu_j}}\right)\frac{(\alpha _1 \alpha _2 \alpha _3 +\alpha _2 \alpha _3 \alpha _5 +\alpha _2 \alpha _3 \alpha _6 +\alpha _1 \alpha _3 \alpha
   _7 +\alpha _2 \alpha _3 \alpha _7)\exp(-\mathcal{F}/\mathcal{U})}{\mathcal{U}^{3}}.\notag
 \end{align}
Using \eqref{eq:dx12schwinger} for $\nu_j=1$ we can write \eqref{eq:YangPDEnoshift} as 
 \begin{align}
\brk[s]*{2u\Duv^{21}+x_{12}^2 \partial_{x_{12}^2}+2x_{12}^2\gen{d}^+ A_1A_3A_7}\phi_{21}=0, 
\\ \notag
\left[2v\Dvu^{21}+x_{14}^2 \partial_{x_{14}^2}+2x_{14}^2\gen{d}^+ A_2A_6A_7\right]\phi_{21}=0.
 \end{align}
Here one should take the $x_{ij}^2$ in the Schwinger parametrisation to be all independent. 
Replacing $x_{12}^2 \partial_{x_{12}^2}\phi(u,v)\to u\partial_u\phi$ and $x_{14}^2 \partial_{x_{14}^2}\phi(u,v)\to v\partial_v\phi$ and dividing by 2 we have finally

\begin{align}\label{eq:WardDoubleLadder2}
\brk[s]*{u\Duv^{21}+\frac{u}{2} \partial_{u}+x_{12}^2\gen{d}^+ A_1A_3A_7}\phi_{21}=0,
\\ \notag
\left[v\Dvu^{21}+\frac{v}{2} \partial_{v}+x_{14}^2\gen{d}^+ A_2A_6A_7\right]\phi_{21}=0.
 \end{align}
Interestingly, \eqref{eq:WardDoubleLadder2} reveals that $F_4(v,u)=x_{12}^2\gen{d}^+ A_1A_3A_7\phi_{21}$, a 6-dimensional conformal integral, can be expressed as a (shifted) Yangian differential operator acting on the double ladder function $\phi_{21}$. 
Both \eqref{eq:WardDoubleLadder1} and \eqref{eq:WardDoubleLadder2} are manifestly conformal representations of the Yangian Ward identities for the double ladder. We find that in general acting with $\partial_u$ on $\phi_{\alpha\beta}$ leads to the most compact Ward identity. However, we still find the representation \eqref{eq:WardDoubleLadder1} useful in the context of separability in two dimensions, see section \ref{sec:separation}.

\subsection{$\ell$-Ladder}
We can write down the Ward identities for the general ladders algorithmically. Recall the first double ladder equation with the insertion of a derivative $\partial_u$ reads
 \begin{align}
 \left[u\Duv^{21}+\frac{u}{2}\partial_u+x_{12}^2\gen{d}^+ A_{1,3,7}\right]\phi_{21}=0,
 \end{align}
 where we abbreviate $A_{j_1,j_2,\dots, j_n}\coloneqq A_{j_1}A_{j_2}\dots A_{j_n} $. We define the generalised triple ladder as
 \begin{align}
 I_{31}^{\nu,D}=\int\frac{\dd^Dx_a}{\pi^{D/2}}\frac{\dd^Dx_{b}}{\pi^{D/2}}\frac{\dd^Dx_{c}}{\pi^{D/2}}\frac{1}{(x_{a1}^{2\nu_1}x_{a3}^{2\nu_7}x_{a4}^{2\nu_8})x_{ab}^{2\nu_9}(x_{b1}^{2\nu_2}x_{b3}^{2\nu_6})x_{bc}^{2\nu_{10}}(x_{c1}^{2\nu_{3}}x_{c2}^{2\nu_4}x_{c3}^{2\nu_5})},
 \end{align}
 which can be represented in diagrammatic form as
 \begin{equation}\label{eq:TripLadderFig}
 \includegraphicsbox{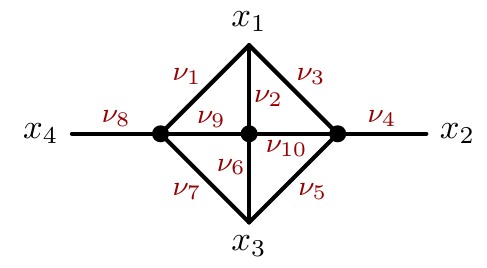} .
 \end{equation}
 We define the conformal function in the unit-propagator four-dimensional limit as
 \begin{equation}
 \phi_{31}(u,v)=\int\frac{\dd^4x_a}{\pi^{2}}\frac{\dd^4x_{b}}{\pi^{2}}\frac{\dd^4x_{c}}{\pi^{2}}\frac{x_{13}^6x_{24}^2}{(x_{a1}^{2}x_{a3}^{2}x_{a4}^{2})x_{ab}^{2}(x_{b1}^{2}x_{b3}^{2})x_{bc}^{2}(x_{c1}^{2}x_{c2}^{2}x_{c3}^{2})}.
 \end{equation}
 Applying the same methods on the Ward identity \eqref{eq:Yangianconstraint2}, the triple ladder equation takes the form
\begin{equation}\label{eq:YangTrip}
\left[u\Duv^{31}
+u\partial_u
+2x_{12}^2\mathrm{d}^+A_{1,4,9,10}
+x_{12}^2\mathrm{d}^+A_{2,4,10}(A_1+A_7+A_8+A_9)\right]\phi_{31}=0.
\end{equation}
The general $\ell$-ladder has $3\ell+1$ propagators, to which we assign propagator powers analogously to \eqref{eq:DoubleLadderFig} and \eqref{eq:TripLadderFig}:
\begin{equation}
I_{\ell 1}^{\nu,D}= \includegraphicsbox{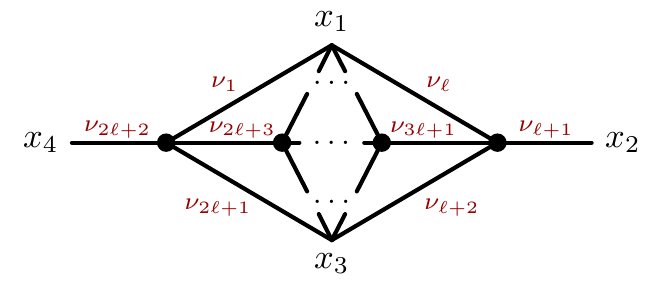} ,
\end{equation}
from which the conformal function is defined as
\begin{equation}
\phi_{\ell 1}= x_{13}^{2\ell} x_{24}^2 I_{\ell 1}^{\nu_j=1, D=4}.
\end{equation}
The quadruple ladder equation takes the form
\begin{align}\label{eq:YangianWardQuad}
&\Big[u\Duv^{41}
+\frac{3}{2}u\partial_u
+3x_{12}^2\mathrm{d}^+A_{1,5,11,12,13}
+2x_{12}^2\mathrm{d}^+A_{2,5,12,13}(A_1+A_9+A_{10}+A_{11})\\ \notag
&+x_{12}^2\mathrm{d}^+A_{3,5,13}\big(A_1 A_2+A_9 A_2+A_{10} A_2+A_{11} A_2+A_1 A_8+A_8 A_9+A_8 A_{10}+\\ \notag&A_1 A_{11}+A_8 A_{11}+A_9 A_{11}+A_{10} A_{11}+A_1 A_{12}+A_9 A_{12}+A_{10} A_{12}+A_{11} A_{12}\big) \Big]\phi_{41}=0.
\end{align}
Note the 15-term expression on the second two lines of \eqref{eq:YangianWardQuad} is the $\mathcal{U}$ polynomial for a double ladder with Feynman parameters $(\alpha_1,\alpha_2,\alpha_8,\alpha_9,\alpha_{10},\alpha_{11},\alpha_{12})$, cf.\ \eqref{eq:DoubleLadderU}. If $\mathcal{U}^{(j)}$ is the $\mathcal{U}$ polynomial for the $j$-ladder where we replace $\alpha\rightarrow A$ then we can rewrite \eqref{eq:YangianWardQuad} as
\begin{equation}\label{eq:YangQuad}
\Big[u\Duv^{41}
+\frac{3}{2}u\partial_u
+3x_{12}^2\mathrm{d}^+A_{1,5,11,12,13}
+2x_{12}^2\mathrm{d}^+A_{2,5,12,13}\mathcal{U}^{(1)}_{1,9,10,11}
+x_{12}^2\mathrm{d}^+A_{3,5,13}\mathcal{U}^{(2)}_{1,2,8,\dots,12} \Big]\phi_{41}=0,
\end{equation}
where $\mathcal{U}^{(1)}_{1,9,10,11}
=A_1+A_9+A_{10}+A_{11}$ is the $\mathcal{U}$ polynomial for the box, or 1-ladder. This pattern persists for the higher ladders
\begin{equation}
\Big[u\Duv^{\ell 1}+\frac{\ell-1}{2}u\partial_u+x_{12}^2 \mathrm{d}^+\sum_{j=0}^{\ell-2}(\ell-1-j)A_{j+1,\ell+1,2\ell+3+j,\dots, 3\ell+1}\mathcal{U}^{(j)}_{1,\dots,j,2\ell-j+2,\dots,2\ell+j+2}\Big]\phi_{\ell 1}=0,
\label{eq:LadderWI}
\end{equation}
where we take $\mathcal{U}^{(0)}=1$. This has been verified explicitly up to the sextuple ladder. One can apply the algorithm presented in section \ref{sec:conformalise} to render the expression manifestly conformal and more compact. For example, the triple ladder equation \eqref{eq:YangTrip} becomes 
\begin{equation}
\left[u\Duv^{31}
+u\partial_u
+3x_{12}^2\mathrm{d}^+A_{1,4,9,10}
+x_{12}^2\mathrm{d}^+A_{1,2,4,10}(x_{13}^3A_{7}+x_{14}^2A_8)\right]\phi_{31}=0,
\end{equation}
and the quadruple ladder equation \eqref{eq:YangQuad} becomes
\begin{align}
&\Big[u\Duv^{41}
+\frac{3}{2}u\partial_u
+6x_{12}^2\mathrm{d}^+A_{1,5,11,12,13}
+3x_{12}^2\mathrm{d}^+A_{1,2,5,12,13}(x_{13}^2 A_9 + x_{14}^2A_{10})\\ \notag
&+x_{12}^2\mathrm{d}^+A_{3,5,13}(x_{13}^2A_{2,9,11}+x_{13}^2A_{1,8,11}+x_{14}^2A_{2,10,11}+x_{13}^4A_{1,2,8,9}+x_{13}^2x_{14}^2A_{1,2,8,10})\Big]\phi_{41}=0.
\end{align}

\subsection{Window}
We also present the Yangian Ward identities for the conformal window integral, given by
\begin{equation}
I_{22}^{\nu, D}= \includegraphicsbox{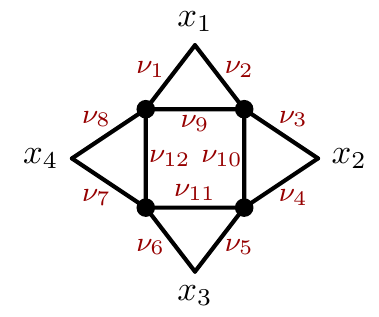},
 \end{equation}
 with the conformal function defined as
 \begin{equation}
 \phi_{22}(u,v)=x_{13}^4x_{24}^4 I_{22}^{\nu_j=1, D=4}.
 \end{equation}
The Yangian equation takes the form
\begin{align}\label{eq:YangianWardWindow}
&[u\Duv^{22}+u\partial_u+2x_{12}^2 \mathrm{d}^+(A_{1,4,11,12}\mathcal{U}^{(1)}_{2,3,9,10}+A_{1,4,9,10}\mathcal{U}^{(1)}_{6,7,11,12})+ \\ \notag
& x_{12}^2 \mathrm{d}^+(\mathcal{U}^{(1)}_{6,7,11,12}(A_{1,3,9}(A_4+A_5+A_{10})+A_{2,4,10}(A_1+A_8+A_9))+\\ \notag
&(A_{1,3}+A_{2,4}) (A_{11}(A_9(A_6+A_7)+A_{12}(A_9+A_{10})))\\ \notag
&+A_{2,4}(A_6+A_7)(A_{10,12}-A_{9,11}))]\phi_{22}=0.
\end{align}
We leave further investigation of this identity for the future.
\subsection{Momentum Space Conformal Anomaly}
\label{sec:momspaceanom}

In this section we  give a brief overview of how the above Yangian Ward identities can be mapped to the momentum space, which is dual to the above $x$-coordinate space. Here the dual $x$- and $p$-coordinates are related by 
\begin{equation}
p_j^\mu=x_j^\mu-x_{j+1}^\mu.
\end{equation}
While the symmetry equations may thereby alternatively be written in momentum space, we emphasise that the Yangian picture was crucial for their derivation. 
To translate the Yangian generators into momentum space it suffices to follow the arguments of \cite{Loebbert:2020glj} with masses set to zero. There it was  shown that the bi-local Yangian level-one generator $\levo{P}^\mu$ as given in \eqref{eq:Phatexpl}, maps to the local special conformal generator in momentum space, i.e.\ on quantities obeying momentum conservation we have
\begin{equation}
\levo{P}^\mu
\simeq
\ihalf \gen{\bar K}^\mu=\ihalf \sum_{j=1}^n \gen{\bar K}^\mu_j.
\end{equation}
Here the $\simeq$ implies that by use of momentum conservation the momentum $p_n$ is eliminated in the quantity acted on, which allows to drop derivatives $\partial_{p_n}$.%
\footnote{
Note that using the definition \eqref{eq:Phatexpl}  of $\levo{P}$, the $n$th momentum is distinguished on the left hand side.}
The generator $\gen{\bar K}^\mu_j$ forms part of the following momentum space representation of the conformal algebra:
\begin{align}
\gen{\bar P}_j^\mu &= p_j^\mu \, , 
&\gen{\bar K}_j^\mu &= p_j^\mu \, \partial_{p_j} \cdot \partial_{p_j}-2\, p_j \cdot \partial_{p_j} \partial_{p_j}^{\mu} - 2 \bar{\Delta}_j\partial_{p_j}^{\mu} \, ,\notag\\
\gen{\bar D}_j &= p_j \cdot \partial_{p_j} + \bar{\Delta}_j \, , 
& \gen{\bar L}_j^{\mu\nu} &=  p_j^\mu \partial_{p_j}^{\nu} -  p_j^\nu \partial_{p_j}^{\mu} \, .
\end{align}
Hence, we can alternatively understand the Yangian Ward identities for $\levo{P}^\mu$ as anomaly equations for a 
momentum space special conformal symmetry. Note that according to the prescription given in  \cite{Loebbert:2020glj}, with \eqref{eq:DeltasAndss} for the Basso--Dixon integrals, we here use%
\footnote{
Note that from \cite{Loebbert:2020glj} we deduce the formula
$
\bar \Delta_j=(\Delta_j+\Delta_{j+1})/{2}+s_j-s_{j+1},
$
which using \eqref{eq:DeltasAndss} yields 
$
\bar \Delta_j=(1+\sfrac{\alpha+\beta}{2},1+\sfrac{\alpha+\beta}{2},1+\sfrac{\alpha+\beta}{2},-3+\sfrac{\alpha+\beta}{2})_j.
$
Eliminating the last momentum using momentum conservation, $\bar \Delta_4$ does not contribute to the action of $ \gen{\bar K}^\mu$ and we can use the homogeneous prescription \eqref{eq:Deltabar}.
}
\begin{equation}
\bar \Delta_j=1+\sfrac{\alpha+\beta}{2}, 
\qquad 
j=1,\dots, 4.
\label{eq:Deltabar}
\end{equation}

\paragraph{Example: Double Ladder.}
The Yangian identities for the double ladder Feynman integral were given in \eqref{eq:DoubleLadderExampleFigures}. Due to the above arguments we can formulate them in momentum space:
\begin{align}
\gen{\bar K}^\mu \includegraphicsbox{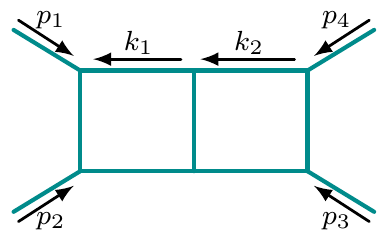} =  +&\includegraphicsbox{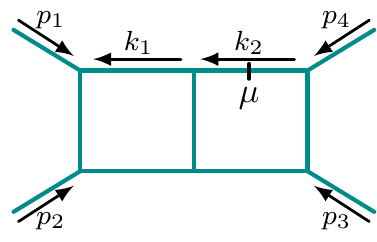}-\includegraphicsbox{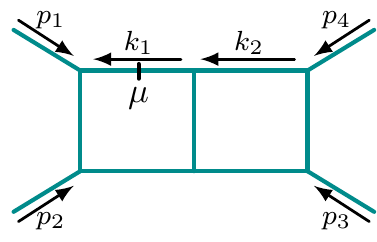}\notag\\
+& \includegraphicsbox{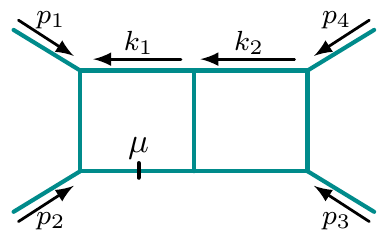}-\includegraphicsbox{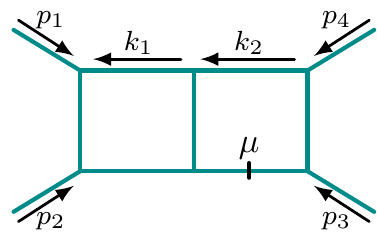},
\label{eq:DoubleLadderExampleFiguresMomSpace}
\end{align}
where the momentum flowing through unlabelled lines can be determined by momentum conservation. Here the respective momentum space Feynman integrals (green lines) are explicitly given by 

\begin{align}
I_{21}
&=\int \frac{\dd^4k_1}{\pi^2}\int \frac{\dd^4k_2}{\pi^2}\frac{1}{k_2^2(k_2+p_{12})^2(k_2+p_{123})^2(k_2-k_1)^2k_1^2(k_1+p_1)^2(k_1+p_{12}^2)},
\\
I_{21}^{\mu,2}
&=\int \frac{\dd^4k_1}{\pi^2}\int \frac{\dd^4k_2}{\pi^2}\frac{k_1^{\mu}}{k_2^2(k_2+p_{12})^2(k_2+p_{123})^2(k_2-k_1)^2k_1^4(k_1+p_1)^2(k_1+p_{12}^2)},
\end{align}
where $p_{ij}\coloneqq p_i+p_j$ and $p_{ijk}\coloneqq p_i+p_j+p_k$.
Note that in order to act with $\gen{\bar K}^\mu $ on the Feynman integral, we eliminate one (arbitrary) momentum using momentum conservation, e.g.\ $p_4$. Then using \eqref{eq:Deltabar} we find the momentum space version of \eqref{eq:pmuhatBD},
\begin{align}
\gen{\bar K}^\mu I_{\alpha\beta}=\frac{4}{p_{12}^{2\alpha}p_{23}^{2\beta}}
\brk[s]*{
\brk*{\frac{p_{1}^\mu}{p_{1}^2}+\frac{p_{3}^\mu}{p_{3}^2}}u \Duv^{\alpha\beta}\phi_{\alpha\beta}(u,v)
+\brk*{\frac{p_{2}^\mu}{p_{2}^2}+\frac{p_{4}^\mu}{p_{4}^2}}v \Dvu^{\alpha\beta}\phi_{\alpha\beta}(u,v)
},
\end{align}
where $p_4$ is understood to be replaced by $-p_1-p_2-p_3$. This can then be compared with the right hand side of \eqref{eq:DoubleLadderExampleFiguresMomSpace}.


\section{Generalisation to Parametric Integrals in $D$ Dimensions}\label{sec:generalD}

Above we have derived Yangian Ward identities for Basso--Dixon integrals with unit propagators in $D=4$. In this section we consider a natural generalisation of that derivation for integrals with generalised propagator powers in $D$ spacetime dimensions. Following the derivation of \eqref{eq:PhatIdentityCorr} naturally leads to a two-parameter family of $D$-dimensional Basso--Dixon integrals. Remarkably, these integrals can be identified with Basso--Dixon correlators in the $D$-dimensional fishnet theory proposed in \cite{Kazakov:2018qbr}, defined by the Lagrangian
 \begin{equation}\label{eq:LFNomegaD}
\mathcal{L}_{\text{FN}}^{\omega D}= N_c\tr \brk[s]*{ -X(-\partial_\mu \partial^{\mu})^{\omega} \bar{X}- Z (-\partial_\mu\partial^{\mu})^{\frac{D}{2}-\omega} \bar{Z}+\xi^2 XZ\bar{X}\bar{Z}},
 \end{equation}
with the anisotropy $\omega\in(0,D/2)$. Like the original fishnet theory, this is a non-unitary theory of scalars $X, Z$. It is also non-local, except for very special choices of $D,\omega$, namely when $D\in 4\mathbb{N}$ for isotropic $\omega=D/4$. Despite this, the theory appears to be integrable, and the corresponding fishnet Feynman integrals still enjoy a Yangian symmetry. At the level of Feynman integrals, the vertical propagator powers are $\omega,$ and the horizontal ones are $D/2-\omega$. 
 
 We present the homogeneous Ward identities for the conformal box in this theory. We then give our derivation of the Ward identities for the generalised double ladder in detail. Moreover we present the identities for the generalised triple ladder and generalised $\ell$-ladder.

\subsection{Generalised Box}
In the theory \eqref{eq:LFNomegaD} we consider the correlator 
\begin{equation}
\left\langle\text{tr}(Z(x_1)\bar{X}(x_2) \bar{Z}(x_3)X(x_4))\right\rangle_{\omega D}= \includegraphicsbox{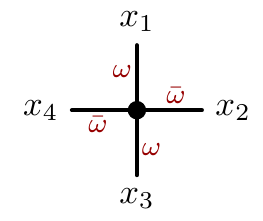},
\label{eq:4ptcorrelatorOmegaD}
\end{equation}
where $\bar{\omega}\coloneqq D/2-\omega$. \eqref{eq:4ptcorrelatorOmegaD} is represented by the Feynman integral
\begin{equation}\label{eq:BoxOmegaD}
\int \frac{\dd^D x_a}{\pi^{D/2}}\frac{1}{x_{a1}^{2\omega}x_{a2}^{D-2\omega}x_{a3}^{2\omega}x_{a4}^{D-2\omega}}=\frac{1}{x_{13}^{2\omega}x_{24}^{D-2\omega}}\phi_{11}^{\omega D}(u,v).
\end{equation}
The homogeneous Yangian Ward identities for the conformal function read
\begin{equation}\label{eq:YangWardBoxOmegaD}
\Duv^{\omega D}\phi_{11}^{\omega D} = \Dvu^{\omega D} \phi_{11}^{\omega D}=0,
\end{equation}
where
\begin{equation}
\Duv^{\omega D} = \omega(\sfrac{D}{2}-\omega)+\left(\sfrac{D}{2}+1\right)v\partial_v +\left (\left(\sfrac{D}{2}+1\right)u-1\right)\partial_u +v^2\partial^2_v+u(u-1)\partial_u^2 +2uv\partial_u\partial_v.
\end{equation}
Equations \eqref{eq:YangWardBoxOmegaD} were solved in terms of Appell hypergeometric $F_4$ functions for slightly more general conformal propagator powers in \cite{Loebbert:2019vcj}.
\subsection{Generalised Double Ladder}

We consider the four-point correlator
\begin{equation}
\left\langle\text{tr}(Z^2(x_1)\bar{X}(x_2) \bar{Z}^2(x_3)X(x_4))\right\rangle_{\omega D}=
\includegraphicsbox{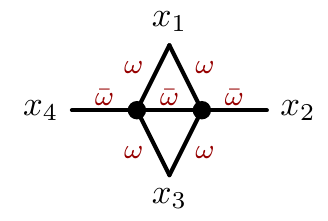},
\label{eq:4ptcorrelatordoubleladdernuD}
\end{equation}
which is represented by the modified double ladder integral
\begin{equation}
I^{\omega D}_{21}=\int\frac{\dd^D x_a}{\pi^{D/2}}\frac{\dd^{D} x_{b}}{\pi^{D/2}}\frac{1}{(x_{a1}^{2\omega}x_{a3}^{2\omega}x_{a4}^{D-2\omega})x_{ab}^{D-2\omega}(x_{b1}^{2\omega}x_{b2}^{D-2\omega}x_{b3}^{2\omega})}=\frac{1}{x_{13}^{4\omega}x_{24}^{D-2\omega}}\phi_{21}^{\omega D}(u,v).
\label{eq:doubleladderOmegaD}
\end{equation}
We use the generic expression \eqref{eq:Phatgen} with $\Delta_a=\omega$ to obtain the action of $\levo{P}^\mu$ on the four-point correlator \eqref{eq:4ptcorrelatordoubleladdernuD}:
\begin{align}
H^\mu_{21,\omega D}\coloneqq&
2i\levo{P}^\mu |_{14}\left\langle\text{tr}(Z^2(x_1)\bar{X}(x_2) \bar{Z}^2(x_3)X(x_4))\right\rangle_{\omega D}
\label{eq:RHSdoubleladdernuD}
\\ \notag
=&+\omega \left\langle\text{tr}(Z(x_1)[\partial_{1}^\mu Z(x_1)]\bar{X}(x_2) \bar{Z}^2(x_3)X(x_4))\right\rangle_{\omega D}
\\ 
&- \omega\left\langle\text{tr}([\partial_{1}^\mu Z(x_1)]Z(x_1)\bar{X}(x_2) \bar{Z}^2(x_3)X(x_4))\right\rangle_{\omega D}
\nonumber\\ \notag
&+\omega\left\langle\text{tr}(Z^2(x_1)\bar{X}(x_2) \bar{Z} (x_3) [\partial_{3}^\mu \bar{Z}(x_3)]X(x_4))\right\rangle_{\omega D}
\\
&-\omega \left\langle\text{tr}( Z^2(x_1)\bar{X}(x_2) [\partial_{3}^\mu \bar{Z} (x_3)] \bar{Z}(x_3)X(x_4))\right\rangle_{\omega D}.
\nonumber
\end{align}
Let us first evaluate the right hand side of this expression.
The correlator appearing in the second line of \eqref{eq:RHSdoubleladdernuD} can be represented by the integral
\begin{align}
2\omega I_{21,\omega D}^\mu&=2\omega\int\frac{\dd^D x_a}{\pi^{D/2}}\frac{\dd^{D} x_{b}}{\pi^{D/2}}\frac{x_{b1}^\mu}{(x_{a1}^{2\omega}x_{a3}^{2\omega}x_{a4}^{D-2\omega})x_{ab}^{D-2\omega}(x_{b1}^{2(\omega+1)}x_{b2}^{D-2\omega}x_{b3}^{2\omega})} \\
&=-\frac{2\omega}{x_{13}^{4\omega}x_{24}^{D-2\omega}}\left(\frac{x_{12}^\mu}{x_{12}^2}F^{\omega D}_2(u,v)+\frac{x_{13}^\mu}{x_{13}^2}F^{\omega D}_3(u,v)+\frac{x_{14}^\mu}{x_{14}^2}F^{\omega D}_4(u,v) \right),
\end{align}
where the conformal vector decomposition \eqref{eq:alphabetavectordecomp} still holds in this case.
The other three correlators are related to this expression by the transpositions $x_1\leftrightarrow x_3$ and $x_2\leftrightarrow x_4$.
Computing $H^\mu_{21,\omega D}$ in full we thus find 
\begin{equation}\label{eq:HnuD}
x_{13}^{4\omega}x_{24}^{D-2\omega}H^\mu_{21,\omega D}=-2\omega^2\left(\frac{x_{12}^\mu}{x_{12}^2}+\frac{x_{34}^\mu}{x_{34}^2}\right)(F^{\omega D}_2(u,v)-F^{\omega D}_4(v,u))-x_1\leftrightarrow x_3.
\end{equation}
In order to obtain the left hand side of \eqref{eq:RHSdoubleladdernuD} we act with $2i\levo{P}^\mu$ on the generic conformal integral
of the form
\begin{equation}
 I_{21}^{\omega D}=\frac{1}{x_{13}^{4\omega}x_{24}^{D-2\omega}}\phi_{21}^{\omega D}(u,v),
\end{equation}
which yields
\begin{align}
2i\levo{P}^\mu I_{21}^{\omega D}=\frac{-4}{x_{13}^{4\omega}x_{24}^{D-2\omega}}
\brk[s]*{
\brk*{\frac{x_{12}^\mu}{x_{12}^2}+\frac{x_{34}^\mu}{x_{34}^2}}u \Duv^{21,\omega D}\phi_{21}^{\omega D}(u,v)
+\brk*{\frac{x_{23}^\mu}{x_{23}^2}+\frac{x_{41}^\mu}{x_{41}^2}}v \Dvu^{21,\omega D}\phi_{21}^{\omega D}(u,v)
}.
\label{eq:pmuhatBDnuD}
\end{align}
 Here the differential operator $\Duv^{21,\omega D}$ takes the form
 \begin{align}
\Duv^{21,\omega D}=&\brk*{\alpha \beta+(\alpha+\beta+1)v \partial_v  +\brk*{(\alpha+\beta+1)u-\gamma} \partial_u +  v^2 \partial^2_v  +(u-1)u \partial^2_u +2 v u \partial_v \partial_u}  ,
\end{align}
with
\begin{equation}
\alpha=2\omega,
\qquad
\beta=\sfrac{D}{2}-\omega,
\qquad
\gamma=1+\sfrac{\omega}{2}.
\end{equation}
Comparing \eqref{eq:HnuD} and \eqref{eq:pmuhatBDnuD} we thus obtain the following Yangian Ward identity for the $(\omega, D)$ generalisation for the double ladder:
\begin{equation}
2u\Duv^{21,\omega D}\phi_{21}^{\omega D}=\omega^2 (F_2^{\omega D}(u,v)-F_4^{\omega D}(v,u)),
\end{equation}
and the same equation with $u\leftrightarrow v$. Performing a tensor decomposition as in \Secref{sec:tensordecomp}, this equation can be written in the form
\begin{equation}
\left(u\Duv^{21,\omega D}+\frac{\omega}{2}u\partial_u+x_{12}^2\gen{d}^+\omega^2\left(\sfrac{D}{2}-\omega\right)^2A_{1,3,7}\right)\phi_{21}^{\omega D}=0.
\label{eq:DnuWI}
\end{equation}
Currently there is no known functional form for the conformal function $\phi_{21}^{\omega D}$.  However, we can use the conformal Feynman parametrisations \eqref{eq:FeynmanParamDoubleLadderOmegaD} and \eqref{eq:FeynmanParamDoubleLadderNuD} to verify \eqref{eq:DnuWI} numerically. There is a further representation of the Ward identity which is analogous to \eqref{eq:WardDoubleLadder1}: 
 \begin{equation}
[2u\Duv^{21,\omega D} - x_{12}^2 \mathrm{d}^+ \omega^2\left(\sfrac{D}{2}-\omega\right)A_{1,2,3}(\omega x_{13}^2 A_5+\left(\sfrac{D}{2}-\omega\right)x_{14}^2A_6)]\phi^{\omega D}_{21}=0,
\label{eq:WardDoubleLadder2nu}
\end{equation}
and the same with $x_2\leftrightarrow x_4$.


\subsection{Generalised Triple Ladder}
The generalised $\omega,D$ triple ladder integral is given by the formula
\begin{align}\label{eq:tripleladderOmegaD}
I_{31}^{\omega D}&=\int\frac{\dd^Dx_a}{\pi^{D/2}}\frac{\dd^Dx_{b}}{\pi^{D/2}}\frac{\dd^Dx_{c}}{\pi^{D/2}}\frac{1}{(x_{a1}^{2\omega}x_{a3}^{2\omega}x_{a4}^{D-2\omega})x_{ab}^{D-2\omega}(x_{b1}^{2\omega}x_{b3}^{2\omega})x_{bc}^{D-2\omega}(x_{c1}^{2\omega}x_{c2}^{D-2\omega}x_{c3}^{2\omega})}
\nonumber
\\
&=\frac{1}{x_{13}^{6\omega}x_{24}^{D-2\omega}}\phi_{31}^{\omega D}(u,v).
\end{align}
The conformal Feynman parametrisation of $\phi_{31}^{\omega D}(u,v)$ is given in \eqref{eq:FeynmanParamTripleLadderOmegaD}.
In analogy to the above, the Yangian equation is obtained in the form
\begin{equation}
\left[u\Duv^{31,\omega D}
+\omega u\partial_u
+2\omega x_{12}^2\mathrm{d}^+\mathcal{A}_{1,4,9,10}
+\omega x_{12}^2\mathrm{d}^+\mathcal{A}_{2,4,10}(\mathcal{A}_1+\mathcal{A}_7+\mathcal{A}_8+\mathcal{A}_9)\right]\phi^{\omega D}_{31}=0,
\end{equation}
where $\mathcal{A}_j=\nu_jA_j$ and the differential operator $\Duv^{31,\omega D}$ takes the form
\begin{align}
\Duv^{31,\omega D}=&\brk*{\alpha \beta+(\alpha+\beta+1)v \partial_v  +\brk*{(\alpha+\beta+1)u-\gamma} \partial_u +  v^2 \partial^2_v  +(u-1)u \partial^2_u +2 v u \partial_v \partial_u}  ,
\end{align}
with
\begin{equation}
\alpha=3\omega,
\qquad
\beta=\sfrac{D}{2}-\omega,
\qquad
\gamma=1+\omega.
\end{equation}


\subsection{Generalised $\ell$-Ladder}

The above Yangian equations for the box, double and triple ladder suggest the following general pattern for the $\ell$-loop ladder:
\begin{align}
\Duv^{\ell 1,\omega D}=&\brk*{\alpha \beta+(\alpha+\beta+1)v \partial_v  +\brk*{(\alpha+\beta+1)u-\gamma} \partial_u +  v^2 \partial^2_v  +(u-1)u \partial^2_u +2 v u \partial_v \partial_u}  ,
\end{align}
with
\begin{equation}
\alpha=\ell \omega,
\qquad
\beta=\sfrac{D}{2}-\omega,
\qquad
\gamma=1+(\ell-1)\sfrac{\omega}{2}.
\end{equation}
The Yangian equation is a generalised form of \eqref{eq:LadderWI}:
\begin{equation}
\Big[u\Duv^{\ell1,\omega D}+\frac{\ell-1}{2}\omega u\partial_u+\omega x_{12}^2 \mathrm{d}^+\sum_{j=1}^{\ell-1}(\ell-j)\mathcal{A}_{j,\ell+1,2\ell+2+j,\dots, 3\ell+1}\tilde{\mathcal{U}}^{(j-1)}_{1,\dots,j-1,2\ell-j+3,\dots,2\ell+j+1}\Big]\phi_{\ell 1}^{\omega D}=0,
\label{eq:LadderWIOmegaD}
\end{equation}
where $\tilde{\mathcal{U}}^{(j)}$ is the $\mathcal{U}$ polynomial for the $j$-ladder where we replace $\alpha_j\rightarrow \mathcal{A}_j=\nu_j A_j$. 
\section{Separation of Variables in 2 Dimensions}\label{sec:separation}

In the previous section we derived the Yangian Ward identities for the ladders in the generalised fishnet theory $\mathcal{L}_{\text{FN}}^{\omega D}$. Here we present the fact that these equations separate in two dimensions. We begin with the Ward identity for the 2D box integral with anisotropy $\omega$ and conformal function $\phi_\omega(z,\bar{z})$, which has an exact Yangian symmetry. The separated Yangian equations for this function represent ordinary differential equations (ODEs), which can be solved straightforwardly. We then present the separated inhomogeneous equations for the two-dimensional double ladder.

\subsection{Separated Ward Identities for the 2D Box}
We consider the 2D Yangian invariant box integral with anisotropy $\omega$:
\begin{equation}\label{eq:Box2dnu}
\includegraphicsbox{FigBoxOmegaD.pdf}=\int \frac{\dd^2 x_a}{\pi}\frac{1}{x_{a1}^{2\omega}x_{a2}^{2-2\omega}x_{a3}^{2\omega}x_{a4}^{2-2\omega}}=\frac{1}{x_{13}^{2\omega}x_{24}^{2-2\omega}}\phi_\omega(z,\bar{z}),
\end{equation}
where here $\bar{\omega}=1-\omega$. The homogeneous Yangian Ward identities can be derived by using \eqref{eq:LadderWIOmegaD} for $\ell=1$. In terms of $z,\bar{z}$ these are
\begin{equation}\label{eq:YangBoxnud}
[\mathcal{D}^\omega_{z,i}-\mathcal{D}^\omega_{\bar{z},i}]\phi_\omega(z,\bar{z})=0, \qquad i=1,2.
\end{equation}
Here the differential operators $\mathcal{D}^\omega_{z,i}$ take the explicit form
\begin{align}
&\mathcal{D}^\omega_{z,1}=\omega(1-\omega)z+z(2z-1)\partial_z+z^2(z-1)\partial_z^2, \\ \notag
&\mathcal{D}^\omega_{z,2}=\omega(1-\omega)z+(2z^2-3z+1)\partial_z+z(z-1)^2\partial_z^2.
\end{align}
These operators can be linearly combined into a separated form
\begin{align}
\mathcal{D}^\omega_z \coloneqq\frac{\bar z-1}{z-\bar z} \brk[s]*{\mathcal{D}_{z,1}^{\omega}-\mathcal{D}_{\bar z,1}^{\omega}}
-\frac{\bar z}{z-\bar z}\brk[s]*{\mathcal{D}_{z,2}^\omega-\mathcal{D}_{\bar z,2}^\omega}
&=\omega(\omega-1)+ (1-2 z)\partial_{z}+ z(1-z)\partial_{ z}^2,
\end{align}
and analogously for $z\leftrightarrow \bar z$.
The function $\phi_\omega$ thus satisfies the ODEs
\begin{equation}
\mathcal{D}^\omega_z \phi_{\omega} =0,
\qquad\quad
 \mathcal{D}^\omega_{\bar{z}}\phi_\omega=0.
 \label{eq:2dBoxSeparated}
\end{equation}


\subsection{Bootstrapping the 2D Box for $\omega=1/2$: Elliptic $K$}

The next goal is to solve the above differential equations, first for $\omega=1/2$ and subsequently for general $\omega$, in order to determine the 2D box integral.
In two dimensions the isotropic box is given by \eqref{eq:Box2dnu} with $\omega=1/2$:
\begin{equation}
\includegraphicsbox{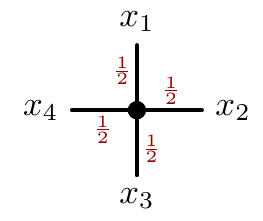}=\int \frac{\dd^2 x_a}{\pi}\frac{1}{|x_{a1}||x_{a2}||x_{a3}||x_{a4}|}=\frac{1}{|x_{13}||x_{24}|}\phi(z,\bar{z}).
\label{eq:2dbox}
\end{equation}
Here we have $|x_{ij}|=\sqrt{x_{ij}^2}$. Yangian invariance implies that $\phi(z,\bar{z})$ satisfies the separated equations \eqref{eq:2dBoxSeparated} with $\omega=1/2$:
\begin{equation}\label{eq:2boxsep}
\mathcal{D}_z \phi = \mathcal{D}_{\bar{z}}\phi = 0,
\qquad
\quad
 \mathcal{D}_z = 1 + 4(2z-1)\partial_z +4z(z-1)\partial_z^2.
\end{equation}
Note that these equations are manifestly symmetric under $z\to 1-z$, $\bar z\to 1-\bar z$. These equations  are ordinary differential equations which can be solved straightforwardly to find
\begin{equation}\label{eq:2boxsepsol}
\phi(z,\bar{z})=f_1(\bar{z})K(z)+f_2(\bar{z})K(1-z),
\end{equation}
and the same with $z\leftrightarrow \bar{z}$. Here the complete elliptic  integral of the first kind is defined by
\begin{equation}
K(z)=\int_0^{\pi/2}\dd\theta \frac{1}{\sqrt{1-z \sin^2 \theta}}.
\end{equation}
Acting on \eqref{eq:2boxsepsol} with $ \mathcal{D}_{\bar{z}}$, and solving the resulting ordinary differential equation for $f_1(\bar z)$ and $f_2(\bar z)$, we conclude that in general $\phi(z,\bar{z})$ is a linear combination of four factorised Yangian invariants:
\begin{equation}
\phi(z,\bar{z})= c_1K(z)K(\bar{z})+c_2K(z)K(1-\bar{z})+c_3K(1-z)K(\bar{z})+c_4K(1-z)K(1-\bar{z}).
\end{equation}


\paragraph{Fixing Constants.}
We can fix the constant prefactors $c_i$ by using reality conditions and the permutation covariance of the integral \eqref{eq:2dbox}. We first note some properties of the function $K(z)$. $K(z)$ has branch points at $z=1$ and $z=\infty$, and we take the usual branch cut on $(1,\infty)$. It satisfies the identities\footnote{We consider the case where $\bar{z}=z^*$ and $\Im(z)\neq 0,$ so that in particular $z\neq \bar{z}.$}
\begin{equation}\label{eq:Kperm1}
K(1-\sfrac{1}{z})=\sqrt{z}K(1-z),
\end{equation}
\begin{equation}\label{eq:Kperm2}
K(\sfrac{1}{z})=\sqrt{z}(K(z)\mp iK(1-z)), 
\end{equation}
where we take the negative sign if $\Im(z)>0$ and the positive sign if $\Im(z)<0$. We also have $K(\bar{z})=K(z)^*$ if $\bar{z}=z^*$. For definiteness we take $z$ to be the solution of $z\bar{z}=u$ and $(1-z)(1-\bar{z})=v$ with positive imaginary part. 

We now proceed to fix the constants $c_i$. Firstly, since \eqref{eq:2dbox} is a Euclidean, real-valued integral we must have $c_2=c_3\coloneqq c$. This is because $K(z)K(1-\bar{z})=(K(1-z)K(\bar{z}))^*$, and this choice ensures that the imaginary parts in these terms cancel. Furthermore, \eqref{eq:2dbox} is invariant under the transposition of points $x_1\leftrightarrow x_3$. This transposition maps $z\rightarrow 1-\bar{z}$ and $\bar{z}\rightarrow 1-z$.  Therefore we should have
\begin{equation}
\phi(z,\bar{z})=\phi(1-\bar{z},1-z),
\end{equation}
which fixes $c_1=c_4\coloneqq \tilde{c}$. The transpositon $x_1\leftrightarrow x_4$ implies the condition
\begin{equation}
\sqrt{u}\phi(z,\bar{z})=\phi(\sfrac{1}{\bar{z}},\sfrac{1}{z}).
\end{equation}
This equation can be expanded
\begin{align}
&\sqrt{z\bar{z}}\brk[s]!{\tilde{c}(K(z)K(\bar{z})+K(1-z)K(1-\bar{z})}
+c\brk[s]!{K(z)K(1-\bar{z})+K(1-z)K(\bar{z}))}\\
&=\tilde{c}\brk[s]!{K(\sfrac{1}{z})K(\sfrac{1}{\bar{z}})+K(1-\sfrac{1}{z})K(1-\sfrac{1}{\bar{z}})}
+c\brk[s]!{K(\sfrac{1}{z})K(1-\sfrac{1}{\bar{z}})+K(1-\sfrac{1}{z})K(\sfrac{1}{\bar{z}})}, \notag
\end{align}
and using \eqref{eq:Kperm1}, \eqref{eq:Kperm2} this gives the constraint
\begin{equation}
\tilde{c}(K(1-z)K(1-\bar{z})+iK(1-z)K(\bar{z})-iK(z)K(1-\bar{z}))=0,
\end{equation}
which is only consistent for generic $z,\bar{z}$ if $\tilde{c}=0$. Overall we have fixed the solution up to an overall constant $c$. We fix the constant $c=4/\pi$ using numerical input, so that the final result is
\begin{equation}
\phi(z,\bar{z})=\frac{4}{\pi}\brk[s]*{K(z)K(1-\bar{z})+K(1-z)K(\bar{z})}.
\label{eq:ResultIsotropicBox2D}
\end{equation}
This result agrees with the one given in \cite{Derkachov:2018rot}, obtained by a direct simplification of hypergeometric functions 
arising from their 
separation of variables approach, see also \cite{Korchemsky:1997fy,Dotsenko:1984nm}. They use a slightly different conformal variable $\eta$, related to our variable $z$ by
\begin{equation}
\eta=1-\frac{1}{z}, \qquad \bar{\eta}=1-\frac{1}{\bar{z}}.
\end{equation}
We chose to use $z$, $\bar z$ since the expression for the conformal function \eqref{eq:ResultIsotropicBox2D} takes a slightly simpler form in terms of these variables.

\paragraph{Cuts and Discontinuities.}

Let us consider discontinuities of the solution \eqref{eq:ResultIsotropicBox2D}, which are related to cuts of the integral \cite{Abreu:2014cla}. Cuts of the integral obey the same Yangian equations \cite{Chicherin:2017cns,Corcoran:2020epz}, and so these discontinuities can be expressed as a linear combination of the four Yangian invariants $K(z)K(\bar{z}), K(1-z)K(\bar{z}), K(z)K(1-\bar{z}),$ and $K(1-z)K(1-\bar{z})$. 

The conformal function \eqref{eq:ResultIsotropicBox2D} is single-valued in $z,\bar{z}$ (in the sense of \cite{Dixon:2012yy}) if $\bar{z}=z^*$, which is true for Euclidean kinematics. It can be thought of as a 2D version of the Bloch--Wigner function \eqref{eq:BlochWigner}, which is similarly single-valued. However for Minkowskian kinematics, $z$ and $\bar{z}$ can also be independent real numbers. 
Conformal invariance is broken globally in this case \cite{Corcoran:2020akn}, and the result for the integral in kinematic regions away from the Euclidean sheet can be obtained by taking discontinuities of \eqref{eq:ResultIsotropicBox2D}. 

We proceed to calculate these discontinuities. For this purpose we only need to calculate the discontinuity of $K(z)$ across the cut on $(1,\infty)$. We have
\begin{equation}
\text{disc}_1 K(z)\coloneqq K(z+i\epsilon)-K(z-i\epsilon)=2i K(1-z), \quad z>1.
\end{equation}
Using this we can fix $\bar{z}\in \mathbb{C}, z>1$ and calculate
\begin{equation}\label{eq:disc1phi}
\text{disc}_1 \phi(z,\bar{z})\coloneqq \phi(z+i\epsilon,\bar{z})-\phi(z-i\epsilon,\bar{z})=-\frac{8}{\pi i}K(1-z)K(1-\bar{z}).
\end{equation}
Similarly fix $z<0$ and $\bar{z}\in\mathbb{C}$. Then
\begin{equation}\label{eq:disc0phi}
\text{disc}_0 \phi(z,\bar{z})\coloneqq \phi(z+i\epsilon,\bar{z})-\phi(z-i\epsilon,\bar{z})=\frac{8}{\pi i}K(z)K(\bar{z}).
\end{equation}
Notably, both $\text{disc}_1 \phi(z,\bar{z})$ and $\text{disc}_0 \phi(z,\bar{z})$ are solutions to the Yangian PDEs \eqref{eq:2boxsep}, as expected.
We can also consider double discontinuities of $\phi(z,\bar{z})$.
Fix now $\bar{z}<0, z>1$. Then we have
\begin{equation} \label{eq:ddisc01phi}
\text{disc}_{\bar{0}} \text{disc}_1 \phi(z,\bar{z})\coloneqq \text{disc}_1 \phi(z,\bar{z}+i\epsilon)-\text{disc}_1 \phi(z,\bar{z}-i\epsilon)=\frac{16}{\pi} K(1-z)K(\bar{z}),
\end{equation}
and similarly if we fix $\bar{z}>1, z<0$, we can find
\begin{equation}\label{eq:ddisc10phi}
\text{disc}_{\bar{1}} \text{disc}_0 \phi(z,\bar{z})=\frac{16}{\pi} K(1-\bar{z})K(z).
\end{equation}
Again, the double discontinuities are solutions to the Yangian PDEs \eqref{eq:2boxsep}.

\paragraph{Transcendentality.} For the isotropic box in
2D we identified four Yangian invariants $g_i$:
\begin{align}\label{eq:invariants2d}
&g_1=K(z)K(\bar{z}),  & g_2&=K(1-z)K(\bar{z}), \\ \notag
&g_3=K(z)K(1-\bar{z}),& g_4&=K(1-z)K(1-\bar{z}).
\end{align}
In comparison, for the 4D conformal box four Yangian invariants $f_i\coloneqq \tilde{f}_i/(z-\bar{z})$ were similarly identified \cite{Loebbert:2019vcj}:
\begin{align}\label{eq:invariants4d}
&\tilde{f}_1=2\text{Li}_2(z)-2\text{Li}_2(\bar{z})+(\log z+\log\bar{z})(\log(1-z)-\log(1-\bar{z})), \\ \notag
&\tilde{f}_2=\log z-\log\bar{z},\\ \notag
&\tilde{f}_3=\log (1-z)-\log(1-\bar{z}), \\ \notag
&\tilde{f}_4= 1.
\end{align}
The respective conformal functions are proportional to the Yangian invariants as follows:
\begin{equation}
 \phi_{\text{2D}}\sim \frac{g_2+g_3}{\pi}, \qquad \phi_{\text{4D}}\sim f_{1}.
 \end{equation}
  We note the differences in the transcendentality of the Yangian invariants in each case. Here $\pi$ is assigned transcendentality 1. While polylogs of order $n$ have transcendentality $n$, the elliptic $K$ integral has been argued in  \cite{Duhr:2019wtr} to have transcendentality $1$, because
\begin{equation}
\lim_{x\to 0}K(x)=\frac{\pi}{2}.
\end{equation}
Therefore each $g_i$ has transcendentality $2$, while the $f_i$ have transcendentality ranging from $0$ to $2$. Furthermore, the conformal functions $\phi_{\text{2D}}$ and $\phi_{\text{4D}}$ have transcendentality $1$ and $2$ respectively. In \tabref{tab:disc} we present the discontinuities of these functions. We notice a curious difference between the two cases. Taking discontinuities of $\phi_{\text{4D}}$ reduces the functional transcendentality by 1 in each case, such that any further discontinuities will simply vanish. However, since all the Yangian invariants for the 2D case have the same transcendentality~2, one can continue to take discontinuities of~$\phi_{\text{2D}}$, and remain within the family of functions $g_i$. This shows, at least in this very simplified setting, a fundamental difference between taking discontinuities of polylogs and elliptic integrals.
\begin{table}
\centering
\renewcommand{\arraystretch}{1.15}
\begin{tabular}{ |c|c|c| } 
 \hline
 & $\phi_{\text{2D}}$ & $\phi_{\text{4D}}$\\ \hline 
 $\text{disc}_1\phi$ & $-\frac{1}{\pi i}g_4$ &$ +\pi i f_2$ \\ 
$\text{disc}_0\phi$ &$ +\frac{1}{\pi i}g_1$&$ +\pi i f_3$ \\
 $\text{disc}_{\bar{0}} \text{disc}_1\phi$ & $+\frac{1}{\pi}g_2$ & $-(\pi i)^2 f_1$\\ 
$\text{disc}_{\bar{1}} \text{disc}_0\phi$ & $+\frac{1}{\pi}g_3$& $+(\pi i)^2 f_1$ \\ 
 \hline
\end{tabular}
\caption{Discontinuities of isotropic box functions $\phi_{\text{2D}}$ and $\phi_{\text{4D}}$ in two and four dimensions, ignoring numerical constants.}
\label{tab:disc}
\end{table}

\subsection{Bootstrapping the 2D Box for Generic $\omega$: Legendre $P$ and $Q$}


For generic $\omega$ the separated differential equations for the box conformal function $\phi_\omega$ read
\begin{align}
0&=\brk[s]*{\omega(\omega-1)+ (1-2 z)\partial_{z}+ z(1-z)\partial_{ z}^2}\phi_\omega,
\\
0
&=\brk[s]*{\omega(\omega-1)+ (1-2\bar z)\partial_{\bar z}+\bar z(1-\bar z)\partial_{\bar z}^2}\phi_\omega,
\end{align}
which are still symmetric under $z\to 1-z$ and $\bar z\to 1-\bar z$, respectively.
The above ODEs are solved by the Legendre functions
\begin{equation}
\phi_\omega(z,\bar{z})=c_1(\bar z) P_{\omega-1}(2z-1)
+
c_2(\bar z)Q_{\omega-1}(2z-1),
\end{equation}
and similarly for $z\leftrightarrow \bar z$. We note that
\begin{align} \label{eq:PtoK}
P_{\omega-1}(2z-1)\big|_{\omega\to \sfrac{1}{2}}
&=
\sfrac{2}{\pi}K(1-z),
\\ \label{eq:QtoK}
Q_{\omega-1}(2z-1)\big|_{\omega\to \sfrac{1}{2}}
&=
K(z),
\end{align}
from which we see that Legendre $P$ has an apparent lower transcendentality than Legendre~$Q$.
With the above solutions for the separated equations, the generic ansatz for the full parametric box reads
\begin{align}
\phi_\omega(z,\bar z)
=
&+c_1P_{\omega-1}(2z-1)P_{\omega-1}(2\bar z-1)
+
c_2 P_{\omega-1}(2z-1)Q_{\omega-1}(2\bar z-1)
\nonumber\\
&+
c_3 Q_{\omega-1}(2z-1)P_{\omega-1}(2\bar z-1)
+
c_4
Q_{\omega-1}(2z-1)Q_{\omega-1}(2\bar z-1).
\end{align}
Using numerical input we can fix these constants, and find
\begin{equation}\label{eq:phinusol}
\phi_{\omega}=\frac{4}{\pi}\brk[s]*{Q_{\omega-1}(\tau)\tilde{P}_{\omega-1}(\bar{\tau})+Q_{\omega-1}(\bar{\tau})\tilde{P}_{\omega-1}(\tau)-2\cot(\pi \omega)\tilde{P}_{\omega-1}(\tau)\tilde{P}_{\omega-1}(\bar{\tau})},
\end{equation}
where $\tau \coloneqq 2z-1$ and 
\begin{equation}
\tilde{P}_{\omega-1}(\tau)\coloneqq\frac{\pi}{2}P_{\omega-1}(\tau).
\end{equation}
 Using \eqref{eq:PtoK} and \eqref{eq:QtoK} this reduces to \eqref{eq:ResultIsotropicBox2D} in the limit $\omega\rightarrow 1/2$. 
This representation differs from the hypergeometric one given in \cite{Derkachov:2018rot}, and this one is perhaps slightly more natural since it expresses a one-parameter integral in terms of one-parameter functions $P_{\omega-1}$ and $Q_{\omega-1}$. 

\paragraph{Discontinuities.}
Again we can verify that the discontinuities of the solution \eqref{eq:phinusol} are given by linear combinations of Yangian invariants. We first give the discontinuities of $\tilde{P}_{\omega-1}(2z-1)$ and $Q_{\omega-1}(2z-1)$:
\begin{align}
&\text{disc}_1 Q_{\omega-1}(2z-1) = 2i \tilde{P}_{\omega-1}(2z-1), \\ \notag
& \text{disc}_1 \tilde{P}_{\omega-1}(2z-1) = 0, \\ \notag
&\text{disc}_0 Q_{\omega-1}(2z-1) = 2i\cos(\pi \omega)( \cos(\pi \omega)\tilde{P}_{\omega-1}(2z-1)-\sin(\pi \omega)Q_{\omega-1}(2z-1)), \\ \notag
& \text{disc}_0 \tilde{P}_{\omega-1}(2z-1) = 2i\sin(\pi\omega)( \cos(\pi \omega)\tilde{P}_{\omega-1}(2z-1)-\sin(\pi \omega)Q_{\omega-1}(2z-1)).
\end{align}
Using these we have
\begin{equation}\label{eq:disc1phinu}
\text{disc}_{1}\phi_\omega(z,\bar{z})=-\frac{8}{\pi i}\tilde{P}_{\omega-1}(2z-1)\tilde{P}_{\omega-1}(2\bar{z}-1),
\end{equation}
and
\begin{align}\label{eq:disc0phinu}
\text{disc}_{0}\phi_\omega(z,\bar{z})
=
\frac{8}{\pi i}\Big[
&\beta_{1,\omega}\tilde{P}_{\omega-1}(2z-1)\tilde{P}_{\omega-1}(2\bar z-1)
+
\beta_{2,\omega} \tilde{P}_{\omega-1}(2z-1)Q_{\omega-1}(2\bar z-1)
\\ 
&+
\beta_{3,\omega} Q_{\omega-1}(2z-1)\tilde{P}_{\omega-1}(2\bar z-1)
+
\beta_{4,\omega}
Q_{\omega-1}(2z-1)Q_{\omega-1}(2\bar z-1)\Big], \notag
\end{align}
where the coefficient functions $\beta_{i,\omega}$ are given by
\begin{equation}
\beta_{1,\omega}=\cos^2( \pi \omega), \qquad 
\beta_{2,\omega}=\beta_{3,\omega}=- \sin (\pi \omega)\cos(\pi\omega), 
\qquad \beta_{4,\omega}=\sin^2( \pi\omega).
\end{equation}
In the limit $\omega\rightarrow 1/2$, equations \eqref{eq:disc1phinu} and \eqref{eq:disc0phinu} reduce to \eqref{eq:disc1phi} and \eqref{eq:disc0phi}, respectively. The double discontinuities take the form
\begin{align}
&\text{disc}_{\bar{0}}\text{disc}_1 \phi_\omega(z,\bar{z})= -\sfrac{16\sin(\pi\omega)}{\pi}\tilde{P}_{\omega-1}(2z-1)\brk[s]*{\cos(\pi\omega)\tilde{P}_{\omega-1}(2\bar{z}-1)-\sin(\pi \omega)Q_{\omega-1}(2\bar{z}-1)},\notag \\ \notag
&\text{disc}_{\bar{1}}\text{disc}_0 \phi_\omega(z,\bar{z})= -\sfrac{16\sin(\pi\omega)}{\pi}\tilde{P}_{\omega-1}(2\bar{z}-1)\brk[s]*{\cos(\pi\omega)\tilde{P}_{\omega-1}(2z-1)-\sin(\pi \omega)Q_{\omega-1}(2z-1)} ,
\end{align}
which reduce to  \eqref{eq:ddisc01phi} and \eqref{eq:ddisc10phi} as $\omega\rightarrow \half$.

\subsection{2D Double Ladder}
Consider the Yangian equations for the double ladder
\begin{equation}
\includegraphicsbox{FigDoubleLadderOmegaD.pdf}
\end{equation}
 as given in
 \eqref{eq:WardDoubleLadder2nu} in $D=2$ with generic $\omega$:
 \begin{align}
 \left[2u\Duv^{21,\omega D}-A_{uv}\mathrm{d}^+ \right]\phi^{\omega D}_{21}=0,
 \label{eq:DblLaddereq1}
 \\
\left[2v\Dvu^{21,\omega D}-A_{vu}\mathrm{d}^+ \right]\phi^{\omega D}_{21}=0. 
 \label{eq:DblLaddereq2}
 \end{align}
 Here we have introduced the shorthand
 \begin{align}
 A_{uv}&:=x_{12}^2 \omega^2(\sfrac{D}{2}-\omega)A_{1,2,3}\brk*{\omega x_{13}^2 A_5+(\sfrac{D}{2}-\omega)x_{14}^2A_6},
 \\
 A_{vu}&:=x_{14}^2 \omega^2(\sfrac{D}{2}-\omega)A_{1,2,6}\brk*{\omega x_{13}^2 A_4+(\sfrac{D}{2}-\omega)x_{12}^2A_3}.
 \end{align}
 Now we switch to $z, \bar z$ such that
 \begin{align}
\brk[s]*{
4\nu(1-\nu)
+(\nu+2)(2z-1)\partial_z
+2z(z-1)\partial_z^2
}\phi(z,\bar z)
 &=
\brk*{\sfrac{1}{1-z}A_{vu}+\sfrac{1}{z} A_{uv}}\mathrm{d}^+\phi(z,\bar z),
\\
\brk[s]*{
4\nu(1-\nu)
+(\nu+2)(2\bar z-1)\partial_{\bar z}
+2\bar z(\bar z-1)\partial_{\bar z}^2
}\phi(z,\bar z)
&=
\brk*{\sfrac{1}{1-\bar z}A_{vu}+\sfrac{1}{\bar z} A_{uv}}\mathrm{d}^+ \phi(z,\bar z).
\end{align}
Again we see that the equations separate into a $z$ and $\bar z$ dependent piece. However, the operators $A_{uv}$ and $A_{vu}$ including the shifts of propagator powers are a priori only defined on a Feynman integral. It is thus not obvious how to solve or interpret these equations as purely mathematical identities for some generic functions in analogy to the above box equations in 2D. We leave the further exploration of this interesting point for future work.

\section{Conclusions and Further Directions}

In this paper we derived the consequences of conformal Yangian symmetry for the Basso--Dixon integrals, and their generalisation to $D$ dimensions. The resulting symmetry equations take the form of inhomogeneous partial differential equations in the cross ratios for the respective conformal functions. The inhomogeneities can be expressed as linear combinations of Basso--Dixon integrals, with shifted propagator powers and dimension $D\rightarrow D+2$. In two dimensions, the Ward identities separate, and we demonstrated a simple application of the Yangian bootstrap for the 2D box integral resulting in products of Legendre functions or elliptic $K$ integrals.

Our findings show that the interpretation of Feynman integrals as correlators or scattering amplitudes in the family of fishnet theories should be taken seriously. In fact, the presented derivation of the Yangian identities was achieved by distinguishing the coordinate and field representation of the symmetry generators. While the resulting symmetry equation can as well be formulated on momentum space Feynman integrals as illustrated in \secref{sec:momspaceanom}, a similar derivation seems obscured within this picture.

There are many directions for further research. Perhaps the most interesting question is whether the inhomogeneous Ward identities presented in this paper can be solved for the conformal functions $\phi_{\alpha\beta}$. Given that the eventual proof of the Basso--Dixon formula did not directly appeal to methods of integrability, it would be interesting if the Ward identities somehow encoded the determinant structure of the Basso--Dixon formula. The main obstacle to the solution of these equations is the inhomogeneity $ \mathcal{A} \mathrm{d}^+\phi_{\alpha\beta}$ with the dimension- and propagator-shift operators $\mathrm{d}^+$ and $\mathcal{A}$, respectively. It is as yet unclear whether these equations can be solved with a series ansatz, as it was possible for the homogeneous case \cite{Loebbert:2019vcj}. One possible direction to proceed would be to try to rewrite the inhomogeneity, for example by using integration by parts or dimensional shifting techniques \cite{Tarasov:1996br,Tarasov:2000sf,Lee:2009dh}, and to understand whether the Yangian equations are actually independent from identities obtained with these methods.

As emphasised above, the derivation of the Ward identities relies on the Yangian invariance of correlation functions in the fishnet theory. Since we have verified the resulting equations on many Feynman integrals, their validity is undoubted. Still it would be important to derive the initial invariance of correlators from first principles, e.g.\ from a Yangian invariance of the Lagrangian. This would require to extend the methods of 
\cite{Beisert:2017pnr,Beisert:2018zxs,Beisert:2018ijg} developed for $\superN=4$ super Yang--Mills theory, such that they can also be applied to the class of fishnet theories. 

Yangian symmetry has also been realised on certain families of massive Feynman integrals~\cite{Loebbert:2020glj,Loebbert:2020hxk,Loebbert:2020tje,Loebbert:2021qef}. Therefore extending the presented Ward identities to massive Basso--Dixon integrals would be natural. In fact, there are very few analytic results available, even for the massive ladder integrals, and as such any analytic constraints on these functions would be valuable.

Notably, Yangian symmetry can be understood as the closure of two different conformal algebras \cite{Drummond:2009fd,Loebbert:2020glj}.
The momentum space interpretation of the level-one momentum Ward identity described in \secref{sec:momspaceanom} suggests to give up on level-zero conformal symmetry in $x$-space and to investigate the momentum space conformal anomaly on its own. 

In \cite{Basso:2021omx} the thermodynamic limit of the Basso--Dixon integrals, and corresponding free energy density $f(k)$, was calculated as a function of the aspect ratio $k$. It would be interesting to take the thermodynamic limit of our Ward identities, and see if they provide any nontrivial constraints on $f(k)$, which could possibly determine it.

Although we focussed in this paper on the limit from the many-point Yangian invariant fishnet graphs to the four-point fishnets, in principle we could consider any coincidence limit of any Yangian invariant Feynman graph. For example, we could investigate coincidence limits of the Yangian invariant $n$-gons in $n$ dimensions, or a window diagram with one integration vertex dropped (broken window).

In two dimensions the Ward identities separate, and for the case of the box the resulting ordinary differential equations can be solved by separation of variables in terms of Legendre functions or elliptic $K$ integrals. We note that in principle it is also possible to solve the Yangian equations in 4D using separation of variables (in a similar sense), although this requires the use of an undetermined separation constant which needs to be integrated over.
Notably, the Basso--Dixon graphs were calculated in 2D using Sklyanin's separation of variables in \cite{Derkachov:2018rot}. We wonder whether there is a link between these facts, and how to make this explicit. There has also been progress on separation of variables in 4D \cite{Derkachov:2019tzo}, with applications to the Basso--Dixon integrals. 

Our Ward identities provide relations between integrals in $D$ and $D+2$ dimensions, with concrete Basso--Dixon formulas available in $D=2$ and $D=4$. It would be interesting to find analogous relations for odd dimensions, for example in $D=3$. In this dimension, the square lattice structure of the fishnet graphs do not appear to be so natural, so perhaps there is a triangular lattice analogue which could be solved exactly. Such graphs could possibly come from the strongly twisted ABJM theory of \cite{Caetano:2016ydc}. Moreover, having the Ward identities for Basso--Dixon integrals at hand should also give us symmetry equations for larger classes of integrals, e.g.\ for the tennis courts, via the magic identities of \cite{Drummond:2006rz}.

Finally, we note that Basso--Dixon integrals also play a role in the context of correlation functions in $\superN=4$ super Yang--Mills theory. In particular they show up in the all-loop bootstrap of the so-called octagon correlator performed in \cite{Coronado:2018cxj}. It should be very interesting to extend the presented Ward identities in the fishnet theory to correlation functions in $\superN=4$ super Yang--Mills theory and other observables entering the AdS/CFT duality.


\subsection*{Acknowledgements}
FL would like to thank Dennis M\"uller for initial collaboration on the idea to generalise the Yangian Ward identities to Basso--Dixon integrals and for important discussions and suggestions.
We thank Thomas Klose and Jos\'e \'Alvarez Roca for many interesting discussions on Basso--Dixon integrals in another context. We also thank Matthias Staudacher for discussions and collaboration on a previous project.
This project has received funding from the European Union's Horizon 2020 research and innovation programme under the Marie Sklodowska-Curie grant agreement  No.\ 764850 `SAGEX'. 
The work of FL is funded by the Deutsche Forschungsgemeinschaft (DFG, German Research Foundation)-Projektnummer 363895012.

\appendix
\section{Conformal Vector Integral Decomposition}\label{app:vectordecomp}
Here we show that the general solution to the inhomogeneous conformal Ward identities for vector integrals \eqref{eq:InhomWard1} and \eqref{eq:InhomWard2}, which we repeat here for convenience,
\begin{align}
\gen{D}I_{\alpha\beta}^{\mu,n} (x_1,\dots,x_4) &= 0,\label{eq:InhomWard1App}\\
\left(\gen{K}^\mu \eta_{\nu\rho} + 2 i (\delta^{\mu}_\nu x_{1,\rho} - \delta^{\mu}_\rho x_{1,\nu})\right)I_{\alpha\beta}^{\rho,n}(x_1,\dots,x_4) &= -2 i \delta^{\mu}_\nu I_{\alpha\beta}(x_1,\dots,x_4),
\label{eq:InhomWard2App}
\end{align}
takes the form
\begin{equation}
x_{13}^{2\alpha}x_{24}^{2\beta}I_{\alpha\beta}^{\mu,n}=-\frac{x_{12}^\mu}{x_{12}^2}F_{2,n}^{\alpha\beta}(u,v)-\frac{x_{13}^\mu}{x_{13}^2}F_{3,n}^{\alpha\beta}(u,v)-\frac{x_{14}^\mu}{x_{14}^2}F_{4,n}^{\alpha\beta}(u,v),
\label{eq:alphabetavectordecompApp}
\end{equation}
with
\begin{align}\label{eq:FsumApp}
F_{2,n}^{\alpha\beta}(u,v) + F_{3,n}^{\alpha\beta}(u,v) + F_{4,n}^{\alpha\beta}(u,v) = I_{\alpha\beta}(u,v).
\end{align}
First of all, Poincar\'e invariance implies that the vector integrals can be parametrised via
\begin{equation}
\label{eq:PoincareParam}
x_{13}^{2\alpha} x_{24}^{2\beta} I_{\alpha \beta}^{\mu, n} = \sum_{j=2}^4 \frac{x_{1j}^\mu}{x_{1j}^2} \tilde{G}_j(x_{12}^2, x_{13}^2, x_{14}^2, x_{23}^2, x_{24}^2, x_{34}^2 ).
\end{equation}
To impose the conformal Ward identities, we first reparamerise the functions $\tilde{G}_j$ in terms of 5 scale-invariant ratios and one $x_{ij}^2$, e.g.%
\footnote{For convenience of the argument, we use different scale-invariant arguments for the different $G_j$. Note that for each $G_j$, one argument is actually equal to 1, hence the $G_j$ and the $\tilde{G}_j$ each depend on the same amount of independent variables.}
\begin{align}
\tilde{G}_j(x_{12}^2, x_{13}^2, x_{14}^2, x_{23}^2, x_{24}^2, x_{34}^2) = G_j\Big(x_{1j}^2,\frac{x_{12}^2}{x_{1j}}, \frac{x_{13}^2}{x_{1j}^2}, \frac{x_{14}^2}{x_{1j}^2}, \frac{x_{23}^2}{x_{1j}^2}, \frac{x_{24}^2}{x_{1j}^2}, \frac{x_{34}^2}{x_{1j}^2}\Big),
\end{align}
which is trivially possible. Then, the dilatation Ward identity \eqref{eq:InhomWard1App} implies a differential equation for the ${G_j}$ with respect to their first variables only. To extract the independent information of this differential equation, we use Poincar\'e invariance to choose a particular configuration of points
\begin{align}
\label{eq:PoincareConfig}
x_1 &= (0,0,0,0), & x_2 &= (\xi,0,0,0), \notag\\
x_3 &= (\eta, \zeta, 0, 0), & x_4 &= (\kappa, \lambda, \rho, 0).
\end{align}
In this parametrisation, \eqref{eq:InhomWard1App} reads
\begin{align}
0 = e_3^\mu \rho G_4' +  e_2^\mu (\lambda G_4'  + \zeta G_3') + e_1^\mu(\kappa G_4' + \eta G_3' + \xi G_2'),
\end{align}
where we omitted the arguments of the $G_j$ and all derivatives are understood to act on the first argument only. The $e_i^\mu$ are the four-dimensional unit vectors. Hence, the Ward identity is equivalent to 
\begin{align}
G_4' = G_3' = G_2' = 0,
\end{align}
and the vector integrals take the form
\begin{align}
\label{eq:ScalingParam}
x_{13}^{2\alpha} x_{24}^{2\beta} I_{\alpha \beta}^{\mu, n} = \sum_{j=2}^4 \frac{x_{1j}^\mu}{x_{1j}^2} \bar{G}_j\Big(\sfrac{x_{12}^2}{x_{1j}}, \sfrac{x_{13}^2}{x_{1j}^2}, \sfrac{x_{14}^2}{x_{1j}^2}, \sfrac{x_{23}^2}{x_{1j}^2}, \sfrac{x_{24}^2}{x_{1j}^2}, \sfrac{x_{34}^2}{x_{1j}^2}\Big).
\end{align}
We now repeat the same form of argument for the dual conformal generator. We first reparametrise the coefficient functions appearing in \eqref{eq:ScalingParam} in terms of two conformal cross ratios and three scale-invariant ratios,
\begin{align}
\bar{G}_j\Big(\sfrac{x_{12}^2}{x_{1j}^2},\sfrac{x_{13}^2}{x_{1j}^2}, \sfrac{x_{14}^2}{x_{1j}^2}, \sfrac{x_{23}^2}{x_{1j}^2}, \sfrac{x_{24}^2}{x_{1j}^2}, \sfrac{x_{34}^2}{x_{1j}^2}\Big) &= G_j\Big(\sfrac{x_{23}^2}{x_{1j}^2}, \sfrac{x_{24}^2}{x_{1j}^2}, \sfrac{x_{34}^2}{x_{1j}^2} , \sfrac{x_{12}^2 x_{34}^2}{x_{13}^2 x_{24}^2}, \sfrac{x_{14}^2 x_{23}^2}{x_{13}^2 x_{24}^2}\Big).
\end{align}
Imposing the special conformal Ward identity \eqref{eq:InhomWard2App} and choosing the special configuration \eqref{eq:PoincareConfig} then implies a $4\times4$ matrix of differential equations for the $G_j$ which for generic values of the parameters is equivalent to 
\begin{align}
&G_j^{(1)} = G_j^{(2)} = G_j^{(3)} = 0, \text{ for all }j,\notag\\
&G_2 + G_3 + G_4 = - I_{\alpha \beta},
\end{align}
where $G_j^{(k)}$ denotes the derivative of $G_j$ with respect to its $k$-th argument. Hence the $G_j$ are really only functions of the conformal cross ratios and the vector integral decomposition takes the final form \eqref{eq:alphabetavectordecompApp}.


\section{Ward Identity Extras}\label{app:ExtraWard}
Here we present the Ward identities in terms of vector coefficients for the triple and quadruple ladder integrals.
\paragraph{Triple Ladder.}
We consider the correlator  \eqref{eq:I4albe} for $\alpha=3, \beta=1$ 
\begin{equation}
\left\langle\text{tr}(Z^3(x_1)\bar{X}(x_2)\bar{Z}^3(x_3)X(x_4))\right\rangle,
\end{equation}
which is represented by the triple ladder integral
\begin{equation}
I_{31}=\int\frac{\dd^4x_a}{\pi^2}\frac{\dd^4x_{b}}{\pi^2}\frac{\dd^4x_{c}}{\pi^2}\frac{1}{(x_{a1}^2x_{a3}^2x_{a4}^2)x_{ab}^2(x_{b1}^2x_{b3}^2)x_{bc}^2(x_{c1}^2x_{c2}^2x_{c3}^2)}.
\label{eq:tripleladder}
\end{equation}
Again specialising \eqref{eq:PhatIdentityCorr} to this case will lead to information about certain vector Feynman integrals, which represent correlators containing a descendant field. In this case there are a priori two independent vector integrals which can appear 
\begin{align}
&I_{31}^{\mu,3}=\int\frac{\dd^4x_a}{\pi^2}\frac{\dd^4x_{b}}{\pi^2}\frac{\dd^4x_{c}}{\pi^2}\frac{x_{c1}^\mu}{(x_{a1}^2x_{a3}^2x_{a4}^2)x_{ab}^2(x_{b1}^2x_{b3}^2)x_{bc}^2(x_{c1}^4x_{c2}^2x_{c3}^2)},\\ 
&I_{31}^{\mu,2}=\int\frac{\dd^4x_a}{\pi^2}\frac{\dd^4x_{b}}{\pi^2}\frac{\dd^4x_{c}}{\pi^2}\frac{x_{b1}^\mu}{(x_{a1}^2x_{a3}^2x_{a4}^2)x_{ab}^2(x_{b1}^4x_{b3}^2)x_{bc}^2(x_{c1}^2x_{c2}^2x_{c3}^2)}.
\end{align}
These are independent in the sense that they cannot be mapped into each other under permutations of the external points. However upon computing 
\begin{equation}
H^\mu_{31}=2i\levo{P}^\mu |_{14}\left\langle\text{tr}(Z^3(x_1)\bar{X}(x_2)\bar{Z}^3(x_3)X(x_4))\right\rangle
\end{equation}
using  \eqref{eq:PhatIdentityCorr} we find that only $I_{31}^{\mu,3}$ contributes
\begin{equation}
H^\mu_{31}=(4I_{31}^{\mu,3}- x_2\leftrightarrow x_4)-x_1\leftrightarrow x_3.
\label{eq:Hmu31}
\end{equation}
We expand $I_{31}^{\mu,3}$ in a vector decomposition \eqref{eq:alphabetavectordecomp}
\begin{equation}
x_{13}^6x_{24}^2I_{31}^{\mu,3}=-\frac{x_{12}^\mu}{x_{12}^2}G_2(u,v)-\frac{x_{13}^\mu}{x_{13}^2}G_3(u,v)-\frac{x_{14}^\mu}{x_{14}^2}G_4(u,v).
\end{equation}  
Comparing \eqref{eq:Hmu31} and \eqref{eq:pmuhatBD} we find the constraint
\begin{equation}
u\Duv^{31} \phi_{31}=G_2(u,v)-G_4(v,u),
\label{eq:Yangianconstraint2}
\end{equation}
and the same equation with $u$ and $v$ swapped. Integral expressions for $G_i(u,v)$ are given in \appref{app:feynmanparam}, which were used to numerically confirm \eqref{eq:Yangianconstraint2}.

\paragraph{Quadruple Ladder.}
For $\alpha=4, \beta=1$ we have the correlator
\begin{equation}
\left\langle\text{tr}(Z^4(x_1)\bar{X}(x_2)\bar{Z}^4(x_3)X(x_4))\right\rangle,
\end{equation}
which is represented by the quadruple ladder integral
\begin{equation}
I_{41}=\int \frac{\dd^4 x_a}{\pi^2}\frac{\dd^4 x_{b}}{\pi^2}\frac{\dd^4 x_{c}}{\pi^2}\frac{\dd^4 x_{d}}{\pi^2}\frac{1}{(x_{a1}^2x_{a3}^2x_{a4}^2)x_{ab}^2(x_{b1}^2x_{b3}^2)x_{bc}^2(x_{c1}^2x_{c3}^2)x_{cd}^2(x_{d1}^2x_{d2}^2x_{d3}^2)}.
\end{equation}
In this case there are also two independent vector integrals which can appear:
\begin{align}
&I_{41}^{\mu,4}=\int \frac{\dd^4 x_a}{\pi^2}\frac{\dd^4 x_{b}}{\pi^2}\frac{\dd^4 x_{c}}{\pi^2}\frac{\dd^4 x_{d}}{\pi^2}\frac{x_{d1}^\mu}{(x_{a1}^2x_{a3}^2x_{a4}^2)x_{ab}^2(x_{b1}^2x_{b3}^2)x_{bc}^2(x_{c1}^2x_{c3}^2)x_{cd}^2(x_{d1}^4x_{d2}^2x_{d3}^2)},\label{eq:Ivec41}
\\
&I_{41}^{\mu,3}=\int \frac{\dd^4 x_a}{\pi^2}\frac{\dd^4 x_{b}}{\pi^2}\frac{\dd^4 x_{c}}{\pi^2}\frac{\dd^4 x_{d}}{\pi^2}\frac{x_{c1}^\mu}{(x_{a1}^2x_{a3}^2x_{a4}^2)x_{ab}^2(x_{b1}^2x_{b3}^2)x_{bc}^2(x_{c1}^4x_{c3}^2)x_{cd}^2(x_{d1}^2x_{d2}^2x_{d3}^2)}.
\label{eq:Ibvec41}\end{align}
We compute 
\begin{equation}
H^\mu_{41}=2i\levo{P}^\mu |_{14}\left\langle\text{tr}(Z^4(x_1)\bar{X}(x_2)\bar{Z}^4(x_3)X(x_4))\right\rangle
\end{equation}
using  \eqref{eq:PhatIdentityCorr} and this time we find that both vector integrals contribute
\begin{align}
H^\mu_{41}&=(6I^{\mu,4}_{41}+2I^{\mu,3}_{41}+(2I^{\mu,3}_{41}+x_2\leftrightarrow x_4)-x_2\leftrightarrow x_4)-x_1\leftrightarrow x_3\notag\\
&=(6I^{\mu,4}_{41}+2I^{\mu,3}_{41}-x_2\leftrightarrow x_4)-x_1\leftrightarrow x_3. \label{eq:Hmu41}
\end{align}
We decompose \eqref{eq:Ivec41} and \eqref{eq:Ibvec41} as  \eqref{eq:alphabetavectordecomp}
\begin{align}
&x_{13}^8x_{24}^2I_{41}^{\mu,4}=-\frac{x_{12}^\mu}{x_{12}^2}V_2(u,v)-\frac{x_{13}^\mu}{x_{13}^2}V_3(u,v)-\frac{x_{14}^\mu}{x_{14}^2}V_4(u,v),\\
&x_{13}^8x_{24}^2 I_{41}^{\mu,3}=-\frac{x_{12}^\mu}{x_{12}^2}\bar{V}_2(u,v)-\frac{x_{13}^\mu}{x_{13}^2}\bar{V}_3(u,v)-\frac{x_{14}^\mu}{x_{14}^2}\bar{V}_4(u,v).
\end{align}
Then comparing \eqref{eq:Hmu41} and \eqref{eq:pmuhatBD} leads to the following constraint between $V_i(u,v)$ and $\bar{V}_i(u,v)$
\begin{equation}
2u\Duv^{41} \phi_{41}=3(V_2(u,v)-V_4(v,u))+\bar{V}_2(u,v)-\bar{V}_4(v,u),
\label{eq:DuvVQuadruple}
\end{equation}
and the same equation with $u$ and $v$ swapped. This was numerically verified using the Feynman parametrisations in  \appref{app:feynmanparam}.


\section{Conformal Feynman Parametrisations}\label{app:feynmanparam}
In this appendix we give explicit Feynman parametrisations for scalar ladder integrals and the vector ladder integrals appearing in \secref{correlatorfeynman} and \appref{app:ExtraWard}. We show some details of the calculation for the vector double ladder integral but otherwise we just state the results.
\subsection{Scalar Ladders}
We give conformal Feynman parametrisations for the scalar ladders $\phi_{\ell 1}$, obtained as specialisations of \eqref{eq:I4albe} and \eqref{eq:ScalarIntegrals}, and cases with more general propagator powers.
\paragraph{Box.}
The Feynman parametrisation for the conformal box function reads
\begin{equation}
\phi_{11}(u,v)=\int_0^{\infty} \dd^2 \alpha \frac{1}{(1+\alpha_1 u + \alpha_2 v )D_{\alpha_1\alpha_2}},
\end{equation}
where $D_{\alpha_1\alpha_2}\coloneqq \alpha_1+\alpha_2+\alpha_1\alpha_2$. For the modified box defined in \eqref{eq:BoxOmegaD} we have
\begin{equation} \label{eq:FeynmanParamBoxOmegaD}
\phi_{11}^{\omega D}(u,v)=\frac{1}{\Gamma_{\omega}\Gamma_{D/2-\omega}}\int_0^\infty \dd^2\alpha \frac{(\alpha_1\alpha_2)^{\omega-1}}{(1+\alpha_1 u+\alpha_2 v)^{D/2-\omega}(D_{\alpha_1\alpha_2})^{\omega}}.
\end{equation}

\paragraph{Double Ladder.}
For the double ladder, the conformal function is Feynman parametrised as
\begin{equation}
\phi_{21}(u,v)=\int_0^\infty \dd^2\beta \int_0^\beta \dd^2\alpha \frac{1}{(1+\beta_1 u+\beta_2 v)D_{\beta_1\beta_2}D_{\alpha_1\alpha_2}},
\end{equation}
where $ \int_0^\beta \dd^2\alpha\coloneqq \int_0^{\beta_1} \dd\alpha_1 \int_0^{\beta_2 }\dd \alpha_2$. For the modified double ladder defined in \eqref{eq:doubleladderOmegaD} we have
\begin{equation} \label{eq:FeynmanParamDoubleLadderOmegaD}
\phi_{21}^{\omega D}(u,v)=\frac{1}{\Gamma^2_{\omega}\Gamma^2_{D/2-\omega}}\int_0^\infty \dd^2\beta \int_0^\beta \dd^2\alpha \frac{[(\beta_1-\alpha_1)(\beta_2-\alpha_2)\alpha_1\alpha_2]^{\omega-1}}{(1+\beta_1 u+\beta_2 v)^{D/2-\omega}(D_{\beta_1\beta_2}D_{\alpha_1\alpha_2})^{\omega}}.
\end{equation}
The double ladder for fully generic conformal propagator powers \eqref{eq:DoubleLadderGenFormula} can be written as
\begin{equation}
I_{21}^{\nu,D}=\includegraphicsbox{FigDoubleLadderFeynmanParameters.pdf}=V_{2,1}^{\nu,D}\phi_{21}^{\nu,D}(u,v),
\end{equation}
where $V_{2,1}^{\nu,D}=x_{12}^{-\nu_1-\nu_2-\nu_3+\nu_4+\nu_5+\nu_6}x_{14}^{-\nu_1-\nu_2-\nu_6+\nu_3+\nu_4+\nu_5}x_{13}^{-\nu_4-\nu_5}x_{24}^{-\nu_3-\nu_4-\nu_5-\nu_6+\nu_1+\nu_2}$. The conformal function can be Feynman parametrised as
 \begin{equation}\label{eq:FeynmanParamDoubleLadderNuD}
\phi_{21}^{\nu, D}(u,v)=c_{21}^{\nu,D}\int_0^\infty \dd^2\beta \int_0^\beta \dd^2\alpha \frac{(\beta_1-\alpha_1)^{\nu_5-1}(\beta_2-\alpha_2)^{\nu_1-1}\alpha_1^{\nu_4-1}\alpha_2^{\nu_2-1}}{(1+\beta_1 u+\beta_2 v)^{\nu_6}(D_{\beta_1\beta_2})^{D/2-\nu_6}(D_{\alpha_1\alpha_2})^{D/2-\nu_7}},
\end{equation}
 where the prefactor is written as
 \begin{equation}
 c_{21}^{\nu,D}=v^{\nu_2-\nu_5}\frac{\Gamma_{D/2-\nu_7}\Gamma_{D/2-\nu_6}}{\prod_{i\neq 6}\Gamma_{\nu_i}}.
 \end{equation}

\paragraph{Triple Ladder.}
The triple ladder Feynman parametrisation is given by
\begin{equation}
\phi_{31}(u,v)=\int_0^{\infty}\dd^2\gamma\int_{0}^\gamma\dd^2\beta\int_{0}^\beta\dd^2\alpha\frac{1}{(1+\gamma_1u+\gamma_2v)D_{\gamma_1\gamma_2}D_{\beta_1\beta_2}D_{\alpha_1\alpha_2}}.
\end{equation}
For the modified triple ladder defined in \eqref{eq:tripleladderOmegaD} we have
\begin{equation}\label{eq:FeynmanParamTripleLadderOmegaD}
\phi_{31}^{\omega D}(u,v)=\frac{1}{\Gamma_\omega^3\Gamma_{\bar{\omega}}^3}\int_0^{\infty}\dd^2\gamma\int_{0}^\gamma\dd^2\beta\int_{0}^\beta\dd^2\alpha\frac{[(\gamma_1-\beta_1)(\gamma_2-\beta_2)(\beta_1-\alpha_1)(\beta_2-\alpha_2)\alpha_1\alpha_2]^{\omega-1}}{(1+\gamma_1u+\gamma_2v)^{\bar{\omega}}(D_{\gamma_1\gamma_2}D_{\beta_1\beta_2}D_{\alpha_1\alpha_2})^\omega},
\end{equation}
where we remind $\bar{\omega}=D/2-\omega$.
\paragraph{$\ell$-ladder.}
The pattern persists for general $\ell$-ladders whose conformal functions are Feynman parametrised by
\begin{equation}
\phi_{\ell 1}(u,v)=\int_0^{\infty} \dd^2 \alpha_n \left(\prod_{i=1}^{n-1}\int_0^{\alpha_{i+1}} \dd^2 \alpha_i\right) \frac{1}{(1+u\alpha_{n,1}+v\alpha_{n,2})(\prod_{j=1}^{n}D_{\alpha_{j,1}\alpha_{j,2}})},
\end{equation}
and in the generalised case we have
\begin{equation}
\phi_{\ell 1}^{\omega D}=\frac{1}{\Gamma_\omega^\ell\Gamma_{\bar{\omega}}^\ell}\int_0^{\infty} \dd^2 \alpha_n \left(\prod_{i=1}^{n-1}\int_0^{\alpha_{i+1}} \dd^2 \alpha_i\right) \frac{\brk[s]*{\prod\limits_{k=2}^{\ell}(\alpha_{k,1}-\alpha_{k-1,1})(\alpha_{k,2}-\alpha_{k-1,2})\alpha_{1,1}\alpha_{1,2}}^{\omega-1}}{(1+u\alpha_{n,1}+v\alpha_{n,2})^{\bar{\omega}}(\prod_{j=1}^{n}D_{\alpha_{j,1}\alpha_{j,2}})^{\omega}}.
\end{equation}


\subsection{Vector Ladders}
Here we provide Feynman parametrisations for the vector ladder integrals.

\paragraph{Double Ladder.} We derive the conformal Feynman parametrisations of the vector integral coefficients of $I_{21}^{\mu,2}$, defined in \eqref{eq:vecdoubleladderdef}. We start with
\begin{align}x_{13}^4x_{24}^2I_{21}^{\mu,2}&=\int \frac{\dd^4 x_a}{\pi^2} \frac{\dd^4 x_{b}}{\pi^2}\frac{x_{13}^4x_{24}^2x_{ b1}^\mu}{x_{a1}^2 x_{a3}^2 x_{a4}^2x_{ab}^2x_{b1}^4x_{b2}^2 x_{b3}^2   }\notag\\&=\int \frac{\dd^4 x_b}{\pi^2}\frac{x_{13}^4x_{24}^2x_{b1}^{\mu}}{x_{b1}^4x_{b2}^2x_{b3}^2}\int \frac{\dd^4 x_{a}}{\pi^2}\frac{1}{x_{ab}^2x_{a1}^2x_{a3}^2x_{a4}^2}.
\end{align}
The second integral is just a scalar box. With Feynman parameters (and subsequently evaluating one of the parameter integrals) it evaluates to \cite{Hodges:2010kq}
\begin{equation}
\int \frac{\dd^4 x_{a}}{\pi^2}\frac{1}{x_{ab}^2x_{a1}^2x_{a3}^2x_{a4}^2}=\int_{0}^{\infty} [\dd^2\alpha ]\frac{1}{(\alpha_1x_{b1}^2+\alpha_3x_{b3}^2+\alpha_4x_{b4}^2)(\alpha_1\alpha_3x_{13}^2+\alpha_1\alpha_4x_{14}^2+\alpha_3\alpha_4x_{34}^2)},
\end{equation}
where $[\dd^2\alpha ]$ denotes a projective integral over $\alpha_1,\alpha_3,\alpha_4$. For example we could take $[\dd^2\alpha ]=\dd\alpha_1\dd\alpha_3\delta(\alpha_4-1)$. The second factor of the integrand does not depend on $x_b$ so we omit it for now. We need to compute
\begin{equation}
\int \frac{\dd^4 x_b}{\pi^2}\frac{x_{b1}^{\mu}}{x_{b1}^4x_{b2}^2x_{b3}^2(\alpha_1x_{b1}^2+\alpha_3x_{b3}^2+\alpha_4x_{b4}^2)},
\label{eq:intvec}
\end{equation}
where we freely exchange loop and parametric integrals. Using Feynman parameters this is
\begin{equation}
\frac{\Gamma_5}{\pi^2}\int [\dd^3\gamma]\int \dd^4x_b\frac{\gamma_1 x_{b1}^{\mu}}{(\gamma_1x_{b1}^2+\gamma_2x_{b2}^2+\gamma_3x_{b3}^2+\gamma_4(\alpha_1x_{b1}^2+\alpha_3x_{b3}^2+\alpha_4x_{b4}^2))^5},
\label{eq:int5}
\end{equation}
The denominator of \eqref{eq:int5} can be written as
\begin{equation}
\sigma^5\left(l^2+\frac{\Delta}{\sigma^2}\right)^5,
\end{equation}
where
\begin{align}
\sigma=& \gamma_1+\gamma_2+\gamma_3+\gamma_4(\alpha_1+\alpha_3+\alpha_4),\\ \notag
l^\mu=& x_b^{\mu}-\frac{x_1^\mu(\gamma_1+\gamma_4\alpha_1)+x_2^\mu\gamma_2+x_3^\mu(\gamma_3+\alpha_3\gamma_4)+x_4^\mu\gamma_4\alpha_4}{\sigma},\\ \notag
\Delta=& (\gamma_1+\gamma_4\alpha_1)\gamma_2x_{12}^2+(\gamma_1+\gamma_4\alpha_1)(\gamma_3+\gamma_4\alpha_3)x_{13}^2+(\gamma_1+\gamma_4\alpha_1)\gamma_4\alpha_4x_{14}^2
\\ \notag& +\gamma_2(\gamma_3+\gamma_4\alpha_3)x_{23}^2+\gamma_2\gamma_4\alpha_4x_{24}^2+(\gamma_3+\gamma_4\alpha_3)\gamma_4\alpha_4x_{34}^2.
\end{align}
The numerator $x_{b1}^\mu$ can be written
\begin{equation}
x_{b1}^\mu=l^\mu-\frac{1}{\sigma}\sum_{i=2}^4B_ix_{1i}^\mu,
\end{equation}
where
\begin{align}
 B_2=\gamma_2, \qquad B_3=\gamma_3+\gamma_4\alpha_3, \qquad B_4=\gamma_4\alpha_4.
\end{align}
The integral over the $l^\mu$ piece vanishes because it is an odd integrand, so \eqref{eq:intvec} reduces to
\begin{equation}
-\sum_{i=2}^4 x_{1i}^\mu \int_0^\infty[\dd^3\gamma]\int_0^\infty \dd l \frac{24\gamma_1 l B_i}{\sigma^6(l^2+\frac{\Delta}{\sigma^2})},
\end{equation}
where we integrate over $l\coloneqq |l^\mu|$. Performing the integral over $l$ leads to 
\begin{equation}
\frac{1}{2}x_{13}^4x_{24}^2I_{21}^{\mu,2}=-\sum_{i=2}^4 x_{1i}^{\mu}\int [\dd^2\alpha][\dd^3\gamma]\frac{x_{13}^4x_{24}^2\gamma_1B_i}{(\alpha_1\alpha_3x_{13}^2+\alpha_1\alpha_4x_{14}^2+\alpha_3\alpha_4x_{34}^2)\Delta^3}.
\end{equation}
We choose to perform the $\gamma_2$ integral, and make the rescaling of Feynman parameters
 \begin{align}
 \alpha_1&\rightarrow x_{34}^2\alpha_1,   
& \alpha_3&\rightarrow x_{14}^2\alpha_3,   
 &\alpha_4&\rightarrow x_{13}^2\alpha_4,\notag\\
 \gamma_1&\rightarrow \frac{x_{34}^2}{x_{13}^2x_{24}^2}\gamma_1,   
 &\gamma_3&\rightarrow \frac{x_{14}^2}{x_{13}^2x_{24}^2}\gamma_3,   
 &\gamma_4&\rightarrow  \frac{1}{x_{13}^2x_{24}^2}\gamma_4.
 \end{align}
This rescaling ensures that the resulting integrands are manifestly conformal invariant \cite{Bourjaily:2019jrk}. We finally de-project $\alpha_4=\gamma_4=1$ and make the change of variables $\gamma_i=\beta_i-\alpha_i$ for $i=1,3$. This leads to the required form
\begin{equation}
x_{13}^4x_{24}^2I_{21}^{\mu,2}=-\frac{x_{12}^\mu}{x_{12}^2}F_2(u,v)-\frac{x_{13}^\mu}{x_{13}^2}F_3(u,v)-\frac{x_{14}^\mu}{x_{14}^2}F_4(u,v)
\label{eq:21vectordecomp2},
\end{equation}
with (we further relabel $\alpha_3,\beta_3\rightarrow \alpha_2,\beta_2$)
\begin{align}
&F_2(u,v)=u\int_0^\infty \dd^2\beta\int_{0}^\beta \dd^2\alpha \frac{\beta_1-\alpha_1}{(1+\beta_1 u + \beta_2 v)^2D_{\beta_1\beta_2}D_{\alpha_1\alpha_2}},\label{eq:F2}\\
&F_3(u,v)=\int_0^\infty \dd^2\beta\int_{0}^\beta \dd^2\alpha \frac{\beta_2(\beta_1-\alpha_1)}{(1+\beta_1 u + \beta_2 v)(D_{\beta_1\beta_2})^2D_{\alpha_1\alpha_2}},\label{eq:F3}\\
&F_4(u,v)=\int_0^\infty \dd^2\beta\int_{0}^\beta \dd^2\alpha \frac{\beta_1-\alpha_1}{(1+\beta_1 u + \beta_2 v)(D_{\beta_1\beta_2})^2D_{\alpha_1\alpha_2}}, \label{eq:F4}
\end{align}
where we remind that $\int_{0}^\beta \dd^2\alpha$ is shorthand for $\int_{0}^{\beta_1} \dd\alpha_1\int_{0}^{\beta_2} \dd\alpha_2$ and $D_{\alpha_1\alpha_2}=\alpha_1+\alpha_2+\alpha_1\alpha_2$. 

\paragraph{Triple Ladder.}
Here we have
\begin{align}
&x_{13}^6x_{24}^2I_{31}^{\mu,3}=-\frac{x_{12}^\mu}{x_{12}^2}G_2(u,v)-\frac{x_{13}^\mu}{x_{13}^2}G_3(u,v)-\frac{x_{14}^\mu}{x_{14}^2}G_4(u,v), \\ \notag
&x_{13}^6x_{24}^2I_{31}^{\mu,2}=-\frac{x_{12}^\mu}{x_{12}^2}\bar{G}_2(u,v)-\frac{x_{13}^\mu}{x_{13}^2}\bar{G}_3(u,v)-\frac{x_{14}^\mu}{x_{14}^2}\bar{G}_4(u,v),
\end{align} 
where
\begin{align}
&G_2(u,v)=\int_0^{\infty}\dd^2\gamma\int_{0}^\gamma\dd^2\beta\int_{0}^\beta\dd^2\alpha\frac{u(\gamma_1-\beta_1)}{(1+u\gamma_1+v\gamma_2)^2D_{\gamma_1\gamma_2}D_{\beta_1\beta_2}D_{\alpha_1\alpha_2}},\\ \notag
&G_3(u,v)=\int_0^{\infty}\dd^2\gamma\int_{0}^\gamma\dd^2\beta\int_{0}^\beta\dd^2\alpha\frac{\gamma_2(\gamma_1-\beta_1)}{(1+u\gamma_1+v\gamma_2)D^2_{\gamma_1\gamma_2}D_{\beta_1\beta_2}D_{\alpha_1\alpha_2}},\\ \notag
&G_4(u,v)=\int_0^{\infty}\dd^2\gamma\int_{0}^\gamma\dd^2\beta\int_{0}^\beta\dd^2\alpha\frac{\gamma_1-\beta_1}{(1+u\gamma_1+v\gamma_2)D^2_{\gamma_1\gamma_2}D_{\beta_1\beta_2}D_{\alpha_1\alpha_2}},
\end{align}
For $\bar{G}_i$ we have
\begin{equation}
\bar{G}_i=\int_0^\infty[\dd^2\alpha][\dd^2\beta][\dd^2\gamma]\frac{E_i}{\Lambda_1\Lambda_2\Lambda_3^2\Lambda_4},
\end{equation}
where
\begin{align}
&E_2=u\gamma\beta_4,&
&E_3=\gamma_3+u\beta_3\gamma+v\alpha_3\bar{\gamma},&
&E_4=v\bar{\gamma}\alpha_2,
\end{align}
and
\begin{align}
&\Lambda_1=\alpha_1\alpha_2+\alpha_1\alpha_3+\alpha_2\alpha_3,\\ \notag
&\Lambda_2=\beta_1\beta_3+\beta_1\beta_4+\beta_3\beta_4,\\ \notag
&\Lambda_3=\gamma_3+v\bar{\gamma}(\alpha_2+\alpha_3)+u\gamma(\beta_3+\beta_4),\\ \notag
&\Lambda_4=(\gamma\beta_1+\bar{\gamma}\alpha_1)F_3+(\gamma_3+u\gamma\beta_3+v\bar{\gamma}\alpha_3)(\bar{\gamma}\alpha_2+\gamma\beta_4)+\gamma\bar{\gamma}\alpha_2\beta_4.
\end{align}

\paragraph{Quadruple Ladder.}
Here we can write
\begin{align}
&x_{13}^8x_{24}^2I_{41}^{\mu,4}=-\frac{x_{12}^\mu}{x_{12}^2}V_2(u,v)-\frac{x_{13}^\mu}{x_{13}^2}V_3(u,v)-\frac{x_{14}^\mu}{x_{14}^2}V_4(u,v), \\ \notag
&x_{13}^8x_{24}^2I_{41}^{\mu,3}=-\frac{x_{12}^\mu}{x_{12}^2}\bar{V}_2(u,v)-\frac{x_{13}^\mu}{x_{13}^2}\bar{V}_3(u,v)-\frac{x_{14}^\mu}{x_{14}^2}\bar{V}_4(u,v),
\end{align} 
where
\begin{align}
&V_2(u,v)=\int_0^\infty\dd^2\delta\int_0^{\delta}\dd^2\gamma\int_{0}^\gamma\dd^2\beta\int_{0}^\beta\dd^2\alpha\frac{u(\delta_1-\gamma_1)}{(1+u\delta_1+v\delta_2)^2D_{\delta_1\delta_2}D_{\gamma_1\gamma_2}D_{\beta_1\beta_2}D_{\alpha_1\alpha_2}},\\ \notag
&V_3(u,v)=\int_0^\infty\dd^2\delta\int_0^{\delta}\dd^2\gamma\int_{0}^\gamma\dd^2\beta\int_{0}^\beta\dd^2\alpha\frac{\delta_2(\delta_1-\gamma_1)}{(1+u\delta_1+v\delta_2)D^2_{\delta_1\delta_2}D_{\gamma_1\gamma_2}D_{\beta_1\beta_2}D_{\alpha_1\alpha_2}},\\ \notag
&V_4(u,v)=\int_0^\infty\dd^2\delta\int_0^{\delta}\dd^2\gamma\int_{0}^\gamma\dd^2\beta\int_{0}^\beta\dd^2\alpha\frac{(\delta_1-\gamma_1)}{(1+u\delta_1+v\delta_2)D^2_{\delta_1\delta_2}D_{\gamma_1\gamma_2}D_{\beta_1\beta_2}D_{\alpha_1\alpha_2}}.
\end{align}
For $\bar{V}_i$ we have
\begin{equation}
\bar{V}_i=\int_0^\infty[\dd^2\alpha][\dd^2\beta][\dd^2\gamma][\dd^2\delta]\frac{\tilde{E}_i}{\tilde{\Lambda}_1\tilde{\Lambda}_2\tilde{\Lambda}_3^2\tilde{\Lambda}_4\tilde{\Lambda}_5},
\end{equation}
where here
\begin{align}
&\tilde{E}_2=uv \bar{\delta}\gamma\beta_2,&
&\tilde{E}_3=v(\delta_3+v(\alpha_3\delta+\bar{\delta}\gamma_3)+u\bar{\delta}\gamma\beta_3),&
&\tilde{E}_4=v^2\delta\alpha_4,
\end{align}
and morevover
\begin{align}
&\tilde{\Lambda}_1=\alpha_1\alpha_3+\alpha_1\alpha_4+\alpha_3\alpha_4,\\ \notag
&\tilde{\Lambda}_2=\beta_1\beta_2+\beta_1\beta_3+\beta_2\beta_3,\\ \notag
&\tilde{\Lambda}_3=u\bar{\delta}\gamma(\beta_2+\beta_3)+v(\alpha_3\delta+\bar{\delta}\gamma_3+\alpha_4\delta)+\delta_3,\\ \notag
&\tilde{\Lambda}_4=\alpha _4 \beta _2 \gamma  \delta  \text{$\bar{\delta} $}+\beta _2 \gamma  \text{$\bar{\delta} $} u \left(\alpha _1 \delta
   +\text{$\bar{\delta} $} \left(\beta _1 \gamma +\gamma _1\right)\right)+\beta _2 \gamma  \text{$\bar{\delta} $}
   \left(\delta _3+\beta _3 \gamma  \text{$\bar{\delta} $} u+\alpha _3 \delta  v+\gamma _3 \text{$\bar{\delta} $}
   v\right)\notag\\ 
   &\qquad
   +\alpha _4 \delta  \left(\delta _3+\beta _3 \gamma  \text{$\bar{\delta} $} u+\alpha _3 \delta  v+\gamma _3
   \text{$\bar{\delta} $} v\right)+\alpha _4 \delta  v \left(\alpha _1 \delta +\text{$\bar{\delta} $} \left(\beta _1 \gamma
   +\gamma _1\right)\right)\notag\\ \notag
   &\qquad
   +\left(\alpha _1 \delta +\text{$\bar{\delta} $} \left(\beta _1 \gamma +\gamma
   _1\right)\right) \left(\delta _3+\beta _3 \gamma  \text{$\bar{\delta} $} u+\alpha _3 \delta  v+\gamma _3
   \text{$\bar{\delta} $} v\right),\\ \notag
&\tilde{\Lambda}_5=u \gamma((\gamma_1+\gamma\beta_1)(\beta_2+\beta_3)+\gamma\beta_2\beta_3)+v \gamma_3(\gamma_1+\gamma\beta_1+\gamma\beta_2).
\end{align}

\renewcommand{\thesection}{$\mathcal{D}$}
\section{Differential Operators in Terms of $z,\bar{z}$}\label{app:zzbdiff}
We provide formulas for some differential operators in terms of $z$ and $\bar{z}$. For instance $\partial_u$ and $\partial_v$ read
\begin{align}
&\frac{\partial }{\partial u}=\frac{-(1-z)\partial_z+(1-\bar{z})\partial_{\bar{z}}}{z-\bar{z}},&
&\frac{\partial }{\partial v}=\frac{-z\partial_z+\bar{z}\partial_{\bar{z}}}{z-\bar{z}}.
\end{align}
Then $\Duv^{\alpha\beta}$ and $\Dvu^{\alpha\beta}$ are given by
\begin{align}
&\Duv^{\alpha\beta}=\frac{\mathcal{D}_{1,z}^{\alpha\beta}-\mathcal{D}_{1,\bar{z}}^{\alpha\beta}}{z-\bar{z}},&
&\Dvu^{\alpha\beta}=\frac{\mathcal{D}_{2,z}^{\alpha\beta}-\mathcal{D}_{2,\bar{z}}^{\alpha\beta}}{z-\bar{z}},
\end{align}
where we have set
\begin{align}
\mathcal{D}_{1,z}^{\alpha\beta}=z\alpha\beta+(z-1)\left(z(1+\alpha+\beta)-\frac{\alpha+\beta}{2}\right)\partial_z+(z-1)^2z\partial^2_z,
\end{align}
\begin{align}
\mathcal{D}_{2,z}^{\alpha\beta}=z\alpha\beta+z\left(z(1+\alpha+\beta)-\frac{\alpha+\beta+2}{2}\right)\partial_z+(z-1)z^2\partial^2_z.
\end{align}

\bibliographystyle{nb}
\bibliography{YangianAndBDGraphs}

\end{document}